\PassOptionsToPackage{unicode}{hyperref}
\PassOptionsToPackage{hyphens}{url}
\PassOptionsToPackage{dvipsnames,svgnames,x11names}{xcolor}
\documentclass[
  12pt]{article}

\usepackage{iftex}
\ifPDFTeX
  \usepackage[T1]{fontenc}
  \usepackage[utf8]{inputenc}
  \usepackage{textcomp} 
\fi
\usepackage{lmodern}
\ifPDFTeX\else  
\fi
\IfFileExists{upquote.sty}{\usepackage{upquote}}{}
\IfFileExists{microtype.sty}{
  \usepackage[]{microtype}
  \UseMicrotypeSet[protrusion]{basicmath} 
}{}
\makeatletter
\@ifundefined{KOMAClassName}{
  \IfFileExists{parskip.sty}{%
    \usepackage{parskip}
  }{
    \setlength{\parindent}{0pt}
    \setlength{\parskip}{6pt plus 2pt minus 1pt}}
}{
  \KOMAoptions{parskip=half}}
\makeatother
\usepackage{xcolor}
\makeatletter
\ifx\paragraph\undefined\else
  \let\oldparagraph\paragraph
  \renewcommand{\paragraph}{
    \@ifstar
      \xxxParagraphStar
      \xxxParagraphNoStar
  }
  \newcommand{\xxxParagraphStar}[1]{\oldparagraph*{#1}\mbox{}}
  \newcommand{\xxxParagraphNoStar}[1]{\oldparagraph{#1}\mbox{}}
\fi
\ifx\subparagraph\undefined\else
  \let\oldsubparagraph\subparagraph
  \renewcommand{\subparagraph}{
    \@ifstar
      \xxxSubParagraphStar
      \xxxSubParagraphNoStar
  }
  \newcommand{\xxxSubParagraphStar}[1]{\oldsubparagraph*{#1}\mbox{}}
  \newcommand{\xxxSubParagraphNoStar}[1]{\oldsubparagraph{#1}\mbox{}}
\fi
\makeatother

\usepackage{longtable,booktabs,array}
\usepackage{calc} 
\usepackage{etoolbox}
\makeatletter
\patchcmd\longtable{\par}{\if@noskipsec\mbox{}\fi\par}{}{}
\makeatother
\IfFileExists{footnotehyper.sty}{\usepackage{footnotehyper}}{\usepackage{footnote}}
\makesavenoteenv{longtable}
\usepackage{graphicx}
\makeatletter
\def\maxwidth{\ifdim\Gin@nat@width>\linewidth\linewidth\else\Gin@nat@width\fi}
\def\maxheight{\ifdim\Gin@nat@height>\textheight\textheight\else\Gin@nat@height\fi}
\makeatother
\setkeys{Gin}{width=\maxwidth,height=\maxheight,keepaspectratio}
\makeatletter
\def\fps@figure{htbp}
\makeatother

\makeatletter
\@ifpackageloaded{caption}{}{\usepackage{caption}}
\AtBeginDocument{%
\ifdefined\contentsname
  \renewcommand*\contentsname{Table of contents}
\else
  \newcommand\contentsname{Table of contents}
\fi
\ifdefined\listfigurename
  \renewcommand*\listfigurename{List of Figures}
\else
  \newcommand\listfigurename{List of Figures}
\fi
\ifdefined\listtablename
  \renewcommand*\listtablename{List of Tables}
\else
  \newcommand\listtablename{List of Tables}
\fi
\ifdefined\figurename
  \renewcommand*\figurename{Figure}
\else
  \newcommand\figurename{Figure}
\fi
\ifdefined\tablename
  \renewcommand*\tablename{Table}
\else
  \newcommand\tablename{Table}
\fi
}
\@ifpackageloaded{float}{}{\usepackage{float}}
\floatstyle{ruled}
\@ifundefined{c@chapter}{\newfloat{codelisting}{h}{lop}}{\newfloat{codelisting}{h}{lop}[chapter]}
\floatname{codelisting}{Listing}

\makeatother
\makeatletter
\@ifpackageloaded{caption}{}{\usepackage{caption}}
\@ifpackageloaded{subcaption}{}{\usepackage{subcaption}}
\makeatother

\ifLuaTeX
  \usepackage{selnolig}  
\fi
\usepackage[]{natbib}
\usepackage{bookmark}

\IfFileExists{xurl.sty}{\usepackage{xurl}}{} 
\hypersetup{
  pdftitle={Title},
  pdfauthor={Author 1; Author 2},
  pdfkeywords={3 to 6 keywords, that do not appear in the title},
  colorlinks=true,
  linkcolor={blue},
  filecolor={Maroon},
  citecolor={Blue},
  urlcolor={Blue},
  pdfcreator={LaTeX via pandoc}}

\usepackage{amsmath,amssymb, amsfonts, amsthm, bm}
\usepackage{thm-restate}
\usepackage{enumitem}
\usepackage{fontawesome}

\newtheorem{theorem}{Theorem}
\newtheorem{lemma}{Lemma}
\newtheorem{proposition}{Proposition}
\newtheorem{definition}{Definition}
\newtheorem{assumption}{Condition}
\newtheorem{remark}{Remark}

\newtheorem{example}{Example}

\def\to{\rightarrow}

\def\D{\mathbf{D}}

\def\H{\mathbf{H}}
\def\h{\mathbf{h}}
\def\I{\mathbf{I}}

\def\P{\mathbf{P}}

\def\Q{\mathbf{Q}}
\def\q{\mathbf{q}}

\def\r{\mathbf{r}}

\def\T{\mathbf{T}}

\def\wt{\widetilde}
\def\wh{\widehat}

\newcommand{\ind}{\perp\!\!\!\!\perp}

\def\mE{\mathbb{E}}

\def\mG{\mathbb{G}}

\def\mH{\mathbb{H}}

\def\mP{\mathbb{P}}

\def\mR{\mathbb{R}}

\def\cF{\mathcal{F}}
\def\cG{\mathcal{G}}
\def\cH{\mathcal{H}}

\def\cL{\mathcal{L}}

\def\cN{\mathcal{N}}

\def\cS{\mathcal{S}}
\def\cT{\mathcal{T}}
\def\cU{\mathcal{U}}

\def\cW{\mathcal{W}}
\def\cX{\mathcal{X}}
\def\cY{\mathcal{Y}}

\def\diag{\mbox{diag}}

\newcommand{\RNum}[1]{\uppercase\expandafter{\romannumeral #1\relax}}

\def\mE{\mathbb{E}}

\allowdisplaybreaks[4]

\newcommand{\anon}{1}

\begin{document}

\def\spacingset#1{\renewcommand{\baselinestretch}%
{#1}\small\normalsize} \spacingset{1}


\if1\anon
{
  \title{\bf Discovering Causal Relationships using Proxy Variables under Unmeasured Confounding}
  \author{Yong Wu$^\star$, Yanwei Fu$^\dagger$, Shouyan Wang$^\star$, Yizhou Wang$^\ddagger$, Xinwei Sun$^\dagger$\\
    ~\\
    $^\star$\textit{Institute of Science and Technology for Brain-Inspired Intelligence,}\\ \textit{Fudan University} \\
    $^\dagger$\textit{School of Data Science, Fudan University} \\
    $^\ddagger$\textit{School of Computer Science, Peking University}}
  \maketitle
} \fi



\bigskip
\begin{abstract}
Inferring causal relationships between variable pairs in the observational study is crucial but challenging, due to the presence of unmeasured confounding. While previous methods employed the negative controls to adjust for the confounding bias, they were either restricted to the discrete setting (\emph{i.e.}, all variables are discrete) or relied on strong assumptions for identification. To address these problems, we develop a general nonparametric approach that accommodates both discrete and continuous settings for testing causal hypothesis under unmeasured confounders. By using only a single negative control outcome (NCO), we establish a new identification result based on a newly proposed integral equation that links the outcome and NCO, requiring only the completeness and mild regularity conditions. We then propose a kernel-based testing procedure that is more efficient than existing moment-restriction methods. We derive the asymptotic level and power properties for our tests. Furthermore, we examine cases where our procedure using only NCO fails to achieve identification, and introduce a new procedure that incorporates a negative control exposure (NCE) to restore identifiability. We demonstrate the effectiveness of our approach through extensive simulations and real-world data from the \emph{Intensive Care Data} and \emph{World Values Survey}.
\end{abstract}

\noindent%
{\it Keywords:} causal hypothesis testing, unmeasured confounders, negative control outcome, integral solving
\vfill

\newpage
\spacingset{1.8} 

\section{Introduction}
\label{sec.intro}

\subsection{Motivation}
\label{sec.intro-motivate}

Discovering causal relationships is fundamentally important in various disciplines, including neurodegenerative disease \citep{young2018uncovering}, clinical care \citep{khetan2021mimicause}, and manufacturing system \citep{marazopoulou2016causal}. The goal is to infer a directed causal graph (DAG) among multiple variables \citep{pearl2009causality,spirtes2001causation}. Although randomized experiments are reliable for establishing causality, they are often expensive, unethical, or infeasible in practice. Consequently, there has been growing interest in uncovering causal relations from purely observational data. A central task in causal discovery is to test the causal null hypothesis \citep{miao2018identifying} of the form $\mH_0: X \ind Y|U$, which assesses whether the exposure $X$ causally influences the outcome $Y$
given the potential confounding set $U$.

Under the Markovian assumption, \emph{i.e.}, there are no unmeasured confounding, the problem reduces to the conditional independence testing. Many tools can be employed for this purpose, including traditional Fisher $Z$-test \citep{fisher1921probable}, the Chi-Square test \citep{tallarida1987chi}, and kernel-based methods \citep{fukumizu2007kernel,zhang2012kernel,cai2022distribution}, and methods based on generative models \citep{bellot2019conditional, shi2021double}. In practice, however, it is often impossible to observe all potential confounders, and the presence of unmeasured confounding can lead to spurious causal discoveries. To mitigate the confounding bias, instrumental variable (IV) methods have been widely adopted \citep{davey2014mendelian, lousdal2018introduction, xue2020inferring, chen2024discovery, li2024discovery}. Yet, these techniques typically depend on restrictive parametric assumptions—such as linearity or Gaussian errors—that seldom hold in complex real-world systems.

Another line of research has explored the use of proxy variables—also known as negative control outcomes (NCOs) or negative control exposures (NCEs)—as substitutes or noisy measurements of latent confounders to test the causal null hypothesis \citep{kuroki2014measurement, miao2018identifying, liu2023causal, miao2023identifying, wubivariate}. Specifically, \citet{miao2018identifying} proposed testing the residuals from linear regressions between probability matrices, establishing the limiting null distribution under the discrete setting. Later, \citet{liu2023causal} extended this approach to continuous variables by discretizing them into bins and applying the same procedure. However, this extension is not sample-efficient since its asymptotic validity requires the number of bins to diverge. Other recent approaches \citep{miao2023identifying, wubivariate} have addressed continuous settings directly, but at the cost of strong identifiability conditions—for instance, assuming that latent confounders are identifiable up to invertible transformations \citep{miao2023identifying}. In summary, these methods are often restricted to specific settings or assumptions. These restrictions highlight the need for a unified and principled framework that remains valid in continuous, discrete, or mixed data and under weaker assumptions.


\subsection{Our contributions}
\label{sec.intro-contribution}

We develop a general non-parametric framework that can efficiently examine the causal null hypothesis in both continuous and discrete settings in the presence of unobserved confounders. Our approach provides a new perspective for identifying and testing causal relationships. We summarize our several major contributions as follows.

First, we establish a novel identification results which only requires a single negative control outcome. The identifiability is based on a newly proposed integral equation \eqref{eq.bridge_Y} between the probability function over the outcome and that over the NCO. We can demonstrate that if the null hypothesis holds—\emph{i.e.}, $Y$ depends on $X$ only through confounders $U$—then the equation admits a square-integrable solution, implying that the variation of the outcome with respect to the exposure can be fully explained by that of the NCO with respect to the exposure. This result enables us to identify causal relationships in the continuous setting, under only completeness conditions and some regularity conditions, without requiring the stronger identifiability assumptions—such as the equivalence condition—imposed in prior work \citep{miao2023identifying}. To the best of our knowledge, this is the first characterization of causal relationships via the solvability of an integral equation. Moreover, our identifiability result holds for settings where variables are discrete, continuous, or of mixed type. In particular, it is compatible to the existing work in the discrete setting \citep{miao2018identifying}, in the sense that the integral equation reduces to the linear equation between probability matrices.  


Second, we propose a general nonparametric testing method called \emph{Proxy Maximum Characteristic Restriction} (PMCR) that can efficiently estimate the solution to the integral equation. Compared to previous first-order moment restriction methods \citep{mastouri2021proximal, kallus2021causal}, our approach can capture information across all-order moments by leveraging the characteristic function, thereby enhancing the power for detecting causal relationships. The proposed restriction leads to a kernel-based estimator in the continuous setting and a least-squares estimator in the discrete setting. We then construct test statistics from the residuals of the restriction equation and proposed a bootstrapped implementation. We establish the asymptotic validity and power properties for our proposed procedure.  



Finally, we study the failure cases of our method for causal identification. Specifically, we investigate the solvability of the integral equation under the alternative hypothesis, which combines with result that the solution exists under $\mH_0$, motivates its use for causal identification. We use the linear Gaussian setting to show that as long as the dependency between outcome and NCO is strong enough, the integral equation may admit a solution under alternatives, making it fail to determine whether the hypothesis holds. To amend this issue, we append our previous procedure with an additional restriction, by incorporating the negative control exposure—commonly used alongside NCOs in the literature \citep{miao2018identifying, eric2024proxyintro}—which can restore identifiability under the above failure cases.

\subsection{Organization}
\label{sec.intro-organize}

The rest of our article is organized as follows. Section \ref{sec.pre} introduces the set-up, notations, and briefly reviews previous methods with proxy variables for causal hypothesis testing. Section~\ref{sec. causal-discovery} establishes a new identification result with a newly proposed integral equation, and shows that it admits a solution under the null hypothesis. It also introduces the PMCR for estimation and constructing the testing statistics. Section \ref{sec. asymptotic_implementations} establishes the asymptotic properties of our statistics and introduces the Bootrapped implementations. Section~\ref{sec.fail} illustrates the non-identifiability relying solely on the NCO-based integral equation, and then introduces an extended procedure by incorporating the additional NCE. Section \ref{sec.exp} and \ref{sec.real_world} respectively demonstrates the validity and effectiveness of our procedures on synthetic data, and real world data from Intensive care data and World Valus Survey data. We conclude with a discussion in section~\ref{sec.con}, while all proofs and additional experiments are provided in the supplementary material.

\section{Set-up and background}
\label{sec.pre}

\begin{figure}[htbp]
\centering
\includegraphics[width=0.7\linewidth]{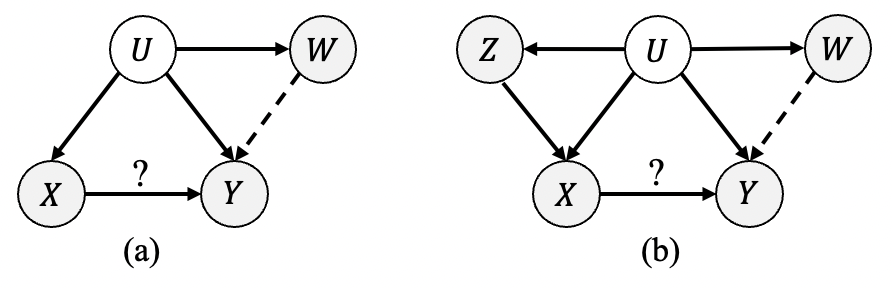}
\caption{Causal diagrams over $X,Y,U,W,Z$. $W$ (\emph{resp.} $Z$) denote the negative control outcome (\emph{resp.} exposure). The dotted line indicates its potential presence or absence.} 
\label{fig.dag}
\end{figure}

Our goal is to examine the causal null hypothesis $\mathbb{H}_0:X\ind Y|U$, where $X, Y, U$ denotes the exposure, outcome, and unmeasured confounders. Similar to proxy-variable methods, we assume the availability of a proxy variable $W$ such that $X\ind W|U$ \citep{kuroki2014measurement}, which also serves as the negative control outcome (NCO) in causal inference. Figure~\ref{fig.dag} (a) shows the causal diagram over $X,Y,U,W$. In some scenarios \citep{miao2018identifying, eric2024proxyintro}, we may also access to an additional proxy variable $Z$ \emph{i.e.}, negative control exposure (NCE), which satisfies $Z \ind (W,Y)| U,X$ as illustrated in Figure~\ref{fig.dag} (b). For technical clarity, we consider two parallel settings: all variables are either continuous or discrete. 

\textbf{Notations.} Suppose $X, Y, U, W, Z$ are random variables defined on the probability space $(\Omega,\mathcal{F},P)$, with state spaces $\mathcal{X}, \mathcal{Y}, \mathcal{U}, \mathcal{W}, \mathcal{Z}$, respectively. For any variable $U$, we denote $\mathcal{L}^2\{F(u)\}$ as the space of square-integrable functions with respect to the cumulative distribution function $F(u)$. For any space $\cW$, let $k_W$ be its positive semi-definite kernel. We denote $\phi_W$ as its associated canonical feature map, \emph{i.e.}, $\phi_W(w):=k_W(w,\cdot)$ for any $w \in \cW$. Besides, we denote $\cH_W$ as the corresponding reproducing kernel Hilbert space (RKHS). For any operator $A:\cH_W \to\cH_X$, we denote $A^*$ as its adjoint operator. For any discrete variables $X,Y$ with respectively $i,j$ categories, we denote $P(y|X):=\left\{P(y|x_1),...,P(y|x_i) \right\}$, the probability matrix $P(Y|X) := \left\{P(y_1|X)^\top,...,P(y_j|X)^\top\right\}^\top$. For any matrix $A$, we use $A^\dagger$ to denote the pseudo-inverse of $A$.

\textbf{Previous methods with proxy variables.} Previous procedures either considered the discrete setting \citep{miao2018identifying} or the continuous setting \citep{miao2023identifying, liu2023causal}, and they suffered from several limitations in either case. Specifically, suppose $X, Y, U, W, Z$ are
discrete variables and $X,Z,W$ respectively take $i,j,k$ categories, \citet{miao2018identifying} proposed to test $\mH_0$ by examining whether $P(W|Z,x)$ can fully explain the variability of $P(y|Z,x)$. Using the conditional independencies $W\ind (Z,X)|U$ and $Y\ind (Z,X)|U$ under $\mH_0$, it follows that for any fixed $(x,y)$,
$$
P(y|Z,x)=P(y|U)P(U|Z,x), \qquad P(W|Z,x)=P(W|U)P(U|Z,x). 
$$
By assuming that $P(W|U)$ is inverse, we can write $P(y|U)=P(y|U)P(W|U)^{-1}P(W|U)$ and obtain
$
P(y|Z,x)=P(y|U)P(W|U)^{-1}P(W|Z,x).
$
Based on this representation, \cite{miao2018identifying} performed a linear regression of $q_y:=\left\{P(y|Z,x_1),...,P(y|Z,x_i)\right\}^\top$ on $Q^\top:=\{P(W|Z,x_1),....,P(W|Z,x_i)\}^\top$, and tested the linearity based on the least-square residues. If $Q^\top$ is the full-column rank and $ij > k$, they derived the null-limiting distribution of the statistics based on the residues. However, this procedure was originally developed for discrete variables and may not generalize easily to the continuous setting. 

For continuous variables, \citet{liu2023causal, miao2023identifying} employed only $W$ for identification. Specifically, \citet{liu2023causal} first noticed that under the discrete case, we can turned to test the following linear relation without $Z$
\begin{equation}
\label{eq.discrete-approximate}
    P(y|x) = P(y| U)P(W| U)^{-1}P(W| x), 
\end{equation}
which allows us to test the linearity between $P(y| X)$ and $P(W| X)$ directly as long as $i>k$. Inspired by this, they first discretized $X, Y, W$ and proposed testing \eqref{eq.discrete-approximate} using the discrete variables. However, this procedure may not be sample efficient, as the asymptotic property was derived under the assumption that the number of bins diverges and there are sufficiently large samples within each bin to well approximate the probability matrix. On the other hand, \citet{miao2023identifying} proposed an integral equation for identification. However, this procedure requires the \emph{equivalence} condition, which means the latent $U$ is identifiable up to invertible transformation and may not hold in general cases. Besides, this procedure may not applicable to discrete cases.

In this paper, we propose a general procedure by investigating the solvability of the integral equation \eqref{eq.bridge_Y} of $p(y|x)$ with respect to $p(w|x)$. Under some completeness conditions, we can show the existence of solution under $\mH_0$, and derive the testing statistics based on the residue for solving this equation. Our identifiability result applies to variables that are continuous, discrete, or of mixed type. 

\section{Hypothesis testing with a single proxy}\label{sec. causal-discovery}
We first consider the scenario when only the proxy $W$ (\emph{i.e.} NCO) is available. To test the causal null hypothesis, we propose to examine whether the integral equation~\eqref{eq.bridge_Y} exists. To this end, section~\ref{sec. null hypothesis} first shows that under $\mH_0$, the solution exists under some completeness conditions. In particular, we will show that the formula derived from the integral equation generalizes the probability matrix formulation used in \cite{miao2018identifying} to the continuous setting. To estimate the solution, section~\ref{sec.equivalent} transforms the equation into a restriction problem, and estimate the solution using a kernel-based method.

\subsection{Solution existence under the null hypothesis}\label{sec. null hypothesis}

We propose to test $\mH_0$ with an integration equation~\eqref{eq.bridge_Y}. We will show that it holds under $\mH_0$. To this end, we require the completeness condition.

\begin{assumption}[Completeness of $P(U|W)$]
\label{assum.completeness}
     For any square-integrable function $g$, we assume $\mathbb{E}\{g(u)|w\} = 0$ almost surely if and only if $g(u) = 0$ almost surely.
\end{assumption}
Completeness is a standard assumption in causal hypothesis testing \citep{miao2018identifying,miao2023identifying,liu2023causal}. This condition is widely applicable, as shown by examples provided in \citep{newey2003instrumental,d2011completeness, hu2018nonparametric,andrews2017examples}. Here, it means $W$ carries all the variability of $U$, which holds generically as long as the dimension of $W$ is no less than that of $U$ \citep{andrews2017examples}. When $W,U$ are discrete with $i,j$ categories $(i>j)$, it means $P(W|U)$ is full-column rank, as used in \cite{liu2023causal} for identification. 
\begin{restatable}{proposition}{uxsolutionsexist}    \label{prop.bridge_U}
    Under condition~\ref{assum.completeness} and regularity conditions \ref{assum.regularity condition}, there exists a $h(w,y) \in \cL^2\{F(w)\}$ for all $y$, such that it solves the following integral equation for all $(y,u)$:
    \begin{equation}
    \label{eq.bridge_U}
        p(y|u) = \int h(w,y)p(w|u)dw. 
    \end{equation}
\end{restatable}
\begin{remark}
\label{remark.icml}
    We put some remarks about the connection of \eqref{eq.bridge_U} to that of existing works. When $W,X,U$ are discrete, the equation reduces to the form of 
    $$
    P(y|U) = P(y|U)P(W|U)^{-1}P(W|U),
    $$ 
    given that $P(W|U)$ is invertible. Besides, \cite{wubivariate} assumed the $p(u|x)=\int g(w,u)p(w|x)dw$ holds under some conditions. However, this equation may not be easy to hold. Specifically, by $p(w|x)=\int p(w|u')p(u'|x)du'$, we can obtain
    $$
    p(u|x) = \int K(u,u') p(u'|x) du', \quad \text{with } K(u,u') := \int g(w,u) p(w|u') dw.
    $$
    By Theorem~\ref{thm:kernel_identity} in Appendix~\ref{sec.explain-remark-1}, we must have $K(u,u')=\delta(u'-u)$. That means, the solution $g(w,u)$ and $p(w|u)$ must form an inverse operator, which is highly restrictive. In Appendix~\ref{appx.Bochner}, we show that this equation never holds under linear Gaussian models.
\end{remark}

The following theorem verifies that \eqref{eq.bridge_Y} admits a square-integrable solution.
\begin{restatable}{theorem}{solutionsexist}
\label{thm.exist_solution}
Suppose conditions in Proposition~\ref{prop.bridge_U} hold. Under $\mathbb H_0$, there exists $h(w,y) \in \cL^2\{F(w)\}$ for all $y$, such that it makes the following integral equation hold for all $(x,y)$: 
\begin{equation}\label{eq.bridge_Y}
    p(y|x) = \int h(w,y) p(w|x) dw.
\end{equation}
\end{restatable}
Intuitively, this equation holds under $\mH_0$ because the absence of the direct effect from $X$ to $Y$ allows $p(w|x)$ to fully explain away the variability of $p(y|x)$. In other words, it suggests rejecting $\mH_0$ when the discrepancy between $p(y|u,x)$ and $p(y|u)$ is sufficiently large to make $p(w|x)$ fail to account for all the variability encoded in $p(y|x)$. Notably, Theorem \ref{thm.exist_solution} is applicable to continuous, discrete, or mixed data type, as long as the completeness condition holds. In particular, when all variables are discrete, \eqref{eq.bridge_Y} reduces to an equation of a probability matrix, as previously established in \cite{miao2018identifying}.

\begin{restatable}{corollary}{DiscreteBridge}\label{cor:discrete_bridge}
Let $X, U, W, Y$ be discrete random variables with finite supports 
$|\mathcal{X}|, |\mathcal{U}|, |\mathcal{W}|, |\mathcal{Y}|$, respectively. We assume that their probability mass functions are strictly positive on their supports. Suppose condition~\ref{assum.completeness} holds. Then, under $\mH_0$, the integral equation in~\eqref{eq.bridge_Y} admits a solution of the form:
\begin{equation}\label{eq.discrete_bridge_Y}
    P(y | X)=\h(W,y)^\top P(W | X),
\end{equation}
where $\h(W,y) = \{P(W | U)^{\dagger} P(y | U)\}^\top$ is a $|\mathcal{W}|$-dimension vector. Moreover, if $P(W | U)$ is a square matrix, the solution is unique. 
\end{restatable}


\textbf{Connection to the tetrad constraint.} The tetrad constraint \citep{spearman1961general} was originally introduced to test whether $(X,Y,Z,W)$ are conditionally independent given $U$. In the classical linear model, this constraint takes the form $\sigma_{XY}\sigma_{ZW} = \sigma_{XZ}\sigma_{YW} = \sigma_{XW}\sigma_{YZ}$. As shown in \cite{ying2025generalized}, the first-moment formulation of \eqref{eq.bridge_Y} is equivalent to this classical tetrad constraint. Moreover, \cite{ying2025generalized} extended the use of this first-moment representation to nonlinear settings, employing it to test conditional independence. Their formulation can be viewed as a special case of our integral equation \eqref{eq.bridge_Y}, which captures the entire distributional relationship rather than only the first-moment information.

\subsection{Testing statistics via integral solving}\label{sec.equivalent}

In this section, we propose \emph{Proxy Maximum Characteristic Restriction} (PMCR) to estimate the solution, and use the residue to construct the testing statistics. This leads to a kernel-based estimator and least-square estimator in the continuous setting and the discrete setting, as will be respectively introduced in section~\ref{sec.test_continuous} and section~\ref{sec.test_discrete}. 



Previous studies considered the first-moment form of \eqref{eq.bridge_Y}, \emph{i.e.},  \emph{Maximum Moment Restriction} (MMR) \citep{mastouri2021proximal, kallus2021causal}. In our scenario, it involves solving $\overline{h}(W)$ from the following moment restriction:
\begin{equation}
\label{eq.bridge_Y_moment}
    \mE_{Y,W}\left\{ Y - \overline{h}(W) |X\right\} = 0.
\end{equation}
However, as it only leverages the first-order moment information, it will lose the testing power, as illustrated by example~\ref{example:linear_gaussian} in Appendix~\ref{appx.existence}, where \eqref{eq.bridge_Y_moment} holds under $\mH_1$. 
To impose more constraints for solving $h$, we leverage the characteristic function to construct the restriction, which can exploit all-order moment information. 

\textbf{Proxy Maximum Characteristic Restriction.} To test whether $p(y|x)$ equals to $\int h(w,y)p(w|x)dw$, we consider the following equation:
\begin{equation}\label{eq.convert_Y}
   \mathbb{E}_{Y,W}\{ \varphi(Y,t) -H(W,t) |X\} =0 \ \forall \thinspace t \in \cT,
\end{equation}
where we set $H(W,t)$ as $\int \varphi(y,t)h(w,y)dy$ to make \eqref{eq.convert_Y} holds. A common choice for $\varphi(Y,t)$ is $\exp(ity)$, where $\cT$ can be an arbitrarily chosen neighborhood around $0$. In this case, $\mE_Y\{\varphi(Y,t)\}$ is the characteristic function, and we hence call \eqref{eq.convert_Y} the \emph{Proxy Maximum Characteristic Restriction}. Since the characteristic function can uniquely determine the probability density and hence all order moments, solving \eqref{eq.convert_Y} offers greater utility to identify causal relations. In practice, we can also set $\varphi(Y,t) = \sin(ty)$ or $\cos(ty)$, and test whether \eqref{eq.convert_Y} holds for these choices.  


Further, corollary~\ref{corollary.exist_solution} further establishes that $H(w,t)$ is square-integrable with respect to $\mathcal{L}^2\{F(w)\}$ for all $t$, thereby guaranteeing that \eqref{eq.convert_Y} admits a solution within $\mathcal{L}^2\{F(w)\}$. 


\begin{restatable}{corollary}{solutioncossexist}
\label{corollary.exist_solution}
Suppose conditions in Theorem~\ref{thm.exist_solution} hold. Assume further that $h: \cY \mapsto \cL^2\{F(w)\}$ is Bochner integrable, \emph{i.e.}, $\int \|h(w,y)\|_{\cL^2\{F(w)\}}dy<\infty$. Then, for any $t$, $H(w,t)$ in \eqref{eq.convert_Y} exists and belongs to $\mathcal{L}^2\{F(w)\}$.
\end{restatable}
\begin{remark}
Intuitively, Bochner integrability (see Definition A.5.20 in \cite{steinwart2008support}) of $h$ guarantees that the Fourier-type transform $H(w,t)$ is well-defined pointwise in $t$ and belongs to $\cL^2\{F(w)\}$. It controls the magnitude of $h(w,y)$ in the $\cL^2\{F(w)\}$ norm, ensuring integrability over $y$. Similar conditions are common in functional data analysis and kernel methods \citep{jeon2020additive,mastouri2021proximal}. The condition is satisfied in a wide range of models. When all variables are discrete, it holds automatically. For continuous variables, Appendix~\ref{appx.Bochner} shows that this condition holds under the linear Gaussian model. 
\end{remark}

\textbf{Integral equation \eqref{eq.convert_Y} in the discrete case.} When all variables are discrete, \eqref{eq.convert_Y} reduces to a finite-dimensional system of linear equations. Specifically, let $\mathcal{X}, \mathcal{W}, \mathcal{Y}$ denote the supports of $X, W, Y$, respectively, \eqref{eq.convert_Y} becomes
\begin{equation}
\label{eq.convert_Y_dis}
\sum_{y\in\mathcal{Y}}\varphi(y,t)P(y|X)= \H^\top(W,t)P(W|X), \ \forall \thinspace t \in \cT,
\end{equation}
where $\H(W,t)=\sum_{y\in\mathcal{Y}}\varphi(y,t)\h(W,y)$, with $\h(W,y)=(h(w,y): w\in\mathcal{W})^\top \in \mR^{|\mathcal{W}|}$. If we set $\varphi(Y,t)$ as a set of functions $\{1(Y=y): y \in \cY\}$, \eqref{eq.convert_Y_dis} corresponds to the linear equation equation in \citet{miao2018identifying}.

In what follows, we will present our test statistics in the continuous and discrete cases \eqref{eq.convert_Y}, respectively. 

\subsubsection{Testing for continuous variables}
\label{sec.test_continuous}

\cite{horowitz2012specification} have shown a impossibility result of achieving uniform consistency by testing the existence of a solution. We hence require some certain smoothness conditions that enable us to solve the equation. Following existing studies \citep{mastouri2021proximal, ghassami2022minimax}, we assume the solution belongs to the reproducing kernel Hilbert space (RKHS) denoted by $\mathcal{H}_W$. 


\begin{assumption}[Smoothness]\label{picard_series}
let $k_W$ be the reproducing kernel for the RKHS $\cH_W$. By spectral theorem, its eigvenvalue decomposition has the form of $k_W(w,w')=\sum_{j=1}^\infty\eta_j\varphi_j(w)\varphi_j(w')$, where $\{\varphi_j\}_j$ is the orthonormal basis of $\cL^2\{F(w)\}$. 
For $H(W, t) $ in \eqref{eq.convert_Y}, we assume 
$$
H(W, t) \in \cH_W:=\left\{H\in \cL^2\{F(w)\}\left|\sum_{i=1}^\infty\frac{\left\langle H,\varphi_i\right\rangle_{\cL^2\{F(w)\}}^2}{\eta_i}<\infty\right\}\right. \ \text{for all $t $.}
$$
This means there exists a solution within the RKHS that satisfies \eqref{eq.convert_Y}.  
\end{assumption}

Let $\cH_{W,0}$ be set of solutions to \eqref{eq.convert_Y}. Our goal is to find the least-norm solution among $\cH_{W,0}$: 
\begin{equation*}
    H^0(W,t) :=\underset{H(W,t) \in \cH_{W,0}}{\mathrm{arg}\min}\|H(W,t)\|_{\cH_W}.
\end{equation*}
To this end, we employ kernel-based methods to estimate from conditional restrictions \citep{zhang2020maximum,mastouri2021proximal,kallus2021causal,ghassami2022minimax}. 
\begin{remark}
    It is worthy to note that $\cH_{W,0}=\{H(\cdot,t)\in\mathcal{H}_W:AH(\cdot,t)(x)=b(x,t)\}=H^0(\cdot,t)+\mathrm{Ker}(A)$, where $A: \cH_W \mapsto \cH_X$ is a compact operator such that 
    \begin{equation}
    \label{eq.Ab}
        AH(\cdot,t)(x) :=\mE\{H(W,t)\phi_X(X)\}, \ b(\cdot,t):=\mE\{\varphi(Y,t)\phi_X(X)\}. 
    \end{equation}
    Apparently, $H^0(\cdot,t)$ is the least-norm solution, since it has no component in the kernel space. To ensure the uniqueness for estimation, previous methods additionally assumed the completeness of $W|X$ to remove those solutions belonging to the kernel space. 
\end{remark}
Formally, for any $g \in \cH_X$, \eqref{eq.convert_Y} implies that $\mathbb{E}_{Y,W,X}[\{\varphi(Y,t)-H(W,t)\}g(X)] = 0$ for almost all $t$.
We define the risk functional as the supremum of the residual moment over the unit ball of $\cH_X$ \citep{mastouri2021proximal},
\begin{equation}
\label{eq.unrestricted}
    R(H)=\underset{g\in \cH_X,\| g\|_{\cH_X} \le 1}{\sup}\left( \mathbb{E} \left[ \{\varphi(Y,t)-H(W,t)\} g(X) \right] \right)^2. 
\end{equation}
Let $\Delta(W,Y,t):=\varphi(Y,t)-H(W,t)$. By \cite{mastouri2021proximal}, the risk is equivalent to the following form:
\begin{equation}\label{eq:mmr_RH}
    R(H)=\mathbb{E} \{\Delta(W,Y,t)\Delta(W',Y',t)k_X(X,X')\},
\end{equation}
where $X',Y',W'$ are independent copies of $X, Y, W$. \cite{zhang2020maximum} showed that under mild conditions on $k_X$, minimizing $R(H)$ ensures us to find the true solution. 
To implement, we consider the empirical risk with Tikhonov regularization:
\begin{equation}\label{eq:mmr-vstat}
\widehat{R}^{\lambda}(H):=\sum_{i,j=1}^n{\frac{\Delta_i\Delta_j}{n^2}K_{X,ij}}+\lambda \|H\|_{\mathcal{H}_W},
\end{equation}
where $\Delta_i:= \varphi(y_i,t)-H(w_i,t)$ and $K_{X,ij}:= k_X(x_i, x_j)$. Using the representer theorem \citep{scholkopf2001generalized}, the estimate is given by $\widehat{H}^{\lambda}(w,t)=\boldsymbol{\alpha}^{\top}\boldsymbol{k}_W(w)$ for any $t$, where $\boldsymbol{k}_W(w):=\left\{ k_W(w_i,w)\right\}_i\in\mathbb{R}^n$ 
Here, $K_X:=\{k_X(x_i,x_j)\}_{ij}\in \mathbb{R}^{n\times n}$, $\boldsymbol{\alpha}:=(K_WK_XK_W+n^2\lambda K_X)^{-1}K_XK_W\varphi(\boldsymbol{y},t)$ with $K_W:=\{k_W(w_i,w_j)\}_{ij}\in \mathbb{R}^{n\times n}$ being Gram matrices, and $\varphi(\boldsymbol{y},t):= (\varphi(y_{1},t),\ldots,\varphi(y_{n},t))^{\top}$. We choose $\lambda$ via cross-validation.

\textbf{Constructing the testing statistics.} We assess the validity of $\mathbb H_0$ by evaluating the residue of the equation. To this end, we employ the conditional moment test procedure \citep{bierens1982consistent,bierens1997asymptotic}. Specifically, we choose a weight function $m(\cdot,s)$ that transforms the conditional restriction to the unconditional one. For power consideration, we can choose characteristic function, exponential function, sine and cosine functions, which enjoy the property \citep{stinchcombe1998consistent} that, for any $U(W,Y,t):=\varphi(Y,t) -H^{0}(W,t)$ with $\mE\{U(W,Y,t)|X\} \neq 0$, the set of $s \in \cT$ such that $\mE\{U(W,Y,t)m(X,s)\} = 0$ has Lebesgue measure zero. Let $\widehat{U}(W,Y,t):=\varphi(Y,t) -\widehat{H}^{\lambda}(W,t)$, we define 
\begin{equation}\label{eq.tn_st}
    T_n(s,t) =\frac{1}{\sqrt{n}}\sum_{i=1}^n{\widehat{U}(w_i,y_i,t)m(x_i,s)}, \ s,t \in \cT. 
\end{equation}
The final statistics for testing $\mathbb{H}_0$ is given by the maximum residue over $\cT$:
\begin{equation}\label{eq.icm-statistics}
   \Delta_{\varphi ,m} =\max_{t\in\mathcal{T}}\int_{\mathcal{T}}{| T_n(s,t)|^2d\mu(s)},
\end{equation}
where $\mu$ denotes the measure of $\cT$ (\emph{e.g.}, Gaussian measure).

\subsubsection{Testing for discrete variables}
\label{sec.test_discrete}
Similar to the continuous case, we first estimate $\wh{\H}(W,t)$ in \eqref{eq.convert_Y_dis} and choose the weight function to construct the testing statistics. Since \eqref{eq.convert_Y_dis} is a linear equation, we can directly solve $H(w,t)$ via least square estimation. Let $(\widehat{\mathbf{q}}_t, \widehat{\mathbf{Q}})$ be consistent estimators of $\mathbf{q}_t := \sum_{y \in \mathcal{Y}} \varphi(y,t) P(y | X) \in \mathbb{R}^{|\mathcal{X}|}$ and $\mathbf{Q} := P(W | X)^{\top} \in \mathbb{R}^{|\mathcal{X}| \times |\mathcal{W}|}$.
Then the least-squares estimator of $\mathbf{H}(W,t)$ is
$
\widehat{\mathbf{H}}(W,t) = (\widehat{\mathbf{Q}}^{\top} \widehat{\mathbf{Q}})^{-1} 
\widehat{\mathbf{Q}}^{\top} \widehat{\mathbf{q}}_t.
$

For the weight function, we can choose indicator functions 
$\{\mathbf{1}\{X = x\} : x \in \mathcal{X}\}$, since we only need to evaluate a finite number of conditional moment equations. 
Therefore, the conditional moment restrictions can be tested by verifying that
$$
\mathbb{E}\{U(W,Y,t) \mathbf{1}(X=x)\} 
= P(X=x) \, \mathbb{E}\{U(W,Y,t) | X=x\} = 0,
\qquad \forall x \in \mathcal{X},
$$
where $U(W,Y,t) := \varphi(Y,t) - H(W,t)$.  
If $\mathbb{E}(U | X) \neq 0$, there exists at least one $x$ to invalidate the above equation, ensuring consistency against general alternatives. Then, we define the testing statistics as:
\begin{equation}\label{eq.tn_st_dis}
    \T_n(t) =\frac{1}{\sqrt{n}}\sum_{i=1}^n{\widehat{U}(w_i,y_i,t)\mathbf{e}(x_i)}, \ t \in \cT,
\end{equation}
where $\mathbf{e}(x) \in \mathbb{R}^{|\mathcal{X}|}$ is the standard basis vector that takes $1$ at the position corresponding to $x$ and zeros elsewhere.  
Aggregating over $t \in \mathcal{T}$ yields a Cram\'er--von~Mises
 statistic
\begin{equation}\label{eq.icm-statistics_dis}
   \Delta_{\varphi} = \int_{\mathcal{T}} \| \T_n(t) \|_2^2 \, d\mu(t).
\end{equation}
While one employ the Chi-square tests in the discrete case \citep{miao2018identifying, miao2023identifying}, we would like to highlight that our proposed integral-equation formulation provies a unified framework, with the discrete case arising as a particular specialization. 


\section{Asymptotic behavior and Implementations}\label{sec. asymptotic_implementations}

We provide the asymptotic level and power for our testing statistics \eqref{eq.icm-statistics}, \eqref{eq.icm-statistics_dis} for the continuous setting and the discrete setting, respectively in section~\ref{sec.asymptotic} and section~\ref{sec. asymptotic_discrete}. A bootstrapped implementation will be introduced in section~\ref{sec.implement}. 


\subsection{Asymptotic behavior for continuous variables}
\label{sec.asymptotic}
We first introduce some regularity conditions. 
\begin{assumption}\label{differentiabilty and integrability}
We assume $\mathbb{E}_X\{m(X,s)|W\}$ and $\mathbb{E}_X\{|m(X,s)| ^2|W\}$ are uniformly bounded for all $s$. 
\end{assumption}
\begin{assumption}\label{assum.bandwidth}
$n\lambda \rightarrow \infty$, $n\lambda^2\rightarrow 0$.
\end{assumption}
\begin{assumption}\label{assum.var}
    For any $s,t \in \cT$, $\mathbb{E}\{U(W,Y,t)^4|X\}<\infty$ and $\mathbb{E}(|m(X,s) -\{A(A^*A)^{-1}g_s\}(X)|^4)<\infty$, where $A$ is defined in \eqref{eq.Ab} and $g_s(\cdot):=\mathbb{E}\{ m(X,s) \phi_W(W)\}(\cdot)$.
\end{assumption}

Conditions~\ref{differentiabilty and integrability}–\ref{assum.bandwidth} are standard in kernel estimation methods \citep{darolles2011nonparametric, babii2020unobservables, beyhum2024testing}. Condition~\ref{differentiabilty and integrability} imposes regularity requirements on the weight function $m$, while condition~\ref{assum.bandwidth} ensures that the regularization bias vanishes asymptotically. Additionally, condition~\ref{assum.var} is required to control the asymptotic variance of the test statistic, which has been similarly assumed in kernel-based methods \citep{vd1998asymptotic,li2003consistent,huang2022unified}.

\begin{restatable}{theorem}{nullhypothesis}
\label{theorem:null-hypothesis}
Let $\eta_{s,t}(O) :=U(W,Y,t)m(X,s)-U(W,Y,t) \{A(A^*A)^{-1}A^*m(\cdot,s)\}(X)$, with $O:=(W,Y,X)$. Suppose conditions \ref{differentiabilty and integrability}--\ref{assum.var}, \ref{ass:y_bounded}--\ref{assum.ISPD kernel}, and \ref{assum.empirical_process}--\ref{assum.source condition} hold. Under $\mathbb H_0$, we have \textbf{(i).} $T_n(s,t)$ converges weakly to $\mathbb{G}(s,t)$ such that $\iint |\mG(s,t)|^2 d\mu(s)d\mu(t) < \infty$, where $\mathbb{G}(s,t)$ is a Gaussian process with zero-mean and covariance:
\begin{align*}
    \Sigma\{(s,t),(s',t')\} =\mathbb{E}\{\eta_{s,t}(O)\eta_{s',t'}(O'))\},
\end{align*}
where $O':=(W',Y',X')$ is an independent copy of $O$; and \textbf{(ii).} $\Delta_{\varphi,m}$ converges weakly to $\underset{t\in \cT}{\max}\int{| \mathbb{G}(s,t)|^2d\mu(s)}$.
\end{restatable}
\begin{remark}
For simplicity, we only present the result for $T_n(s,t)$ being a real-valued function, or as the real and imaginary parts of a complex-valued function, since the result can be trivially extended to complex-valued functions.
\end{remark}
    

\textbf{Power analysis.} We consider the power performance under two alternatives, where \eqref{eq.convert_Y} has no solution. First, we consider the global alternative that has been similarly considered in proximal causal discovery \citep{liu2023causal}. That is, for any $H(w,t) \in \cH_W$ for all $t$, the global alternative $\mH_1^{\mathrm{fix}}$ satisfies the following:
\begin{equation*}
    \mH_1^{\mathrm{fix}}: \mathbb{E}\{\varphi(Y,t)-H(W,t)|X\} \neq 0 \text{ for some $t \in \cT$}. 
\end{equation*}
We also consider a sequence of local alternatives $\mH^\alpha_{1n}$. There exists $H^0(w,t) \in \cH_W$ for all $t$, such that: 
\begin{equation*}
    \mathbb H^\alpha_{1n}: \mathbb{E}\{\varphi(Y,t)|X\}= \mathbb{E}\{H^0(W,t)|X\}+\frac{r(X,t)}{n^\alpha}, \ \forall \thinspace t
\end{equation*}
where $0 < \alpha \leq \frac{1}{2}$ and $r(X,t)\in\cH_X$. To be a valid alternative, $r(X,t)/n^\alpha$ can not be written as $\mE\{H - H^0|X\}$ for any $H \in \cH_W$. Theorem~\ref{theorem:alternative-hypothesis} suggests that our statistics has asymptotic power of one under $\mH_1^{\mathrm{fix}}$ and $\mH^\alpha_{1n}$ when $\alpha < \frac{1}{2}$, and has nontrivial power when $\alpha = \frac{1}{2}$. 

\begin{restatable}{theorem}{powerhypothesis}
\label{theorem:alternative-hypothesis}
Suppose conditions in Theorem~\ref{theorem:null-hypothesis} hold. Besides, we assume $\mE\{r(X,t)^4\}<\infty$. Then, we have:
\begin{itemize}[noitemsep,topsep=0pt]
    \item[\textbf{\emph{(i)}}] \textbf{Global alternative}.  $\lim_{n \to \infty} \max_{t \in \cT}  |T_n(s,t)| = \infty$ for almost all $s$ under $\mH_1^{\mathrm{fix}}$. 
    \item[\textbf{\emph{(ii)}}] \textbf{Local alternative} ($\alpha < \frac{1}{2}$). $\lim_{n \to \infty} \max_{t \in \cT}$ $|T_n(s,t)| = \infty$ for a.s. $s$ under $\mH^\alpha_{1n}$. 
    \item[\textbf{\emph{(iii)}}] \textbf{Local alternative} ($\alpha = 1/2$). $T_n(s,t)$ converges weakly to $\mathbb{G}(s,t)+\mu(s,t)$ such that $\iint |\mathbb{G}(s,t)+\mu(s,t)|^2 d\mu(s)d\mu(t) < \infty$ under $\mH^\alpha_{1n}$, where $\mathbb{G}(s,t)$ is defined in Theorem~\ref{theorem:null-hypothesis} and $\mu(s,t):=\mathbb{E}(r(X,t)[m(X,s)-\{A(A^*A)^{-1}A^*m(\cdot,s)\}(X)])$.
\end{itemize} 
\end{restatable}

\subsection{Asymptotic behavior for discrete variables}
\label{sec. asymptotic_discrete}
Next, we give the asymptotic properties of $\Delta_{\varphi}$ \eqref{eq.icm-statistics_dis} in the discrete setting. 
\begin{restatable}{theorem}{ProjCltNull}
\label{thm:proj-clt_null}
Denote $\D:=\mathrm{diag}\{P(x^{(1)}),...,P(x^{(|\mathcal{X}|})\}$ and $\P:=\Q(\Q^\top\Q)^{-1}\Q^\top$. Suppose conditions~\ref{assum.completeness} and \ref{ass:cardinality} hold. Under $\mathbb H_0$, we have \textbf{(i).} $\T_n(t)$ converges weakly to $\mathbb{G}(t)$ such that $\int \|\mG(t)\|_2^2 d\mu(t) < \infty$, where $\mathbb{G}(t)$ is a Gaussian process with zero-mean and covariance
$$
\Sigma(t,t')=\D(\I-\P)\Sigma'(t,t')(\I-\P)\D,
$$
where $\Sigma'(t,t')$ is the block-diagonal kernel with diagonal blocks
$$
\Sigma'_{k k}(t,t')=\frac{1}{P(x^{(k)})}\mathrm{Cov}(\varphi(Y,t),\varphi(Y,t')| X=x^{(k)})
\quad\text{and}\quad \Sigma_{k k'}(t,t')=0\;\;(k\neq k').
$$
\textbf{(ii).} $\Delta_{\varphi}$ converges weakly to $\int{\| \mathbb{G}(t)\|_2^2d\mu(t)}$.
\end{restatable}

Similar to the continuous case, we consider global alternatives and local alternatives. For any $\H(W,t)$, the global alternative $\mH_1^{\mathrm{fix}}$ satisfies the following:
\begin{equation*}
\sum_{y\in\mathcal{Y}}\varphi(y,t)P(y|x)= \H^\top(W,t)P(W|x) \neq 0 \text{ for some $t \in \cT$ and some $x\in \mathcal{X}$}. 
\end{equation*}
We also consider a sequence of local alternatives $\mH^\alpha_{1n}$, with $0 < \alpha \leq 1/2$. Formally, there exists $\H^0_t := \left(H^0(w^{(i)}, t): 1 \leq i \leq |\mathcal{W}| \right)^\top \in \mR^{|\mathcal{W}|}$, such that: 
\begin{equation*}
    \mathbb H^\alpha_{1n}: \sum_{y\in\mathcal{Y}}\varphi(y,t)P(y|x)= \H_0^\top(W,t)P(W|x) + \frac{r(x,t)}{n^\alpha}, \forall \thinspace t
\end{equation*}
where $0 < \alpha \leq \frac{1}{2}$. To be a valid alternative, $r(X,t)/n^\alpha$ can not be written as $\H^\top(W,t)P(W|X) - \H_0^\top(W,t)P(W|X)$ for any $\H$; besides, there exists $t$ and $x$ such that $|r(x,t)| \neq 0$. We define $\r_t := [\r(x^{(1)},t), ..., \r(x^{(|\mathcal{X}|)},t)]^\top$. 
\begin{restatable}{theorem}{ProjCltAlt}
\label{thm:proj-clt_alt}
Suppose conditions in Theorem~\ref{thm:proj-clt_null} hold. Then, we have:
\begin{itemize}[noitemsep,topsep=0pt]
    \item[\textbf{\emph{(i)}}] \textbf{Global alternative}.  $\lim_{n\rightarrow \infty} \max_{t\in \cT} \|\{\T_n(t)\|_{\infty} =\infty$ under $\mH_1^{\mathrm{fix}}$. 
    \item[\textbf{\emph{(ii)}}] \textbf{Local alternative} ($\alpha < 1/2$). $\lim_{n\rightarrow \infty} \max_{t\in \cT} \|\{\T_n(t)\|_{\infty} =\infty$ under $\mH^\alpha_{1n}$. 
    \item[\textbf{\emph{(iii)}}] \textbf{Local alternative} ($\alpha = 1/2$). $\T_n(t)$ converges weakly to $\mathbb{G}(t)-\mu(t)$ such that $\int |\mathbb{G}(t)-\mu(t)|^2 d\mu(t) < \infty$ under $\mH^\alpha_{1n}$, where $\mathbb{G}(t)$ is defined in Theorem~\ref{theorem:null-hypothesis} and $\mu(t):=\D(\I-\P)\r_t$.
\end{itemize}
\end{restatable}

\subsection{Implementations}\label{sec.implement}

We present the implementation details for computing $\Delta_{\varphi,m}$, $\Delta_{\varphi}$, and corresponding critical values. For brevity, we only introduce the procedure, as the implementation for $\Delta_{\varphi}$~\eqref{eq.icm-statistics_dis} follows similarly. Because $\Delta_{\varphi,m}$~\eqref{eq.icm-statistics} involves integration, we approximate it using Monte Carlo methods. Furthermore, as the limiting distribution of $\Delta_{\varphi,m}$ lacks a closed-form expression, we estimate the critical value via the Bootstrap.



\textbf{Monte-Carlo methods for approximating $\Delta_{\varphi,m}$.} We set $m(\cdot,s)$ to the characteristic function and $\mu$ to be symmetric around the origin (\emph{e.g.}, Lebesgue measure), since such a setting enables the integration to be computed in closed form. By \cite{stinchcombe1998consistent}, setting $m$ to the characteristic function can preserve power when transforming the conditional restriction to the unconditional one. To approximate the maximal value of $\int_{\cT} |T_n(s,t)|^2d\mu(s)$ over $\cT$, we evaluate the process at a grid of equi-distant indices $\{t_i,i\in [K]\}$ and estimate $\wh{\Delta}_{\varphi,m}:=\max_{k\in [K]} \int_{\cT} |T_n(s,t_k)|^2d\mu(s)$. Corollary~\ref{coro:hypothesis} shows that when $K$ is sufficiently large, $\wh{\Delta}_{\varphi,m}$ converges to $\max_{t \in \cT} \int_{\cT}  |\mG(s,t)|^2 d\mu(s)$. 


\textbf{Estimate the critical value via bootstrap.} Since it is difficult to obtain the explicit form of $\mG(s,t)$, we employ the residue-based wild bootstrap procedure for approximation under the null-limiting distribution. We repeat the procedure for $B$ times. For the $b$-th time, we first employ the empirical process $\wh{T}^b_n(s,t) = \frac{1}{\sqrt{n}} \sum_{i=1}^n \omega^b_i \wh{U}(w_i,y_i,t)m(x_i,s)$ to approximate $T_n(s,t)$ for each $(s,t)$, where $\{\omega^b_i\}_{i=1}^n$ is a sequence of zero-mean, unit variance variables. Here, we follow \cite{enno1993} to set 
$\mP(\omega_i = 1- \kappa) = \kappa/\sqrt{5}$ and $\mP(\omega_i = \kappa) = 1-\kappa/\sqrt{5}$ with $\kappa = \frac{\sqrt{5}+1}{2}$. The bootstrapped statistic is given by:
\begin{equation}
\label{eq.bootstrap}
    \wh{\Delta}^b_{\varphi,m} = \max_{k\in [K]} \int_{\cT} |\wh{T}^b_n(s,t_k)|^2d\mu(s). 
\end{equation}
Given the level of significance $\alpha$, the critical value is computed as the $(1-\alpha)$-quantile of $\left\{\wh{\Delta}^1_{\varphi,m},...,\wh{\Delta}^B_{\varphi,m}\right\}$, denoted by $\wt{\Delta}^{1-\alpha}_{\varphi,m}$. We then reject the null hypothesis if $\wh{\Delta}_{\varphi,m} \geq \wt{\Delta}^{1-\alpha}_{\varphi,m}$. Corollary~\ref{coro:hypothesis} shows that the bootstrap statistics $\wh{\Delta}^b_{\varphi,m}$ converges to $\max_{t \in \cT} \int_{\cT}  |\mG(s,t)|^2 d\mu(s)$.

\begin{restatable}{corollary}{girdhypothesis}
\label{coro:hypothesis}
Suppose conditions in Theorem~\ref{theorem:null-hypothesis} hold. If $\varphi(y,t)$ is continuous with respect to $t$ for each $y$, then $\widehat{\Delta}_{\varphi ,m}$ is weakly convergent to $\max_{t \in \cT}\int_{\mathcal{T}}{| \mathbb{G}(s,t)|^2d\mu(s)}$ under $\mH_0$, as $n, K \to \infty$. Besides, conditional on the original sample $\{y_i,w_i,x_i\}_{i=1}^n$, the bootstrapped statistics \eqref{eq.bootstrap} is also weakly convergent to the $\max_{t \in \cT}\int_{\mathcal{T}}{| \mathbb{G}(s,t)|^2d\mu(s)}$. 
\end{restatable}
\begin{remark}
   Since the characteristic function holds for any $t$ the restricted choice of $[K]$ in the experiment may lead to a loss of power.
\end{remark}

\section{Nonidentifiability with integral equation}
\label{sec.fail}

In the power analysis, we assume that the integral equation~\eqref{eq.bridge_Y} has no solution. In this section, we examine the failure case where this condition is violated, resulting in non-identifiability of the causal relationship. Next, section~\ref{sec.two_proxy} introduces a new procedure that can restore identifiability, when an NCE is additionally available.




\subsection{Failure case for identifying $\mH_1$}

The following proposition presents an impossibility result to identify $\mH_1$ under the linear Gaussian case. 

\begin{proposition}
\label{prop.fail}
Suppose $U,X, Y, W$ follow from the linear Gaussian model, \emph{i.e.} $U=\varepsilon_U,X =\alpha_UU + \alpha_0+\varepsilon_X,W =\beta_U U +\beta_0 + \varepsilon_W, Y=\gamma_U U + \gamma_X X + \gamma_X W+\gamma_0 +\varepsilon_Y,$
where $\varepsilon_U,\varepsilon_X,\varepsilon_W,\varepsilon_Y\sim \cN(0,1)$. When $\gamma_W = 0$, as long as $|\gamma_X| > g_X(\alpha_U,\beta_U,\gamma_U)$, the integration equation \eqref{eq.bridge_Y} has no solution. Further, if $|\gamma_W| > g_W(\alpha_U,\beta_U, \gamma_U)$\footnote{We leave the detailed form of $g_X,g_W$ in Appendix~\ref{appx.two_fail}.}, \eqref{eq.bridge_Y} has a solution. 
\end{proposition}
\begin{remark}
    We derive some additional results during the proof. For example, we show that the dependency between $W$ and $U$ (\emph{i.e.}, $\beta_U$) must be sufficiently strong to ensure the existence of a solution under $\mH_0$. Besides, we extend these results to settings where $W$ and $Y$ share a non-causal dependence, that is, when there exists an unobserved $U_1$ such that $U_1 \to W$ and $U_1 \to Y$. More details can be found in  Appendix~\ref{appx.two_fail}. 
\end{remark}
Proposition~\ref{prop.fail} demonstrates that when the dependence between $Y$ and $W$ (\emph{i.e.}, $\gamma_W$) is sufficiently strong, a solution exists even in the presence of a strong direct effect from $X$ to $Y$. Intuitively, the additional dependence on $Y$ provides $p(w|x)$ with greater variability to explain the variability in $p(y|x)$. Formally, this corresponds to the convergence of $\sum_{n=1}^\infty \lambda_n^{-2} |\langle p(y|x),\phi_n \rangle|^2$ (similar to condition~\ref{assum.regularity condition} (2)), 
which indicates that $p(y|x)$ can be completely represented in the basis $\{\phi_n\}$, the eigenfunctions of the conditional expectation operator $T: \cL^2\{F(w)\} \to \cL^2\{F(x)\}$ defined by $Tf := \mE\{f(W)| X \}$. To illustrate, consider the following example. 

\begin{figure}[htbp]
\centering
\includegraphics[width=0.95\linewidth]{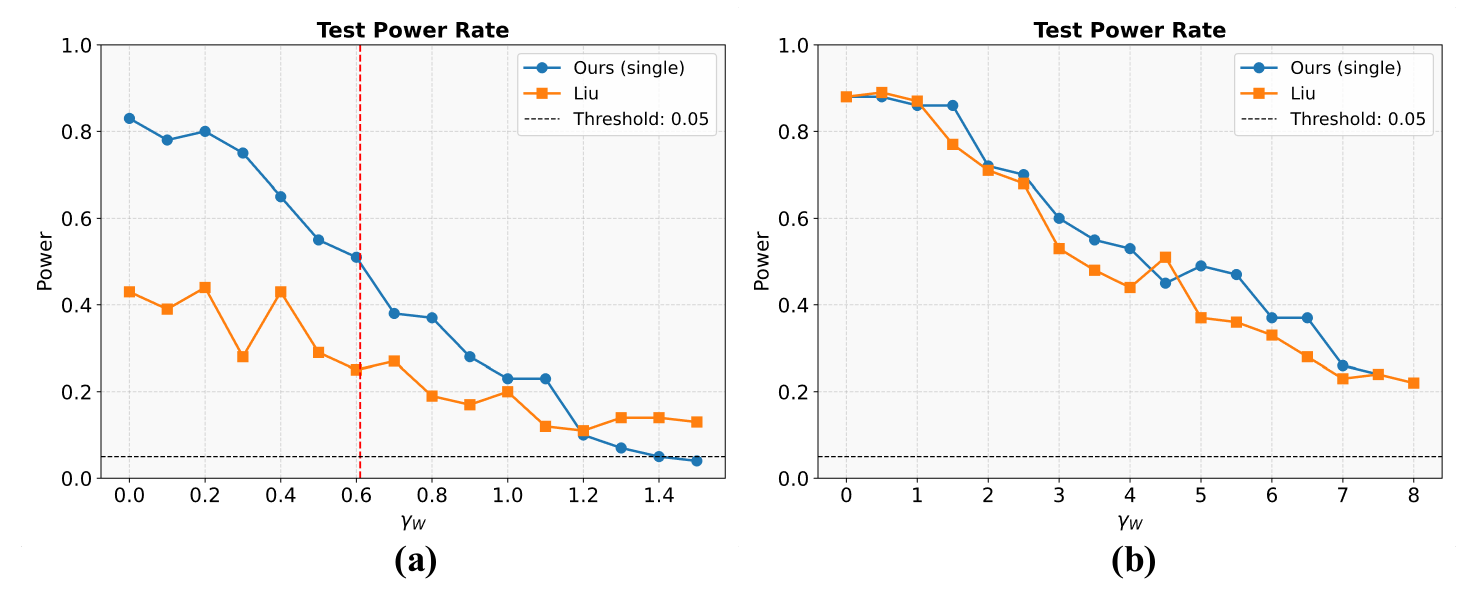}
\caption{The change of power across $\gamma_W$ in example~\ref{example:linear_gaussian_two_proxy} (left) and in the nonlinear example (right).}
\label{fig.power_fail}
\end{figure}

\begin{restatable}{example}{examplelineargaussian}
\label{example:linear_gaussian_two_proxy}
Suppose that $X, U, W$ satisfy the linear Gaussian model, \emph{i.e.} $U=\varepsilon_U,X =2U + \varepsilon_X,W =-2U +\varepsilon_W$. Let $X', W'$ denote the standarlized version of $X, W$, \emph{i.e.}, $X' = \frac{X}{\sqrt{\mathrm{Var}(X)}}$, $W' = \frac{W}{\sqrt{\mathrm{Var}(W)}}$. With $X',W'$, the structural equation of $Y$ is $Y= X' + U+\gamma_W W'+\varepsilon_Y,$
where $\varepsilon_U,\varepsilon_Y,\varepsilon_W,\varepsilon_X \sim \cN(0,1)$. The integral equation \eqref{eq.bridge_Y} has a solution if and only if $\gamma_W>\frac{-15+36\sqrt{5}}{72+16\sqrt{5}}\approx0.61$. Besides, the series $\sum_{n=1}^\infty \lambda_n^{-2} |\langle p(y|x'),\phi_n \rangle|^2$ converges if and only if $\gamma_W>\frac{-15+36\sqrt{5}}{72+16\sqrt{5}}\approx0.61$, where $(\lambda_{n},\varphi_{n},\phi_{n})_{n=1}^{\infty}$ denote a singular value decomposition of the conditional expectation operator $T: \cL^2\{F(w)\} \mapsto \cL^2\{F(x)\}$ defined by $Tf := \mE\{f(W)| X \}$. 
\end{restatable}
This example shows that the key reason for non-identifiability of $\mH_1$ lies in the convergence of the series $\sum_{n=1}^\infty \lambda_n^{-2} |\langle p(y|x'),\phi_n \rangle|^2$. As illustrated in Fig.~\ref{fig.power_fail} (a), the power significantly drops as $\gamma_W$ surpasses the threshold. Besides, we can observe similar phenomena under the nonlinear case (details can be found in Appendix \ref{appx.two_fail}), as illustrated in Figure~\ref{fig.power_fail} (b). 


\subsection{A new procedure with two proxies}\label{sec.two_proxy}

To identify $\mH_1$ when $W$ is strongly dependent on $Y$, we impose another restriction introduced by the NCE $Z$, which, together with $W$, has been widely used in proximal causal inference \citep{miao2018identifying, cui2024semiparametric}. For simplicity, we only discuss the continuous case. 

The key idea is to characterize the property of $h(w,y)$ that satisfies  \eqref{eq.bridge_Y} under $\mH_0$, and imposes the restriction via $Z$ to examine this property. To this end, we require some completeness conditions, which are standard in the literature on proximal causal inference \citep{miao2018identifying, liu2023causal, eric2024proxyintro}. 
\begin{assumption}[Completeness]
\label{assum.complete_U_X_Zx}
    For any square-integrable function $g$, we assume 
    \begin{enumerate}
        \item $\mathbb E\{g(u)|x\} = 0$ almost surely if and only if $g(u) = 0$ almost surely; 
        \item for any fixed $x$, $\mathbb E\{g(u)|z,x\} = 0$ almost surely if and only if $g(u) = 0$ almost surely.
    \end{enumerate}
\end{assumption}

The following theorem elaborates a property of the solution under $\mH_0$. 
\begin{restatable}{theorem}{solutionsexisttwoproxies}
\label{thm.exist_solution_twoproxies}
Suppose condition~\ref{assum.complete_U_X_Zx} holds and that $Y\ind Z|U$. For any $h(w,y)$ that satisfies \eqref{eq.bridge_Y}, $\mathbb H_0$ holds if and only if $h(w,y)$ also satisfies the following equation for all $z$ and $x$:
\begin{equation}
\label{eq.bridge_Y_new}
    p(y|z,x) = \int h(w,y)p(w|z,x)dw.
\end{equation}
\end{restatable}

It is worth to note that solving $h(w,y)$ from \eqref{eq.bridge_Y_new} is different from solving $h(w,y,x)$ in \cite{miao2018identifying}, 
$$p(y|z,x) = \int h(w,y,x)p(w|z,x)dw,
$$
where the bridge function $h(w,y,x)$ that additionally depends on $x$, is used to compute $p\{y|do(x)\}=\int h(w,y,x)p(w)dw$. 
In our context, the goal is to test whether $X$ directly affects $Y$, which requires $h$ to be independent of $x$ while still ensuring that \eqref{eq.bridge_Y_new} holds as $x$ varies. 

\begin{remark}
One might argue that when both $W$ and $Z$ are available, the average causal effect is identifiable from the above formula, rendering our analysis unnecessary. However, as elaborated in Appendix~\ref{appx.example_intervention}, the causal hypothesis testing is conceptually distinct from causal effect estimation. In particular, we provide an example where a causal relationship exists even though the average causal effect is zero.
\end{remark}

Inspired by Theorem~\ref{thm.exist_solution_twoproxies}, we can use the residue in \eqref{eq.bridge_Y_new} to construct the testing statistics. The procedure is similar to section~\ref{sec.implement}. Specifically, if $\wh{H}^\lambda$ can well approximate the solution of ~\eqref{eq.convert_Y}, the equation 
\begin{align}
\label{eq.convert_YZ}
   \mathbb{E}_{Y,W}\{ \varphi(Y,t) -H(W,t) |Z,X\} =0 \ \forall \thinspace t \in \cT,
\end{align}
also approximately holds for all $t \in \cT$. This allows us to assess the validity of $\mH_0$ via the residual process $\wh{U}(W,Y,t):=\varphi(Y,t) - H(W,t)$. We then define the test statistics:
\begin{align}
    T^{(Z)}_{n}(s,t) &=\frac{1}{\sqrt{n}}\sum_{i=1}^n{\widehat{U}(w_i,y_i,t)m(x_i,z_i,s)}, \ s,t \in \cT \nonumber \\
    \Delta^{(Z)}_{\varphi, m}&=\max_{t\in\mathcal{T}}\int_{\mathcal{T}}{|T^{(Z)}_{n}(s,t)|^2d\mu(s)}, \label{eq.icm_statistics_two}
\end{align}
where $m(Z,X,s)$ is a weight function over $(Z,X)$. Similarly, the asymptotic behavior of $\Delta^{(Z)}_{\varphi, m}$ can be established, which can be found in Appendix~\ref{appx.two_asymptotic}.

\section{Simulation}
\label{sec.exp}

In this section, we evaluate our methods on synthetic data. In section~\ref{sec.exp-single}, we consider the single proxy setting, where only the NCO, \emph{i.e.}, $W$ is available and $W \ind Y|U$. We report the type-I error and recall of the statistics \eqref{eq.icm-statistics} and \eqref{eq.icm-statistics_dis} on the continuous and discrete data, respectively. In section~\ref{sec.exp-two}, we additionally evaluate our two-proxy procedure when the NCE, \emph{i.e.}, $Z$ is also available, under the case when $W$ is dependent on $Y$ given $U$. Code is available at \url{https://anonymous.4open.science/r/proximal_causal_discovery_cv-F364}.


\textbf{Compared baselines.} For the continuous case, we compare our methods with: \textbf{(i) Liu} \citep{liu2023causal} that designed a discretization method for bivariate causal discovery over continuous variables; \textbf{(ii) KCI} (Kernel-based Conditional Independence test) \citep{zhang2012kernel} that tested the null hypothesis of $X\ind Y|W$ using kernel matrices. For the discrete case and two-proxy setting, we also conduct \textbf{(iii) Miao} \citep{miao2018identifying} that was designed for causal hypothesis testing over discrete variables using $W$ and $Z$.

\textbf{Implementation details.} We set the significance level $\alpha$ to $0.05$. We choose $\varphi$ and $m$ to be complex exponential functions. Under continuous setting, for PMCR estimation, we set $K = 100$ and follow \citep{mastouri2021proximal} to select the optimal $\lambda$ from a sequence ranging from $4.9 \times 10^{-6}$ to $0.25$, with a step size chosen to ensure the sequence contains 50 values. Besides, we use Gaussian kernels with the bandwidth parameters being initialized using the median distance heuristic. Under discrete setting, we use the OLS of section~\ref{sec:discrete_integral} and set $K = 100$. For the procedure of \textbf{Liu}, we follow its implementation to set the bin numbers of $W$ and $X$ to $l_X = 14$, $l_W = 12$, respectively.  For the procedure described in \textbf{Miao}, we implement the R code released in the paper and set $l_X = 3, l_W = 2, l_Z = 2$ by default under continuous setting. Besides, we set $l_X = |\mathcal{X}|, l_W = |\mathcal{W}|$ under discrete setting. For \textbf{KCI}, we adopt the implementations provided in the $\mathrm{causallearn}$ packages \url{https://causal-learn.readthedocs.io/}.

\subsection{Single proxy with $W \not\to Y$}\label{sec.exp-single}

In this section, we consider the setting where only $W$ is available, where the results for the continuous case and the discrete case are recorded in section~\ref{sec.exp_continuous} and section~\ref{sec.exp_discrete}, respectively. 

\subsubsection{Continuous setting}
\label{sec.exp_continuous}

\textbf{Data generation.} We follow \cite{liu2023causal} to generate data of $V \in \{X,Y,U,W\}$ via $V=f_V(\mathbf{PA}_V)+\varepsilon_V$, where $\mathbf{PA}_V$ and $\varepsilon_V$ respectively denotes the parent set and the noise of $V$. For the variable $V$, $f_V$ is randomly selected from $\{\mathrm{linear},\mathrm{tanh},\mathrm{sin},\mathrm{sqrt}\}$. Besides, the distribution of $\varepsilon_V$ is randomly chosen from $\{\mathrm{Gaussian},\mathrm{uniform},\mathrm{exponential},\mathrm{gamma}\}$. To mitigate the effect of randomness, we repeat the process $20$ times. At each time, we generate 100 replications under each $\mathbb{H}_0$ and $\mathbb{H}_1$, and record the type-I error rate and power rate.

\begin{figure}[htbp]
\centering
\includegraphics[width=0.95\linewidth]{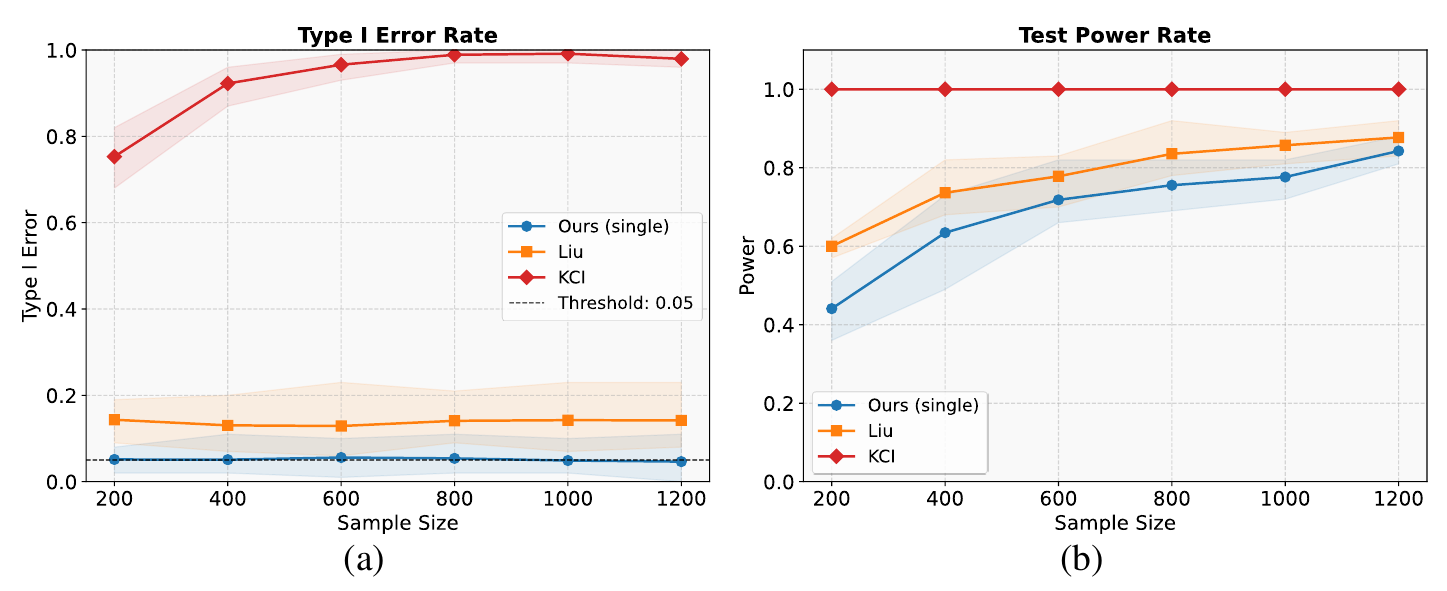}
\caption{Type-I error rate (left) and power rate (right) of our testing procedure and baseline methods in the single-proxy setting. The solid line reports the average value over 20 times, and the shaded area denotes the region $(\mathrm{mean}-\mathrm{std},\mathrm{mean}+\mathrm{std})$.}
\label{fig.cmp_baselines_single}
\end{figure}

\textbf{Type-I error and power.} In Figure~\ref{fig.cmp_baselines_single}, we report the average type-I error rate and power rate for our testing procedure and others. As shown, the type-I error rate of our method closely approximates $\alpha=0.05$ as $n$ increases, while other methods fail to control the type-I error. Specifically, conditioning on the proxy $W$, \textbf{KCI} cannot eliminate the confounding bias, leading to uncontrollable type-I errors; while the additional error in \textbf{Liu} \citep{liu2023causal} may arise from discretization errors with a finite bin number or probability estimation error due to limited sample size. Besides, our power approximates to one as $n$ increases. Compared to previous baselines \textbf{Liu}, these results demonstrate the utility and its ability to make better use of available data in causal discovery.

\begin{figure}[htbp]
\centering
\includegraphics[width=0.95\linewidth]{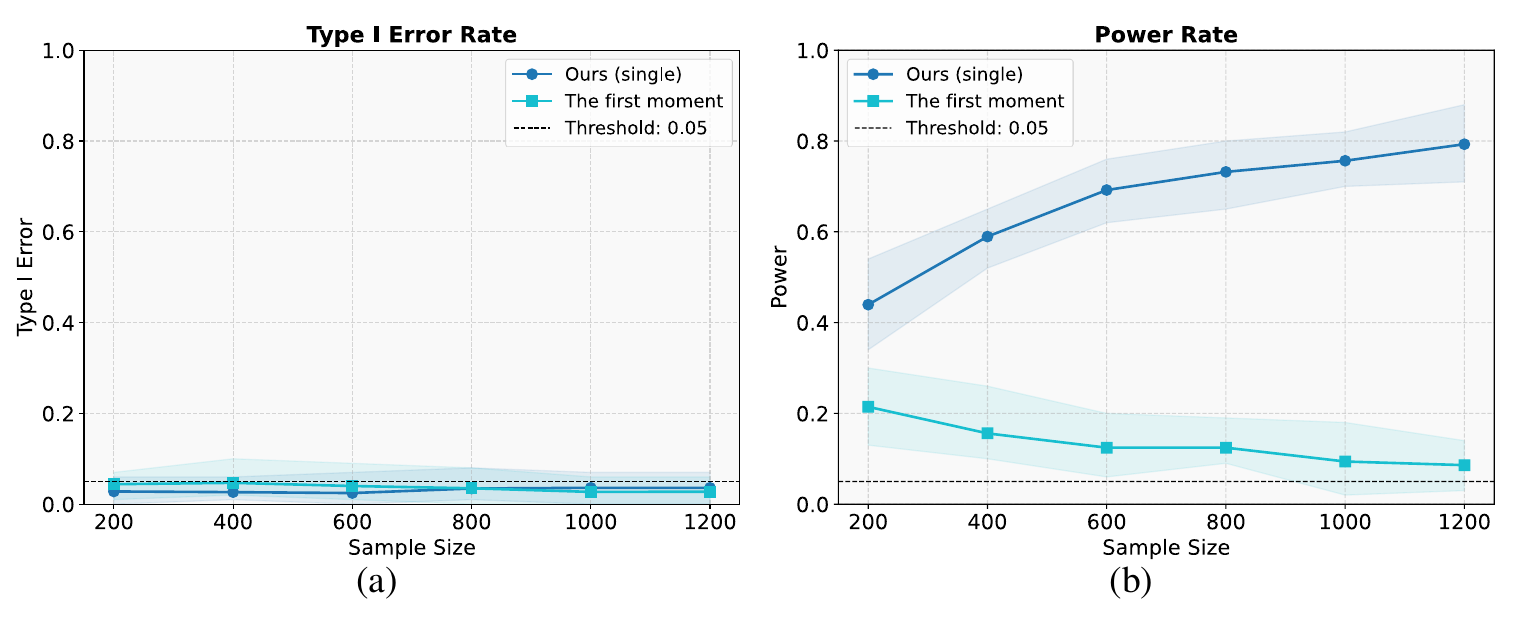}
\caption{Type-I error rate (left) and power rate (right) of our procedure with PMCR and the first-moment method in example~\ref{example:linear_gaussian}.}
\label{fig.moment}
\end{figure}

\textbf{Comparisons with MMR.} To further demonstrate the effectiveness of our estimation method (\emph{i.e.}, PMCR) over the traditional first-order moment restriction method (\emph{i.e.}, MMR), we apply both methods to the data generated in example~\ref{example:linear_gaussian}, where we have shown that the solution of the first-moment equation exists under the alternative hypothesis. As shown in Figure~\ref{fig.moment}, although both methods can asymptotically control the type-I error as $n \to \infty$, the power of our procedure approaches $1$ while the MMR still lies around $\alpha=0.05$ under $\mH_1$.

\subsubsection{Discrete setting}
\label{sec.exp_discrete}

\textbf{Data generation.} Following \cite{miao2018identifying}, we generate discrete random variables $X,Y,U,W$. Specifically, the distributions of $X$, $U|X$, and $W|U$ are specified as:
{\renewcommand{\arraystretch}{0.65}$$
P(X) = 
\left(\!\!
\begin{array}{c}
3 \\ 3 \\ 4
\end{array}
\!\!\right) / 10,
\quad
P(U | X) = 
\left(\!\!
\begin{array}{ccc}
3 & 6 & 5 \\
7 & 4 & 5
\end{array}
\!\!\right) / 10,
\quad
P(W | U) = 
\left(\!\!
\begin{array}{cc}
8 & 3 \\
2 & 7
\end{array}
\!\!\right) / 10.
$$}
Under $\mH_1$, the conditional distribution of $Y$ given $(U,X)$ is further specified as
{\renewcommand{\arraystretch}{0.65}$$
P(Y | U, x_1) =
\left(\!\!
\begin{array}{cc}
5 & 4 \\
3 & 2 \\
2 & 4
\end{array}
\!\!\right) / 10,
\quad
P(Y | U, x_2) =
\left(\!\!
\begin{array}{cc}
4 & 6 \\
2 & 3 \\
4 & 1
\end{array}
\!\!\right) / 10,
\quad
P(Y | U, x_3) =
\left(\!\!
\begin{array}{cc}
3 & 2 \\
4 & 5 \\
3 & 3
\end{array}
\!\!\right) / 10.
$$}
Under $\mH_0$, $\mP(Y | U,x)$ does not depend on $x$, \emph{i.e.},
\renewcommand{\arraystretch}{0.65}$$
P(Y | U, x_1) = P(Y | U, x_2) = P(Y | U, x_3) =
\left(\!\!
\begin{array}{cc}
5 & 4 \\
3 & 5 \\
2 & 1
\end{array}
\!\!\right) / 10.
$$
Similar to the continuous case, we repeat the process $20$ times, where each time we generate 100 replications under each $\mathbb{H}_0$ and $\mathbb{H}_1$. 

\begin{figure}[htbp]
\centering
\includegraphics[width=0.95\linewidth]{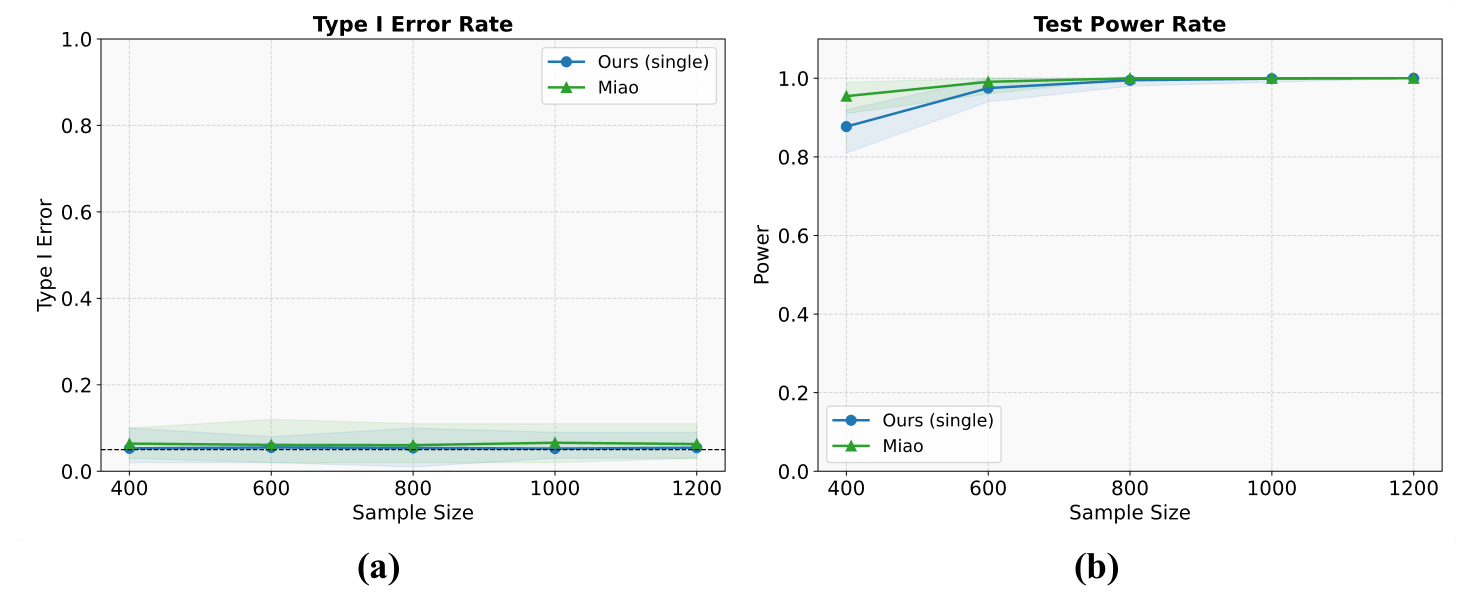}
\caption{Type-I error rate (left) and power rate (right) of our procedure and the Miao's method in the discrete setting.}
\label{fig.miao}
\end{figure}

\textbf{Type-I error and power.} As shown in Figure~\ref{fig.miao}, our procedure is comparably effective to that in \cite{miao2018identifying}. Specifically, the average type-I error rate of our method is very close to $\alpha=0.05$ when $n = 400$. Moreover, our power approximates to one as $n$ increases. However, since we only considered a finite number of $t$ values when computing $\Delta_\varphi$ \eqref{eq.icm-statistics_dis}, our method exhibits a slight loss of power relative to \textbf{Miao}, especially when the sample size is small. This problem can be mitigated as we increase the number of $t$, as shown in Appendix \ref{appx.discrete}. 

\subsection{Two proxies with $W \to Y$}\label{sec.exp-two}

In this section, we apply our two-proxy procedure in section~\ref{sec.fail} to the setting when $W \to Y$, where the single-proxy procedure may fail as the integral equation may admit a solution under $\mH_1$.

\textbf{Data generation.} Following example~\ref{example:linear_gaussian_two_proxy}\footnote{We also consider a nonlinear setting, as detailed in Appendix~\ref{appx.add_experiment}.}, we set $\gamma_W=1$, which implies there exists $h$ that satisfies the integral equation \eqref{eq.bridge_Y}. Similar to the single-proxy setting, we repeat the process 20 times, where at each time we generate 100 replications under $\mH_0$ and $\mH_1$.


\begin{figure}[htbp]
\centering

\includegraphics[width=1.0\linewidth]{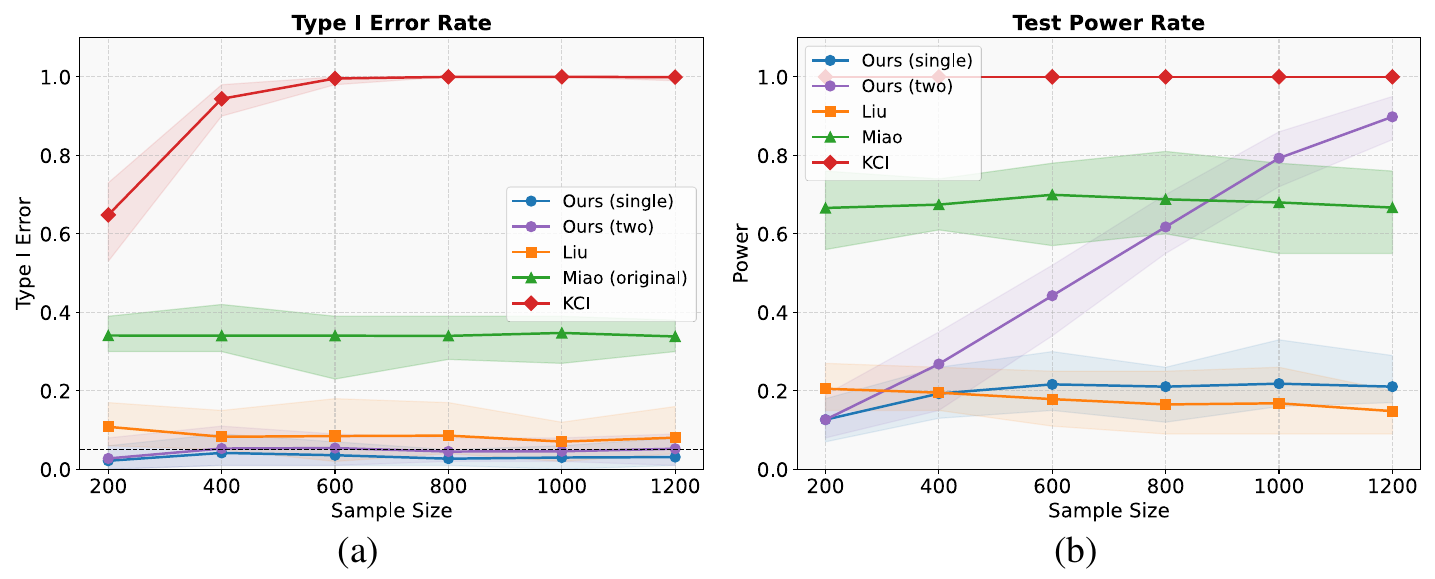}
\caption{Type-I error rate (left) and power rate (right) of our procedure and baselines on synthetic data with two proxies. }
\label{fig.cmp_baselines_two}

\end{figure}

\textbf{Type-I error and power.} We report the average results in Figure~\ref{fig.cmp_baselines_two}. As shown, although our single-proxy procedure can control the type-I error, it suffers from low power in identifying the causal relation, due to the existence of solution under $\mH_1$ in this example. With additional information provided by $Z$, the power significantly improves and approaches one as $n$ increases. This verifies our findings in section~\ref{sec.fail}, and demonstrates the utility of employing $Z$ (\emph{i.e.}, NCE) in discovering the causal relation when the effect of $W$ on $Y$ is strong enough to invalidate the procedure with only $W$. 



\section{Real-world experiments}
\label{sec.real_world}

In this section, we evaluate our methods on real data. Section~\ref{sec.sepsis} applies our approach to Intensive Care (MIMIC-III) data to examine the effectiveness of antibiotics to the antibiotics. Section~\ref{sec.exp_real} introduces the result on \emph{World Values Survey} (WVS) data, where the goal is to examine the causal relationship between moral attitudes and dishonest behaviors. 

\subsection{Application to Intensive Care Data}\label{sec.sepsis}

Following \cite{liu2023causal}, we apply the proposed method to the Medical Information Mart for Intensive Care (MIMIC-III) database \citep{johnson2016mimic}\footnote{The data are available at \url{https://physionet.org/content/mimiciii}.} to investigate whether the antibiotics are effective against sepsis. We extracted data for 3,251 patients diagnosed with sepsis during their ICU stays from MIMIC-III.

We examine two potential causal relationships: \textit{Vancomycin} $\to$ \textit{White Blood Cell count (WBC)} and \textit{Morphine} $\to$ \textit{WBC}. In both cases, the patient’s underlying health status is a plausible unmeasured confounder that may jointly affect medication use and WBC levels. Among the patients, 1,888 received vancomycin and 559 received morphine. To adjust for the latent health status, we follow \cite{liu2023causal} and use blood pressure as the NCO (\emph{i.e.}, $W$). According to \cite{rybak2009therapeutic, dowell2022cdc}, blood pressure is not expected to directly influence the prescription of vancomycin or morphine, as these medications are primarily administered in response to infection or pain rather than hemodynamic conditions. This supports the plausibility of the conditional independence assumption underlying the use of $W$ as a valid proxy.

Table \ref{tab:pvalues_mimic_rct} reports the $p$-values obtained from three different causal discovery tests for the two medication–WBC pairs in the context of sepsis. As shown, both our method and Liu's method yield $p$-values significantly above the significance level for testing Vancomycin $\to$ WBC, indicating no causal relationship, and small $p$-values to examine Morphine $\to$ WBC, suggesting a potential causal relationship. In contrast, the KCI test produces $p$-values above the significance level for both pairs. 


\begin{table}[h]
    \centering
    \caption{$p$-values of Different Methods for Sepsis-Related Causal Pairs Compared to RCT.}
    \label{tab:pvalues_mimic_rct}
    \begin{tabular}{ccc}
        \toprule
        \textbf{Method} & \textbf{Vancomycin$\to$WBC} & \textbf{Morphine$\to$WBC} \\
        \midrule
        KCI & 0.2990 & 0.4891 \\
        Liu & 0.9201 & 0.0217 \\
        Our(single) & 0.8980 & 0.0095 \\
        \midrule 
        \textbf{RCT} & \faCheck &\faTimes \\ 
        \bottomrule
    \end{tabular}
\end{table}

Our results are consistent with the conclusions drawn from two randomized controlled trials (RCTs), which serve as the gold standard for causal discovery. Prior RCTs studies have shown that vancomycin administration alters WBC \citep{rosini2015randomized}, whereas morphine has no such causal impact \citep{anand2004effects}. Overall, our proposed procedure successfully recovers causal relations that align with the RCTs evidence, demonstrating its validity and practical utility.

\subsection{Application to the World Values Survey}
\label{sec.exp_real}
Following the empirical strategy in the study by \cite{ying2025generalized}, we utilize data from the World Values Survey (WVS) Wave 7 date \citep{haerpfer2012world}\footnote{The data are available at \url{https://www.worldvaluessurvey.org/WVSDocumentationWV7.jsp}.} to examine whether moral attitudes toward dishonest behaviors are conditionally independent. Specifically, we focus on responses collected in Canada, in which data are collected from $N=3,997$ participants. The WVS includes several survey items asking respondents to evaluate the extent to which certain morally questionable actions can be justified. A possible underlying latent factor that governs their evaluations is personal honesty.

In our analysis, we examine whether attitudes toward two specific dishonest behaviors—cheating on government benefits $(X)$ and fare dodging $(Y)$—are conditionally independent given a latent honesty trait $U$ and a set of observed covariates $V$, \emph{e.g.} gender, age, highest educational level, and income level. Formally, our goal is to test the conditional independence $\mH_0:X\ind Y|U,V$. We follow \citet{ying2025generalized} and use responses to two additional questions—regarding tax evasion and bribe acceptance—as proxies, denoted by $Z$ and $W$, respectively. Previous studies \citep{halla2008taxes,chabova2017measuring} found that these proxies capture distinct behavioral domains. Specifically, the question on tax evasion (\emph{i.e.}, $X$) and the target behavior of benefits cheating (\emph{i.e.}, $W$) both capture fiscal compliance, whereas the question on bribe acceptance (\emph{i.e.}, $Z$) and the target behavior of fare dodging (\emph{i.e.}, $Y$) both capture attitudes toward corruption in public-service contexts. This supports the plausibility of the conditional independence assumption underlying the use
of $W$ and $Z$ as valid proxies.
\begin{table}[htbp]
\centering
\caption{Proxies for the latent honesty trait $U$.}
\begin{tabular}{ll}
\toprule
\textbf{Variable} & \textbf{Survey Question} \\
\midrule
$W$ & Cheating on taxes if you have a chance \\
$Z$ & Someone accepting a bribe in the course of their duties \\
\bottomrule
\end{tabular}
\label{tab:proxies}
\end{table}

We follow the same implementation for our procedure, and that in KCI on synthetic data. Table \ref{tab:pvalues_wvs} reports the $p$-values obtained from three different tests. Since the implementation of \textbf{Liu} does not support covariate adjustment, we omit it from the comparison. Among them, all two proxy-based methods—\emph{Our (single)} and \emph{Our (two)}—yield $p$-values that are much larger than the significance level (\emph{i.e.}, 0.05). In contrast, the $p$-value of the \textbf{KCI} test is nearly zero. This discrepancy likely stems from the fact that \textbf{KCI} fails to account for the confounding effect introduced by the latent variable ``honesty" which biases its results. 
\begin{table}[ht]
\centering
\caption{$p$-values of different methods for WVS.}
\begin{tabular}{lccc}
\toprule
 & Our(single) & Our(two)  & KCI \\
\midrule
\textbf{$p$-values} & 0.7975 &  0.7535  & 0.000\\
\bottomrule
\end{tabular}
\label{tab:pvalues_wvs}
\end{table}

These findings provide empirical support for the hypothesis that the observed relationship between attitudes toward cheating on government benefits and fare dodging is not causal but is rather driven by an individual's underlying honesty. Our results are consistent with the conclusions drawn by \citet{ying2025generalized}, which demonstrated that a single latent factor (together with the covariates) can effectively explain joint variations across multiple dishonesty-related behaviors in the WVS dataset.

\section{Conclusions and discussions}\label{sec.con}

This paper develops a general nonparametric framework for causal hypothesis testing in the presence of unmeasured confounding. We introduce the integral equation that links the outcome and NCO, and investigate the solvability of the equation for identifying the causal relation. A kernel-based procedure called PMCR is proposed for estimating the solution and constructing the test based on the residue. We then derive the asymptotic null distribution and
power properties of the test, and perform a bootstrapped implementation for computing the critical value. Within the linear Gaussian setting, we show that the causal relation may not be identifiable using only NCO, and demonstrate that additionally incorporating a NCE can effectively amend this problem. 

Several important directions remain for future research. First, while our current framework demonstrates favorable performance with low-dimensional covariates (see Appendix~\ref{appx.observed}), it remains an important direction to extend it to high-dimensional covariate, where nonparametric estimation becomes challenging. This is because our testing procedure is based on conditional moment restrictions, whose statistical power may degrade as the dimensionality of the conditioning variables increases \citep{tan2022integrated}. Addressing this limitation may involve incorporating dimension-reduction \citep{stute2002model} or projection-based strategies \citep{lavergne2008breaking}, which may improve the power of our method to modern high-dimensional problems.

Second, although our proposed framework accommodates settings in which all variables are either continuous or discrete, it is interesting to extend our method to handle mixed data types. As long as condition~\ref{assum.completeness} and regularity conditions \ref{assum.regularity condition}, the integral equation \eqref{eq.bridge_Y} remains valid even in mixed-type settings. When condition~\ref{assum.completeness} does not hold, it is unknown whether our current procedure is valid. This is because the completeness condition does not allow these variables are mixed in arbitrary forms. For example, when the proxy $W$ is discrete but the confounder $U$ is continuous, $W$ may fail to adequately capture the variation in $U$, making the completeness condition difficult to satisfy. Thus, it becomes necessary to develop improved nonparametric estimators that can accommodate mixed data structures. In the current estimation, we employ kernel-based estimators for continuous variables; however, it may not be applicable in the presence of mixed variables. In this case, we can employ neural networks-that can accommodate mixed data types-for optimizing the risk \eqref{eq.unrestricted} that transforms the conditional restrictions to unconditional ones \citep{dikkala2020minimax,wudoubly2024}. In particular, analyzing the estimation errors of such models and their asymptotic impact on the proposed hypothesis testing procedure would be crucial for ensuring valid inference.

Third, it is unknown whether the solution also exists under the alternative when NCO is strongly dependent on the outcome, although we have verified empirically that the power also drops as the dependency gets stronger. Besides, for our two-proxy identification strategy, the completeness \ref{assum.complete_U_X_Zx} requires that the treatment $X$ and the latent confounder $U$ have the same dimension. Since $X$ is typically univariate, this condition restricts $U$ to be effectively one-dimensional (such as discrete variables), which may limit its applicability when multiple latent continuous confounders are present. Addressing this limitation may involve incorporating dimension-reduction. 

Last but not least, our framework relies on a unidirectional assumption with known causal directions, which allows us to distinguish between negative control outcomes and negative control exposures. Recently, some studies \cite{li2024discovery} have shown that causal effects can be identified even in the presence of bidirectional relationships by leveraging invalid instrumental variables. Likewise, when causal directions are unknown, the NCOs we employ may not be valid \citep{yang2025double}, motivating future research on leveraging invalid NCOs for causal identification. This line of investigation also offers insights into integrating causal hypothesis testing into multivariate causal discovery algorithms \citep{spirtes2001causation} under latent confounding.


\bibliography{JASA-template/reference.bib}

\newpage
\pagenumbering{arabic}
\appendix
\bigskip
\begin{center}
{\large\bf SUPPLEMENTARY MATERIAL}
\end{center}

\section{Notations}

We introduce notations used throughout the appendix. 

{\renewcommand{\arraystretch}{1.5}
\begin{longtable}{ >{\centering\arraybackslash}m{3.4cm} | >{\centering\arraybackslash}m{12.2cm} } 
\caption{Notations.}
\renewcommand*{\arraystretch}{1.5}
\\ 
\hline\hline
Notation & Definition \\ 
\hline
\endfirsthead 

\hline
Notation & Definition \\ 
\hline
\endhead 

\hline
\endfoot 

\hline\hline
\endlastfoot 

 $Z,W, U$ & Negative control exposure, negative control outcome, and unobserved confounder \\ 
 
 $P(X)$ & $\left\{P(x_1),...,P(x_k)\right\}^\top$ for any discrete variables $X$ with $k$ categories \\

 $P(Y|X) $ &  $\begin{pmatrix}
	P( y_1|x_1)&		\cdots&		P( y_1|x_k)\\
	\vdots&		\ddots&		\vdots\\
	P( y_l|x_1)&		\cdots&		P( y_l|x_k)
\end{pmatrix}$ for any discrete variables $Y,X$ with $l,k$ categories \\

 $P(Y=y|X,Z)$ &  $ \begin{pmatrix}
	P( y|x_1,z_1)&		\cdots&		P( y|x_1,z_m)\\
	\vdots&		\ddots&		\vdots\\
	P( y|x_k,z_1)&		\cdots&		P( y|x_k,z_m )
\end{pmatrix} $ for any discrete variables $X,Z$ with $k,m$ categories \\

\hline
$\cH_W,\cH_X$ & The reproducing kernel Hilbert spaces (RKHS) defined on the domains of $W$ and $X$\\

$\phi_W(w),\phi_X(x)$ & The canonical feature map defined on the domains of $W$ and $X$ \\

$k_W(w,w'),k_X(x,x')$ & The reproducing kernel of the RKHS $\cH_W$ and $\cH_X$, respectively \\

$R(H)$ & The population loss function defined in \eqref{eq:mmr-sup-h} \\

$\wh{R}^{\lambda}(H)$ & The regularized empirical risk \eqref{eq:mmr-vstat} \\

$A, b_t(x):=b(x,t)$ & The operator, and the target \eqref{eq.operator_defination} \\

$\wh{A}, \wh{b}_t(x): =\wh{b}(x,t)$ & The plugging operator, and the target \eqref{eq.estimate_f_K} \\

$A^*, \wh{A}^*$ & The adjoint operator of $A$ and $\wh{A}$ that are respectively defined in \eqref{eq.operator_defination_adjoin_copy} and \eqref{eq.operator_defination_adjoin} \\

$(\lambda_{j},\varphi_{j},\phi_{j})_{j}$ &  The singular value decomposition of the operator $A$ \\

$\cH_{W,0}$ & The set of all
solutions defined in \eqref{eq. all_solution} \\

$H^{\lambda}_t(w):=H^{\lambda}(w,t)$ & The population Tikhonov
regularization solution \eqref{eq:h_pop} \\

$\wh{H}^{\lambda}_t(w):=\wh{H}^{\lambda}(w,t)$ & The empirical Tikhonov
regularization solution \eqref{eq:h_emp} \\

$H^0_t(w):=H^0(w,t)$ & Least norm solution in \eqref{eq.convert_Y} \\

$\mathrm{Ker}(A)$ & Null space of the operator $A$, \emph{i.e.}, $\mathrm{Ker}(A):=\{H:AH=0\}$ \\

$\mathrm{Ran}(A)$ & Range space of the operator $A$, \emph{i.e.}, $\mathrm{Ran}(A)=\{f:AH=f\}$ \\

\hline

$\cL^2\{F(w)\},\cL^2\{F(x)\}$ & The space of square-integrable functions with respect to the cumulative distribution function $F(w)$ and $F(x)$, respectively \\

$\mathcal{L}^{2}\{\cS \times \cT,\mu \times \mu\}$ & We say $\mG(s,t) \in \mathcal{L}^{2}\{\cS \times \cT,\mu \times \mu\}$ if $\iint |\mG(s,t)|^2 d\mu(s)d\mu(t) < \infty$ \\

$\varphi(\cdot,t),m(\cdot,s)$ & The weight function in section \ref{sec.equivalent}, below formula \eqref{eq.convert_Y} and in section \ref{sec.equivalent}, above formula \eqref{eq.tn_st}   \\

$g_s$ & $g_s=\mathbb{E} \{ m(X,s)\phi_W(W) \}$  \\

$U(W,Y,t),\wh{U}(W,Y,t)$ & The residual $\varphi(Y,t)-H^0(W,t)$ and estimated version $\varphi(Y,t)-\wh{H}^{\lambda}(W,t)$ \\

$T_n(s,t)$ & The statistics  defined in \eqref{eq.tn_st}\\

$\Delta_{\varphi,m}$ & The statistics defined in \eqref{eq.icm-statistics}\\

\hline

\hline
$\mathbb{E}(\cdot)$ & The expectation with respect to both a random variable and data\\

$\mathbb{P}(\cdot)$ & The expectation with respect to  a random variable alone\\

$\mathbb{P}_n(\cdot)$ & The empirical expectation with respect to a random variable given data\\

$\|\cdot\|_{\cF}$ & The norm with respect to space $\cF$\\
   
\end{longtable}
}
\section{Solution existence with a single proxy}
\label{appx.existence}
Let $\mathcal{L}^2\{F(x)\}$ denote the space of all square-integrable functions of $x$ with respect to a cumulative distribution function $F(x)$, which is a Hilbert space with inner product $\langle g_1,g_2\rangle=\int g_1(x)g_2(x)p(x)dx$. Let $T$ denote the operator: $\mathcal{L}^2\{F(w)\}\to \mathcal{L}^2\{F(u)\}$ such that  $Tg=\mathbb{E}\{g(W)|U=\cdot\}$ for any $g \in \mathcal{L}^2\{F(w)\}$, and let $(\lambda_{n},\varphi_{n},\phi_{n})_{n=1}^{\infty}$ denote a singular value decomposition of $T$. 

\begin{assumption}\label{assum.regularity condition}
     Assume the following conditions for all $y$: 
\begin{itemize}
   \item[(1)] $\iint p(u|w)p(w|u)dwdu<\infty $ and $ \int\{p(y|u)\}^2p(u)dx<\infty $;
    \item[(2)] $\sum_{n=1}^{\infty}\lambda_{n}^{-2}|\langle p(y|u),\phi_{n}\rangle_{\mathcal{L}^2\{F(u)\}}|^{2}<\infty$.
    \end{itemize} 
\end{assumption}

Condition \ref{assum.regularity condition} imposes integrability and smoothness conditions on the density $p(y|u)$. The first part ensures that the conditional expectation operator $T$ is compact. The second part requires that the Fourier coefficients of $p(y|u)$ converge sufficiently rapidly relative to the eigenvalues of $T$. These conditions are standard in the literature on inverse problems and proximal causal inference \citep{carrasco2007linear,miao2018identifying,liu2023causal}. As illustrated in example \ref{exm.linear2}, these conditions hold automatically in the linear Gaussian setting.

\subsection{Proof of Theorem~\ref{thm.exist_solution}}
We first show that under conditions in  Theorem~\ref{thm.exist_solution}, there exists a solution $h(w,y) \in \cL^2\{F(w)\}$ for all $u$, such that $p(y|u) =\int{h(w,y) p(w|u) dw}$. Our proof is based on Picard's theorem as stated in Lemma \ref{lemma.Picard}.

\uxsolutionsexist*
\begin{proof}
Note that for any fixed $y$, the mapping $h(w,y) \to \int h(w,y)p(w|u)dw$ can be regarded as a conditional expectation operator. Hence, our objective is to establish the existence of a solution $g$ to the operator equation $Tg=p(y|u)$, where 
$$
T: \mathcal{L}^2\{F(w)\}\rightarrow  \mathcal{L}^2\{F(u)\}: \ Tf =\mathbb{E}\{ f(W)|U=\cdot\}, \ f \in \cL^2\{F(w)\}.
$$
For convenience, we also consider another operator
$$
 S:  \mathcal{L}^2\{F(u)\}\rightarrow  \mathcal{L}^2\{F(w)\}: \ Sg =\mathbb{E} \{ g(U)|W=\cdot \}, \ g \in \cL^2\{F(u)\}.
$$
By Lemma~\ref{lemma.Picard} and condition~\ref{assum.regularity condition} (2) for $p(y|u)$, the desired conclusion follows if we can verify that $T$ is compact, $S$ is the adjoint operator of $T$, and that $p(u|x) \in \mathrm{Ker}(S)^{\perp}$.

\noindent\textbf{(i). $S$ is the adjoint operator of $T$.} 

For the operator $T$, for all $f\in \mathcal{L}^2\{F(w)\}$ and $g\in \mathcal{L}^2\{F(u)\}$, we compute
\begin{align*}
	\langle Tf,g \rangle_{\mathcal{L}^2\{F(u)\}}&=\mathbb{E}_U[ \mathbb{E} \{ f(W) |U \} g(U) ]=\mathbb{E}\{f(W)g(U)\}.
\end{align*}
Similarly, for the operator $S$, 
\begin{align*}
	\langle f,Sg \rangle_{\mathcal{L}^2\{F(w)\}}&=\mathbb{E}_W[ f(W) \mathbb{E} \{ g(U) |W \} ]=\mathbb{E}\{f(W)g(U)\}.
\end{align*}
Therefore, we obtain the adjoint relation
$$
\langle Tf,g \rangle_{\mathcal{L}^2\{F(x)\}}=\langle f,Sg \rangle_{\mathcal{L}^2\{F(w)\}}.
$$

\noindent\textbf{(ii). $T$ is compact.}

We define the integral kernel
\begin{equation}\label{eq.kernel}
    K(w,u) =\frac{p(w,u)}{p(w) p(u)}.
\end{equation}
For the operators introduced above, this yields the representations
\begin{align}
    Tf &=\int{K(w,u) f(w)dF (w)}=\mathbb{E}\{ f(W) |U \}, \ f\in \mathcal{L}^2\{F(w)\}, \label{eq.T}\\
	Sg &=\int{K(w,u) g(u)dF(u)}=\mathbb{E}\{ g(U) |W \}, \ g\in \mathcal{L}^2\{F(u)\}. \label{eq.S}
\end{align}
By the definition of $K$ in \eqref{eq.kernel}, we obtain
\begin{equation*}
    \iint |K(w,u)|^2p(w)p(u)dwdu = \iint p(w|u)p(u|w)dwdu \overset{(1)}{<} \infty,
\end{equation*}
where ``(1)" arises from condition~\ref{assum.regularity condition} (1). This implies the square-integrability of $K$. Hence, by Lemma \ref{lem:hs-kernel}, the operator $T$ is a Hilbert-Schmidt. It then follows from Lemma \ref{lem:hs-compact} that $T$ is compact.

\noindent\textbf{(iii). $p(y|u)\in \mathrm{Ker}(S)^{\perp}$ for any $y$.}

By the completeness assumption of $P(U|W)$, we have $\mathbb{E}\{g(U)|W\}=0$ if and only if $g(U)=0$, which means that $\mathrm{Ker}(S)=\{g(u)=0\}$. Therefore, we can obtain $\mathrm{Ker}(S)^{\perp}=\mathcal{L}^2\{F(u)\}$. Since $p(y|u)\in \mathcal{L}^2\{F(u)\}$, we have $p(y|u)\in \mathrm{Ker}(S)^{\perp}$. Combining the above three steps together, we obtain the conclusion.
\end{proof}

\solutionsexist*
\begin{proof}
    By Proposition~\ref{prop.bridge_U}, $h(w,y)$ satisfies the integral equation $ p(y|u) =\int{h(w,y) p(w|u) dw}$. Then, under $\mathbb{H}_0$ and $W\ind X|U$, we have
    \begin{align*}
    	p(y|x) &=\int{p(y|u) p(u|x) du}\\
    	&=\iint{ h(w,y) p(w|u)   p(u|x) dwdu}\\
        &= \int h(w,y)p(w|x)dw
    \end{align*}
   If $h(w,y)$ is square integrable with respect to $F(w)$, it is the solution to~\eqref{eq.bridge_Y}.  By Proposition~\ref{prop.bridge_U}, $h(w,y)\in \cL^2\{F(w)\}$, which means that $h(w,y)$ is square integrable with
respect to $F(w)$. Thus, we obtain $h(w,y)$ solves the integral equation \eqref{eq.bridge_Y}.
\end{proof}

\subsection{Proof of Corollary~\ref{corollary.exist_solution}}
\solutioncossexist*
\begin{proof}
    \textbf{(i).} We first prove that $H(w,t)$ is well-defined. To be specific, we take $\varphi$ to be the complex exponential function $e^{ity}$. By definition, $H(w,t) =\int{\varphi (y,t)h(w,y)dy}$. By the Cauchy-Schwarz inequality, for any fixed $y$,
    $$
    \int |h(w,y)|p(w)dw \le \left\{\int |h(w,y)|^2 p(w) dw\right\}^{1/2}\cdot\left\{\int  1^2 p(w)dw\right\}^{1/2}\le \left\{\int |h(w,y)|^2 p(w) dw\right\}^{1/2}.
    $$
Thus, since  $\int \|h(w,y)\|_{\cL^2\{F(w)\}}dy<\infty$, we have
$$
\iint |h(w,y)|p(w)dwdy\le \int \|h(w,y)\|_{\cL^2\{F(w)\}}dy<\infty.
$$
By Fubini’s theorem, we can obtain $\int |h(w,y)|dy<\infty$ for any $w$, which implies that
$$
\left| \int{e^{ity}h(w,y)dy} \right|\le \int{|e^{ity}|\cdot |h(w,y)|dy}<\infty.
$$

\textbf{(ii).} We prove that $H(w,t)\in \mathcal{L}^2\{F(w)\}$. By the Cauchy-Schwarz inequality, 
\begin{equation}\label{eq.Ht_in_L2}
    \begin{aligned}
	\int{| H(w,t) |^2p(w) dw}&=\int\left| \int{h(w,y) e^{ity}dy} \right|^2p(w) dw\\
    &= \int\left\{ \int{h(w,y_1) e^{ity_1}dy_1} \right\}\left\{ \int{h(w,y_2) e^{-ity_2}dy_2} \right\}p(w) dw\\
    &=\iint \left\{\int h(w,y_1) e^{ity_1} h(w,y_2) e^{-ity_2} p(w) dw\right\} dy_1dy_2\\
    &\le\iint \left\{\sqrt{\int |h(w,y_1)|^2p(w)dw }\sqrt{\int |h(w,y_2)|^2p(w)dw }\right\}dy_1dy_2\\
    &=\left(\int \|h(w,y)\|_{\cL^2\{F(w)\}}dy \right)^2<\infty.
\end{aligned}
\end{equation}
We complete the proof. 
\end{proof}

\subsection{Counter-example to the solvability of the first-order moment equation under $\mH_1$}

\begin{example}\label{example:linear_gaussian}
Suppose that $X, Y, U, W$ satisfy the linear Gaussian model, \emph{i.e.} $U=\varepsilon_U,X =\alpha_U U + \alpha_0+\varepsilon_X,W =\beta_U U +\beta_0 + \varepsilon_W, Y=\gamma_U U + \gamma_X X + \gamma_0 +\varepsilon_Y$, 
where $\varepsilon_U,\varepsilon_X,\varepsilon_W,\varepsilon_Y$ are Gaussian noises. Then, there exists $h(W)=b_w W+b_0$ such that $\mathbb{E}(Y|X)=\mathbb{E}\{h(W)|X\}$ holds, where $b_w = \frac{(\alpha_U^2+1)\gamma_X+\gamma_U\alpha_U}{\beta_U\alpha_U}$ and $b_0 = \gamma_0+\gamma_X\alpha_0-\frac{(\alpha_U^2+1)\gamma_X+\gamma_U\alpha_U}{\beta_U\alpha_U}\beta_0$. 
\end{example}
\begin{proof}
The goal is to solve $(b_w,b_0)$ in the following integral equation:
$$
\mathbb{E}(Y |X) =\mathbb{E}(b_w W+b_0|X).
$$
We note that
$$
\mathbb{E}(U|X)=\mathbb{E}(U)+\frac{\mathrm{Cov}(U,X)}{\mathrm{Var}(X)}\{X-\mathbb{E}(X)\}=\frac{\alpha_U(X-\alpha_0)}{\alpha_U^2+1}.
$$
For the left-hand side,
\begin{align*}
	\mathbb{E} (Y|X)&=\gamma _0+\gamma _XX+\gamma _U\mathbb{E} (U|X)\\
	&=\gamma _0+\gamma _XX+\gamma _U\frac{\alpha _U(X-\alpha _0)}{\alpha _{U}^{2}+1}\\
	&=\left( \gamma _0-\frac{\gamma _U\alpha _U\alpha _0}{\alpha _{U}^{2}+1} \right) +\left( \gamma _X+\frac{\gamma _U\alpha _U}{\alpha _{U}^{2}+1} \right) X.
\end{align*}
For the right-hand side, 
\begin{align*}
	\mathbb{E} \{g(W)|X\}&=\mathbb{E} (b_0+b_wW|X)\\
	&=b_0+b_w\mathbb{E} (\beta _0+\mu _UU|X)\\
	&=b_0+b_w\left\{ \beta _0+\mu _U\frac{\alpha _U(X-\alpha _0)}{\alpha _{U}^{2}+1} \right\}\\
	&=\left( b_0+b_w\beta _0-\frac{b_w\mu _U\alpha _U\alpha _0}{\alpha _{U}^{2}+1} \right) +\frac{b_w\mu _U\alpha _U}{\alpha _{U}^{2}+1}X.
\end{align*}
Equating coefficients of the constant and linear terms in $X$, we obtain the system
$$
\begin{cases}
\dfrac{b_w\beta_U\alpha_U}{\alpha_U^2+1}
= \gamma_X+\dfrac{\gamma_U\alpha_U}{\alpha_U^2+1}, \\[1.2ex]
b_0+b_w\beta_0-\dfrac{b_w\beta_U\alpha_U\alpha_0}{\alpha_U^2+1}
=\gamma_0-\dfrac{\gamma_U\alpha_U\alpha_0}{\alpha_U^2+1}.
\end{cases}
$$
Solving this equation yields
$$
b_w=\frac{(\alpha_U^2+1)\gamma_X+\gamma_U\alpha_U}{\beta_U\alpha_U},\qquad
b_0=\gamma_0+\gamma_X\alpha_0-\frac{(\alpha_U^2+1)\gamma_X+\gamma_U\alpha_U}{\beta_U\alpha_U}\beta_0.
$$
\end{proof}

\subsection{Verification of Bochner integrability in Corollary~\ref{corollary.exist_solution}}\label{appx.Bochner}

\begin{example}\label{exm.linear2}
Suppose that $X, Y, U, W$ satisfy the linear Gaussian model, \emph{i.e.} $U=\varepsilon_U,X =\alpha_U U + \alpha_0+\varepsilon_X,W =\beta_U U +\beta_0 + \varepsilon_W, Y=\gamma_U U + \gamma_0 +\varepsilon_Y$, where $\varepsilon_X, \varepsilon_W, \varepsilon_Y, \varepsilon_U$ are standard normal. Then if $1-\gamma_U^2/\beta_U^2>0$, the solution of \eqref{eq.bridge_Y} is given by:
$$
h(w,y)=\frac{1}{\sqrt{1-\left( \frac{\gamma_U}{\beta_U} \right)^2}}\phi \left( \frac{y-\frac{\gamma_U}{\beta_U}w+\frac{\gamma_U}{\beta_U}\beta_0-\gamma_0}{\sqrt{1-\left( \frac{\gamma_U}{\beta_U} \right)^2}} \right). 
$$
Besides, we have
$$
\int \left\{\int |h(w,y)|^2 p(w) dw\right\}^{1/2} dy =\left(\frac{\beta_U^2+\gamma_U^2+2\beta_U^2\gamma_U^2}{\beta_U^2-\gamma_U^2}\right)^{1/4}.
$$
\end{example}
\begin{proof}
Based on the data generation structure, we can obtain the joint distribution
\begin{equation}\label{eq:joint_normal}
    \left( \begin{array}{c}
	U\\
	X\\
	W\\
	Y\\
\end{array} \right) \sim \cN\left\{ \left( \begin{array}{c}
	0\\
	\alpha_0\\
	\beta_0\\
	\gamma_0\\
\end{array} \right) ,\left( \begin{matrix}
	1&		\alpha_U&		\beta_{U}&		\gamma_U\\
	\alpha_U&		1+\alpha_U^{2}&		\alpha_U\beta_{U}&		\alpha_U\gamma_U\\
	\beta_{U}&		\alpha_U\beta_{U}&		1+\beta_{U}^{2}&		\beta_U\gamma_U\\
	\gamma_U&		\alpha_U\gamma_U&		\beta_U\gamma_U&		\gamma_U^2+1
\end{matrix} \right) \right\}.
\end{equation}
We first get the conditional distributions $p(y|u)$ and $p(w|u)$. By standard Gaussian conditioning formulas, we have
$$
\begin{aligned}
	W|U=u&\sim \cN\left\{ \mu_W+\frac{\mathrm{Cov}(W,U)}{\mathrm{Var}(U)}(u-\mu_{U}),\mathrm{Var}(W)\left( 1-\frac{\mathrm{Cov}^2(W,U)}{\mathrm{Var}(U)\cdot \mathrm{Var}(W)} \right) \right\}\\
	&\sim \cN\left\{ \beta_0+\beta_U u,1\right\} 
\end{aligned}
$$
$$
\begin{aligned}
    Y|U=u&\sim \cN\left\{ \mu_Y+\frac{\mathrm{Cov}(Y,U)}{\mathrm{Var}(U)}(u-\mu_{U}),\mathrm{Var}(Y)\left( 1-\frac{\mathrm{Cov}^2(Y,U)}{\mathrm{Var}(U)\cdot \mathrm{Var}(Y)} \right) \right\} \\
	&\sim \cN\left\{ \gamma_0+\gamma_U u, 1\right\},
\end{aligned}
$$
Applying Lemma~\ref{lemma.miao-example}, we have
$$
h(w,y)=\frac{1}{\sqrt{1-\left( \frac{\gamma_U}{\beta_U} \right)^2}}\phi \left( \frac{y-\frac{\gamma_U}{\beta_U}w+\frac{\gamma_U}{\beta_U}\beta_0-\gamma_0}{\sqrt{1-\left( \frac{\gamma_U}{\beta_U} \right)^2}} \right). 
$$

Next, we compute the conditional distributions $p(w|x)$ and $p(y|x)$. By standard Gaussian conditioning formulas, we have
\begin{align*}
	W|X=x&\sim \cN\left\{ \mu_W+\frac{\mathrm{Cov}(W,X)}{\mathrm{Var}(X)}(x-\mu_X),\mathrm{Var}(W)\left( 1-\frac{\mathrm{Cov}^2(W,X)}{\mathrm{Var}(X)\cdot \mathrm{Var}(W)} \right) \right\}, \\
	&\sim \cN\left\{ \frac{\alpha_U\beta_U}{\alpha_U^{2}+1}x-\frac{\alpha_U\beta_U}{\alpha_U^{2}+1}\alpha_0+\beta_0,\beta_{U}^{2}+1-\frac{\left( \alpha_U\beta_U \right)^2}{\alpha_U^{2}+1} \right\}
\end{align*}
\begin{align*}
	Y|X=x&\sim \cN\left\{ \mu_W+\frac{\mathrm{Cov}(Y,X)}{\mathrm{Var}(X)}(x-\mu_X),\mathrm{Var}(Y)\left( 1-\frac{\mathrm{Cov}^2(Y,X)}{\mathrm{Var}(X)\cdot \mathrm{Var}(Y)} \right) \right\}, \\
	&\sim \cN\left\{ \frac{\alpha_U\gamma_U}{\alpha_U^{2}+1}x-\frac{\alpha_U\gamma_U}{\alpha_U^{2}+1}\alpha_0+\gamma_0,\gamma_{U}^{2}+1-\frac{\left( \alpha_U\gamma_U \right)^2}{\alpha_U^{2}+1} \right\}
\end{align*}
Applying Lemma~\ref{lemma.miao-example}, we have
$$
h(w,y)=\frac{1}{\sqrt{1-\left( \frac{\gamma_U}{\beta_U} \right)^2}}\phi \left( \frac{y-\frac{\gamma_U}{\beta_U}w+\frac{\gamma_U}{\beta_U}\beta_0-\gamma_0}{\sqrt{1-\left( \frac{\gamma_U}{\beta_U} \right)^2}} \right). 
$$

Finally, we verify Bochner integrability. Define $\rho = \gamma_U/\beta_U$ and $\sigma_W^2=1+\beta_U^2$. Note that
$$
|h(w,y)|^2 =\frac{1}{2\pi(1-\rho^2)} \exp\left\{- \left(\frac{y - \rho w + \rho\beta_0 - \gamma_0}{\sqrt{1-\rho^2}}\right)^2 \right\}
$$
$$
p(w) = \frac{1}{\sqrt{2\pi\sigma_W^2}} \exp\left\{-\frac{(w-\beta_0)^2}{2\sigma_W^2}\right\}
$$
Thus, we have
\begin{align*}
    \int |h(w,y)|^2 p(w) dw &= \frac{1}{2\pi (1-\rho^2)\sqrt{2\pi \sigma_W^2}}\int \exp\left\{-\frac{(y - \rho w + \rho \beta_0 - \gamma_0)^2}{1-\rho^2}\right\} \cdot  \exp\left\{-\frac{(w-\beta_0)^2}{2\sigma_W^2}\right\} dw\\
    &= \frac{1}{2\pi (1-\rho^2)\sqrt{2\pi \sigma_W^2}}\int \exp\left(-\frac{1}{2}Aw^2+Bw+C\right) dw,
\end{align*}
where 
$$
A = \frac{2\rho^2}{1 - \rho^2} + \frac{1}{\sigma_W^2}\, ,B = \frac{2\rho(y + \rho\beta_0 - \gamma_0)}{1 - \rho^2} + \frac{\beta_0}{\sigma_W^2}\,  ,C = -\frac{(y + \rho\beta_0 - \gamma_0)^2}{1 - \rho^2} - \frac{\beta_0^2}{2\sigma_W^2}
$$
Applying the standard Gaussian integral identity, \emph{i.e.},
$$
\int \exp\left(- \frac{1}{2} A w^2 + B w\right) dw = \sqrt{\frac{2\pi}{A}} \exp\left( \frac{B^2}{2A} \right),\quad A>0,
$$
we have
\begin{align*}
    \left\{\int |h(w,y)|^2 p(w) dw\right\}^{1/2}&= \frac{1}{\sqrt{2\pi (1-\rho^2)} ( \sigma_W^2)^{1/4} A^{1/4}} \exp\left( \frac{B^2}{4A} + \frac{C}{2} \right)\\
    &=  \frac{1}{\sqrt{2\pi (1-\rho^2)} ( \sigma_W^2)^{1/4} A^{1/4}}\exp\left\{ -\frac{(y-\gamma_0)^2}{2+\rho^2(4\sigma_W^2-2)}\right\}.
\end{align*}
Since $1-\rho^2>0$ and $1+\beta_U^2>0$, it follows that $A>0$. Next, applying the Gaussian integral
$$
\int \exp\left\{- \alpha(y-\mu)^2\right\} dy = \sqrt{\frac{\pi}{\alpha}} ,\quad \alpha>0,
$$
with $\alpha=\frac{1}{2+\rho^2(4\sigma_W^2-2)}=\frac{1}{2+\rho^2(4\beta_U^2+2)}>0$ and $\mu = \gamma_0$, we obtain
$$
\int \exp\left\{ -\frac{(y-\gamma_0)^2}{2+\rho^2(4\sigma_W^2-2)}\right\} dy =\sqrt{2\pi}\sqrt{1+\rho^2(2\sigma_W^2-1)}.
$$
Combining all terms, we have
$$
\int \left\{\int |h(w,y)|^2 p(w) dw\right\}^{1/2} dy =\left(\frac{1+\rho^2+2\gamma_U^2}{1-\rho^2}\right)^{1/4} < \infty. 
$$
We complete the proof. 
\end{proof}
 
\begin{lemma}
    The integral equation $p(u|x)=\int g(w,u)p(w|x)dw$ has no solution in the linear Gaussian setting, as introduced in example~\ref{exm.linear2}. 
\end{lemma}

\begin{proof}
From \eqref{eq:joint_normal}, we have
\begin{align*}
U | X = x 
&\sim \mathcal{N}\left\{ 
\beta_U + \frac{\mathrm{Cov}(U,X)}{\mathrm{Var}(X)}(x - \mu_X),\;
\mathrm{Var}(U)\left(1 - \frac{\mathrm{Cov}^2(U,X)}{\mathrm{Var}(X)\mathrm{Var}(U)}\right)
\right\} \\
&= \mathcal{N}\left\{
\frac{\alpha_U}{\alpha_U^2 + 1}(x - \alpha_0),\;
\frac{1}{\alpha_U^2 + 1}
\right\}=\cN(\gamma^0_{UX}+\gamma^1_{UX}X,\sigma_{UX}^{2}).
\end{align*}
Similarly,
\begin{align*}
W| X = x 
&\sim \mathcal{N}\left\{
\mu_W + \frac{\mathrm{Cov}(W,X)}{\mathrm{Var}(X)}(x - \mu_X),\;
\mathrm{Var}(W)\left(1 - \frac{\mathrm{Cov}^2(W,X)}{\mathrm{Var}(X)\mathrm{Var}(W)}\right)
\right\} \\
&= \mathcal{N}\left\{
\frac{\alpha_U \beta_U}{\alpha_U^2 + 1}(x - \alpha_0) + \beta_0,\;
1 + \frac{\beta_U^2}{\alpha_U^2 + 1}
\right\}=\mathcal{N}(\beta_{WX}^0+\beta_{WX}^1X,\sigma_{WX}^{2}).
\end{align*}
By Lemma~\ref{lemma.miao-example}, the solution $g(w,u)$, if it exists, must take a Gaussian form. Hence, its variance parameter $\sigma^2$ must be positive. However, direct computation yields
\begin{align*}
    \sigma^2&=\frac{1}{\alpha_U^2 + 1}-\frac{\alpha_U^2/(\alpha_U^2+1)^2}{\alpha_U^2\beta_U^2/(\alpha_U^2+1)^2}\left(1 + \frac{\beta_U^2}{\alpha_U^2 + 1}\right)\\
    &=-\frac{1}{\beta_U^2}<0,
\end{align*}
which is impossible. Therefore, the solution $g(w,u)$ does not exist.
\end{proof}

\subsection{Explaination of remark~\ref{remark.icml}}
\label{sec.explain-remark-1}
\begin{theorem}\label{thm:kernel_identity}
Let $p(u|x)$ and $p(w|x)$ be a conditional probability density. Define the kernel
$$
K(u,u') := \int g(w,u)  p(w|u')  dw.
$$
Suppose the integral equation $p(u|x) = \int K(u, u') p(u'|x)  du'$ holds for a dense set of probability densities (\emph{e.g.}, sequences approximating Dirac deltas, such as Gaussians with vanishing variance), denoted by $\mathcal{F} = \{ p(\cdot|x) | x \in \mathcal{X} \} $. Then, $ K(u, u') = \delta(u - u') $, and this kernel is unique.
\end{theorem}
\begin{proof}
Define the integral operator $ T $ with kernel $ K $:
\begin{equation}
\label{eq.tf}
    (Tf)(u) = \int K(u, u') f(u')  du',
\end{equation}
where $ f(u) = p(u|x) \in \mathcal{F} $. The given equation implies that
$$
Tf = f \quad \text{for all } f \in \mathcal{F},
$$
i.e., $ T $ acts as the identity operator on $ \mathcal{F} $. To determine $K$, consider a sequence of probability densities $\rho_\epsilon(u) \in \mathcal{F}$ approximating the Dirac delta:
$$
\rho_\epsilon(u) = \frac{1}{\sqrt{2\pi \epsilon}} \exp\left( -\frac{u^2}{2\epsilon} \right),
$$
which satisfies $ \rho_\epsilon(u) \geq 0 $, $ \int\rho_\epsilon(u)  du = 1 $, and converges to $\delta(u)$ in the distributional sense:
$$
\lim_{\epsilon \to 0^+} \int \rho_\epsilon(u) \phi(u)  du = \phi(0)
$$
for any continuous, bounded test function $ \phi $. Let $ f(u') = \rho_\epsilon(u' - v) $. By \eqref{eq.tf}, we have
$$
\rho_\epsilon(u - v) = \int K(u, u') \rho_\epsilon(u' - v)  du'.
$$
Substitute $u' = v + \sqrt{\epsilon} t$, $du' = \sqrt{\epsilon} dt$, and denote $\rho(t) := \frac{1}{\sqrt{2\pi}} e^{-t^2/2}$. The equation becomes
$$
\rho_\epsilon(u - v) = \int K(u, v + \sqrt{\epsilon} t) \rho(t)  dt.
$$
Test with a continuous, bounded function $ \phi(u) $:
$$
\int \rho_\epsilon(u - v) \phi(u)  du = \int \left\{ \int K(u, v + \sqrt{\epsilon} t) \rho(t)  dt \right\} \phi(u)  du.
$$
For the left-hand side, change variables $u = v + \sqrt{\epsilon} s$, $du = \sqrt{\epsilon} ds$:
$$
\int \rho_\epsilon(u - v) \phi(u)  du = \int \rho(s) \phi(v + \sqrt{\epsilon} s)  ds \to \phi(v) \quad \text{as } \epsilon \to 0^+.
$$
For the right-hand side, applying dominated convergence,
$$
\int \left\{ \int K(u, v + \sqrt{\epsilon} t) \rho(t)  dt \right\} \phi(u)  du \to \int K(u, v) \phi(u)  du.
$$
This gives us
$\phi(v) = \int K(u, v) \phi(u)  du$ for any bounded and continuous function $\phi$. This implies that the kernel $K(u,v)$ acts as the Dirac delta distribution, \emph{i.e.}, $K(u,v)=\delta(u-v)$. By Theorem 1.3.1 in \cite{friedlander1998introduction}, which establishes the uniqueness of distributions satisfying such an identity for a suitable class of test functions, we conclude that $K(u,v)=\delta(u-v)$.
\end{proof}
\section{Hypothesis testing with discrete variables}\label{sec:discrete_integral}

In this section, we introduce how to test the null hypothesis in the discrete case. Section \ref{appx.corollary} provide the proof of Corollary \ref{cor:discrete_bridge}. Section~\ref{sec:estimation} gives a detailed introduction to the least-squares estimation described in the main text. Finally, Section~\ref{sec:asymptotic_properties} establishes the asymptotic validity, including level and power, of the proposed statistics.

\subsection{Proof of Corollary \ref{cor:discrete_bridge}}\label{appx.corollary}


\DiscreteBridge*
\begin{proof}
\noindent\textbf{Step 1: Completeness and full column rank.}
By definition, completeness of $W$ relative to $U$ means that, for any function $g:\mathcal{U}\to\mathbb{R}$,
$$
\sum_{\ell=1}^{|\mathcal{U}|} g(u_\ell)P(U=u_\ell | W=w_k) = 0, 
\quad \forall k=1,...,|\mathcal{W}|
\quad \implies \quad g(u_\ell) = 0,\ \forall \ell.
$$
This means $P(U | W)$ is full row-rank and $|\cW| \geq |\cU|$. By Bayes' rule, we have 
\begin{align*}
    P(W|U) = \mathrm{diag}\left\{ P(w^{(1)}),...,P(w^{(|\cW|)}) \right\} P(U|W)^\top \mathrm{diag}\left\{ \frac{1}{P(u^{(1)})},...,\frac{1}{P(u^{(|\cU|)})} \right\},
\end{align*}
which means $P(W|U)$ is full-column rank since $P(u_i)$ and $P(w_j)$ is positive for each $i \leq k$ and $j \leq l$. 

\noindent\textbf{Step 2: Bridge function for $P(U | X)$.}
Note that $P(W|X)=P(W|U)P(U|X)$, since $P(W | U)$ has full column rank with $|\mathcal{W}|\ge|\mathcal{U}|$, it is left invertible. That is, there is a $|\mathcal{U}|\times |\mathcal{W}|$ matrix denoted as $P(W | U)^{+}$ such that $P(W | U)^{+}P(W | U)=\I_{|\mathcal{U}|}$. Thus, we obtain that:
\begin{equation}\label{eq:bridge_U_dis}
    P(W | U)^{+} P(W | X) = P(U | X).
\end{equation}

\noindent\textbf{Step 3: Bridge function for $P(Y | X)$.}
Under $\mH_0$, we have the factorization $P(y | X) = P(y | U)P(U | X)$. 
Substituting~\eqref{eq:bridge_U_dis} yields
\begin{equation}\label{eq:bridge_Y_dis}
    P(y | X) = P(y | U) P(W | U)^{+} P(W | X),
\end{equation}
which is the discrete counterpart of~\eqref{eq.bridge_Y}. 
This shows the existence of a valid bridge representation.

\noindent\textbf{Step 4: Uniqueness in the square case.}
If $|\mathcal{W}| = |\mathcal{U}|$, then $P(W | U)$ is a square, full-rank matrix and hence invertible. 
In this case, the solution to~\eqref{eq:bridge_U_dis} is unique, and we obtain explicitly
$$
\h(W,y)  = \{P(W | U)^{-1}P(y | U)\}^\top.
$$
This completes the proof.
\end{proof}

To test whether $P(y|X)$ equals to $\h(W,y)^\top P(W | X)$, we consider the following equation:
$$
\sum_{y\in\mathcal{Y}}\varphi(y,t)P(y|X)= \sum_{y\in\mathcal{Y}}\varphi(y,t)\h(W,y)^\top P(W|X). \ \forall \thinspace t \in \cT,
$$
where $\varphi(Y,t)$ can be chosen as $\exp(ity)$, where $\cT$ can be an arbitrarily chosen neighborhood around $0$. Define $\H(W,t)=\sum_{y=1}^{|\mathcal{Y}|}\varphi(y,t)\h(W,y)$, which can be rewritten as the vector of length $|\mathcal{W}|$ given by $[H(w^{(1)}, t), ..., H(w^{(|\mathcal{W}|)}, t)]^\top$. Then, we have \eqref{eq.convert_Y_dis}.
In practice, we can set $\varphi(Y,t) = \sin(ty)$ and $\cos(ty)$, and test whether \eqref{eq.convert_Y_dis} holds for these choices. Finally, we provide the assumptions required for estimation and hypothesis testing, which is similar to \cite{miao2018identifying}.
\begin{assumption}\label{ass:cardinality}
    We assume $|\mathcal{X}|>|\mathcal{W}|$ and $P(W|X)$ has full row rank.
\end{assumption}
\begin{remark}
    Condition \ref{ass:cardinality} has been similarly made in \cite{miao2018identifying}, which ensures that $P(W|U)$ is invertible.
\end{remark}

\subsection{Estimation}\label{sec:estimation}
Below we present \textbf{(i)} how to compute the conditional-estimator $\sum_{y \in \mathcal{Y}} \varphi(y, t) P(y | x)$ and \textbf{(ii)} closed-form estimator $\wh{\H}_t$. Define the cell counts as
$n(x) := \#\{i:x_i=x\}$, $n(x,w) := \#\{i:x_i=x, w_i=w\}$, for $x\in\mathcal{X}$ and $w\in\mathcal{W}$.

\noindent\textbf{(i) Empirical Conditional-Frequency Estimator.}

The functional equation \eqref{eq.convert_Y_dis} implies that, for each $x \in \mathcal{X}$,
$$
q(x,t) := \sum_{y \in \mathcal{Y}} \varphi(y, t) P(y | x) = \mathbb{E}[\varphi(Y, t) | X = x].
$$
The empirical conditional-frequency estimator is the sample analogue:
$$
\wh{q}(x,t) :=  \frac{1}{n(x)} \sum_{i:x_i=x} \varphi(y_i,t).
$$
This is an unbiased and consistent estimator of $q(x,t)$ under standard moment conditions. Then, we can denote $\wh{\q}_t$ as $\wh{\q}_t := \left\{ \wh{q}(x^{(1)},t), ..., \wh{q}\left(x^{(|\mathcal{X}|)},t \right) \right\}^\top$. 

\noindent\textbf{(ii). Closed-Form Estimator $\wh{\H}_t$.}

The empirical conditional probability matrix has entries $\wh{P}(w | x) = n(x, w) / n(x)$, yielding the matrix $\wh{P}(W | X)$ of dimension $|\mathcal{W}| \times |\mathcal{X}|$ with $(j, k)$-th entry $\wh{P}(w_j | x^{(k)})$. Define $\wh{Q} := \wh{P}(W | X)^\top$, a matrix of dimension $|\mathcal{X}| \times |\mathcal{W}|$.

The functional equation \eqref{eq.convert_Y_dis} in matrix form is $\q_t = \Q \mathbf{H}_t$, where $\mathbf{H}_t := [H(w^{(1)}, t), ..., H(w^{(|\mathcal{W}|)}, t)]^\top$. The plug-in estimator solves the empirical linear system
$$
\wh{\Q} \wh{\mathbf{H}}_t = \wh{\q}_t.
$$
Since $\wh{\Q}$ has full column rank, the closed-form solution via ordinary least squares is
$$
\wh{\mathbf{H}}_t = (\wh{\Q}^\top \wh{\Q})^{-1} \wh{\Q}^\top \wh{\q}_t.
$$

\begin{theorem}\label{thm:consistency}
Under conditions~\ref{assum.completeness} and \ref{ass:cardinality}, as $n\to\infty$,
$$
\widehat{\mathbf{q}}_t \xrightarrow{p} \mathbf{q}_t,\qquad
\widehat{\mathbf{Q}} \xrightarrow{p} \mathbf{Q},\qquad
\widehat{\mathbf{H}}_t \xrightarrow{p} (\mathbf{Q}^\top\mathbf{Q})^{-1}\mathbf{Q}^\top\mathbf{q}_t.
$$
\end{theorem}
\begin{proof}
    \noindent\textbf{Step 1: Element-wise convergence of $\widehat{\mathbf{q}}_t$.}

For fixed $x\in\mathcal{X}$, since $\varphi(Y,t)$ is uniformly bounded for any $Y$ and $t$, we can obtain $\mathbf{1}\{X=x\}\varphi(Y,t)$ is integrable. By the weak law of large numbers (WLLN)
$$
\frac{1}{n}\sum_{i=1}^n \mathbf{1}\{X_i=x\}\varphi(Y_i,t) \xrightarrow{p} \mathbb{E}[\mathbf{1}\{X=x\}\varphi(Y,t)]
= P(x)\mathbb{E}[\varphi(Y,t)| X=x].
$$
Similarly $\frac{1}{n}\sum_{i=1}^n \mathbf{1}\{X_i=x\}\xrightarrow{p} P(x)>0$. By the continuous mapping theorem, we have:
$$
\widehat{q}(x,t)\xrightarrow{p}\mathbb{E}[\varphi(Y,t)| X=x]=q(x,t).
$$
Since $\mathcal{X}$ is finite, the convergence holds jointly for all $x$, hence $\widehat{\mathbf{q}}_t\xrightarrow{p}\mathbf{q}_t$.

\noindent\textbf{Step 2: Elementwise convergence of $\widehat{\mathbf{Q}}$.}

By the WLLN,
$$
\frac{1}{n}\sum_{i=1}^n \mathbf{1}\{X_i=x,W_i=w\}\xrightarrow{p}P(X=x,W=w),
\qquad
\frac{1}{n}\sum_{i=1}^n \mathbf{1}\{X_i=x\}\xrightarrow{p}P(x)>0,
$$
and by the continuous mapping theorem, the ratio converges in probability to
$P(W=w| X=x)$. Since $\mathcal{X}\times\mathcal{W}$ is finite, the convergence is entrywise for $\widehat{\mathbf{Q}}$, so $\widehat{\mathbf{Q}}\xrightarrow{p}\mathbf{Q}$.

\noindent\textbf{Step 3: Consistency of $\widehat{\mathbf{H}}_t$.}

From Step 2 we have $\widehat{\mathbf{Q}}\xrightarrow{p}\mathbf{Q}$. Hence $\widehat{\mathbf{Q}}^\top\widehat{\mathbf{Q}}\xrightarrow{p}\mathbf{Q}^\top\mathbf{Q}$. By condition \ref{ass:cardinality}, $\mathbf{Q}^\top\mathbf{Q}$ is nonsingular, and inversion is continuous in a neighbourhood of an invertible matrix. Therefore, we have $(\widehat{\mathbf{Q}}^\top\widehat{\mathbf{Q}})^{-1}\xrightarrow{p}(\mathbf{Q}^\top\mathbf{Q})^{-1}$. Combining this with $\widehat{\mathbf{Q}}^\top\xrightarrow{p}\mathbf{Q}^\top$ and $\widehat{\mathbf{q}}_t\xrightarrow{p}\mathbf{q}_t$ from Step 1, and using the continuous mapping theorem for matrix multiplication, we obtain
$$
\widehat{\mathbf{H}}_t
=
(\widehat{\mathbf{Q}}^\top\widehat{\mathbf{Q}})^{-1}\widehat{\mathbf{Q}}^\top\widehat{\mathbf{q}}_t
\xrightarrow{p}
(\mathbf{Q}^\top\mathbf{Q})^{-1}\mathbf{Q}^\top\mathbf{q}_t.
$$
We complete the proof. 
\end{proof}

\subsection{Asymptotic properties}
\label{sec:asymptotic_properties}

\ProjCltNull*
\begin{proof}
The proof contains four steps.

\noindent\textbf{(i). Equivalent transformation.}

Recall that $n(x)=\sum_{i=1}^n\mathbf 1\{x_i=x\}$ and $n(x,w):=\sum_{i=1}^n\mathbf 1\{x_i=x,w_i=w\}$. For fixed $x^{(k)}\in\cX$ and $t\in\cT$, we have
\begin{align*}
    \{\wh{\q}_t-\wh{\Q} \wh{\mathbf{H}}_t\}_k&=\wh{q}(x^{(k)},t)- \sum_{w \in \mathcal{W}} H(w, t)\wh{P}(w | x^{(k)}) \\
    &= \frac{1}{n(x^{(k)})} \sum_{i:x_i=x^{(k)}} \varphi(y_i,t)-\sum_{w \in \mathcal{W}} \wh{H}(w, t)\frac{n(x^{(k)},w)}{n(x^{(k)})}\\
    &=\frac{1}{n(x^{(k)})} \sum_{i:x_i=x^{(k)}} \varphi(y_i,t)-\frac{1}{n(x^{(k)})}\sum_{i:x_i=x^{(k)}} \wh{H}(w_i, t)\\
    &=\frac{1}{n(x^{(k)})}\sum_{i=1}^n \{\varphi(y_i,t)- \wh{H}(w_i, t)\}\mathbf{1}\{x_i=x^{(k)}\}.
\end{align*}
Define $\wh{\P}:= \wh{\Q}(\wh{\Q}^\top\wh{\Q})^{-1}\wh{\Q}^\top$ and  $\wh{\D}:=\diag\{n(x^{(1)})/n,...,n(x^{|\mathcal{X}|})/n\}$. Since $\wh{\mathbf{H}}_t = (\wh{\Q}^\top \wh{\Q})^{-1} \wh{\Q}^\top \wh{\q}_t$, we have $\wh{\q}_t-\wh{\Q} \wh{\mathbf{H}}_t=(\I-\wh{\P})\wh{\q}_t$. Therefore,
$$
\T_n(t) = \sqrt{n}\wh{\D}(\wh{\q}_t-\wh{\Q}\wh{\H}_t) =\sqrt{n}\wh{\D}(\I-\wh{\P})\wh{\q}_t, 
$$
and each component $k$ equals
$$
\T_n^{(k)}(t) = \sqrt{n}\frac{n(x^{(k)})}{n}\cdot \frac{1}{n(x^{(k)})}\sum_{i=1}^n\{\varphi(Y_i,t)-\wh{H}(W_i,t)\}\mathbf{1}\{x_i=x^{(k)}\}.
$$

\noindent\textbf{(ii). A functional CLT for $\sqrt{n}(\wh{\q}_t-\q_t)$.}

Note that
$$
\wh{\q}_t-\q_t=\left( \begin{array}{c}
	 \frac{1}{n(x^{(1)})} \sum_{i:x_i=x^{(1)}} \varphi(y_i,t)-\mathbb{E}\{\varphi(Y,t)|X=x^{(1)}\}\\
	\vdots\\
	\frac{1}{n(x^{|\mathcal{X}|})} \sum_{i:x_i=x^{|\mathcal{X}|}} \varphi(y_i,t)-\mathbb{E}\{\varphi(Y,t)|X=x^{(|\mathcal{X}|)}\}
\end{array} \right).
$$
We aim to prove the convergence of the $k$-th component of the sequence $\sqrt{n}(\wh{\mathbf{q}}_t - \mathbf{q}_t)$. Notice that
\begin{align*}
    \sqrt{n}\{\wh{\q}^{(k)}_t-\q^{(k)}_t\}&=\frac{\sqrt{n}}{n(x^{(k)})} \sum_{i:x_i=x^{(k)}} \varphi(y_i,t)-\mathbb{E}\{\varphi(Y,t)|X=x^{(k)}\}\\
    &=\left\{\frac{n(x^{(k)})}{n}\right\}^{-1}\frac{1}{\sqrt{n}} \sum_{i=1}^n \left[\varphi(y_i,t)-\mathbb{E}\{\varphi(Y,t)|X=x^{(k)}\}\right]\mathbf{1}(x_i=x^{(k)})\\
    & \overset{\mathrm{def}}{=} \left\{\frac{n(x^{(k)})}{n}\right\}^{-1}\frac{1}{\sqrt{n}} \sum_{i=1}^n Z_i(t).
\end{align*}
We will prove that $\frac{1}{\sqrt{n}} \sum_{i=1}^n Z_i^{(k)}(t)$ converges weakly to a zero-mean Gaussian process by applying Lemma~\ref{ central limit theorem}. We first verify the $k$-th component $Z_i^{(k)}(t)$ of $Z_i(t)$ is zero mean,
\begin{align*}
    \mathbb{E}\{Z_i^{(k)}(t)\} &= \mathbb{E}[\mathbf{1}\{x_i = x^{(k)}\} \varphi(y_i, t)]-\mathbb{E}[\mathbf{1}\{x_i = x^{(k)}\}\mathbb{E}\{\varphi(Y,t)|X=x^{(k)}\}] \\
    &= P(x^{(k)}) \cdot [\mathbb{E}\{\varphi(Y,t)|X=x^{(k)}\}-\mathbb{E}\{\varphi(Y,t)|X=x^{(k)}\}] = 0.
\end{align*}
Next, we verify the integrability condition
\begin{equation}
\label{eq.var-infinity_dis}
    \mathbb{E} \left( \| Z_i^{(k)} \|_{\cL^2(\cT, \nu)}^2 \right) <\infty, 
\end{equation}
where $\|\cdot\|_{\cL^2(\cT, \nu)}^2 = \int_{\cT} (\cdot)^2  d\nu(t)$ and $\nu$ is the measure on $\cT$. Since $\varphi(Y,t)$ is uniformly bounded for any $Y$ and $t$ (say, $|\varphi(Y,t)| \leq M < \infty$), it follows that $|\mathbb{E}\{\varphi(Y,t)|X=x^{(k)}\}| \leq M$ and $|Z_i^{(k)}(t)| \leq 2M \cdot \mathbf{1}\{x_i = x^{(k)}\} \leq 2M$. As long as the measure $\nu(\cT)$ is chosen to be finite, we have 
$$
\mathbb{E} \left( \|Z_i^{(k)}\|_{\cL^2(\cT, \nu)}^2 \right) = \int_{\cT} \mathbb{E}\left\{ Z_i^{(k)}(t)^2 \right\}  d\nu(t) = P(x^{(k)}) \int_{\cT} \operatorname{Var}\left\{ \varphi(Y, t) | X = x^{(k)} \right\}  d\nu(t) < \infty,
$$
since the integrand is bounded by $4M^2$. By Lemma~\ref{ central limit theorem}, $\frac{1}{\sqrt{n}} \sum_{i=1}^n Z_i^{(k)}(t)$ converges weakly to $\mathbb{G}'(t)$ in $\cL^2(\cT, \nu)$, where $\mathbb{G}'(t)$ is a zero-mean Gaussian process with covariance kernel
$$
\mathbb{E}[Z_i^{(k)}(t) Z_i^{(k)}(t')] = P(x^{(k)}) \cdot \operatorname{Cov}\left\{ \varphi(Y, t), \varphi(Y, t') | X = x^{(k)} \right\}.
$$
Since $n(x^{(k)})/n \xrightarrow{p} \mathbb{P}(x^{(k)})$, by Slutsky's theorem,
$$
\sqrt{n}\{\wh{\q}^{(k)}_t-\q^{(k)}_t\} \xrightarrow{d} \frac{1}{\mathbb{P}(x^{(k)})} \mathbb{G}(t).
$$
The limiting process $\frac{1}{\mathbb{P}(x^{(k)})} \mathbb{G}(t)$ is zero-mean Gaussian with covariance kernel
\begin{align*}
    \mathbb{E}\left\{ \frac{\mathbb{G}(t)}{\mathbb{P}(x^{(k)})} \cdot \frac{\mathbb{G}(t')}{\mathbb{P}(x^{(k)})} \right\} &= \frac{1}{\mathbb{P}(x^{(k)})^2} \cdot \mathbb{P}(x^{(k)}) \cdot \operatorname{Cov}\left\{ \varphi(Y, t), \varphi(Y, t') | X = x^{(k)}) \right\} \\
    &= \frac{1}{\mathbb{P}(x^{(k)})} \operatorname{Cov} \left\{ \varphi(Y, t), \varphi(Y, t') | X = x^{(k)} \right\}.
\end{align*}
For the vector-valued process over all $k = 1, ..., |\mathcal{X}|$, the components are asymptotically independent because the indicators $\mathbf{1}\{x_i = x^{(k)}\}$ and $\mathbf{1}\{x_i = x^{(k')}\}$ are mutually exclusive for $k \neq k'$, leading to zero cross-covariances. Thus, $\sqrt{n} (\wh{\q}_t - \q_t)$ converges weakly to a zero-mean vector-valued Gaussian process with block-diagonal covariance structure $\Sigma'$, where the $k$-th block is $\frac{1}{\mathbb{P}(x^{(k)})} \operatorname{Cov}\left\{\varphi(Y, t), \varphi(Y, t') | X = x^{(k)} \right\}$.

\noindent\textbf{(iii). Continuous mapping to the statistic.}

Given that $\wh{\Q}\xrightarrow{p} \Q$ and $\wh{\D}\xrightarrow{p} \D$ in probability by theorem \ref{thm:consistency}, and that $\sqrt{n} (\wh{\q}_t - \q_t)$ converges weakly to a zero-mean vector-valued Gaussian process, apply Slutsky’s theorem, we
have $\sqrt{n}\wh{\D}(\I-\wh{\P})\wh{\q}_t-\sqrt{n}\D(\I-\P)\q_t$ converges weakly to a zero-mean vector-valued Gaussian process, where covariance kernel $\D(\I-\P)\Sigma(\I-\P)^\top \D^\top $. Besides, since we have $(\I-\P)\q_t=0$ under $\mathbb{H}_0$, which implies that $\T_n(t)=\sqrt{n}\wh{\D}(\I-\wh{\P})\wh{\q}_t$ converges weakly to a zero-mean vector-valued Gaussian process, where covariance kernel is $\D(\I-\P)\Sigma(\I-\P)^\top \D^\top $.

\noindent\textbf{(iv). Asymptotic behavior of $\Delta_{\varphi}$.}

For any fixed $t$ and $\T_n(t)\in \cL^2\{\cT, \mu \}$, we use the continuous mapping theorem (Theorem 1.3.6 of \cite{wellner2013weak}) to obtain 
$$
\int{\|\T_n(t)\|_2^2d\mu(t)}\xrightarrow{d}\int{\|\mathbb{G} (t)\|_2^2d\mu(t)},
$$
by the continuity of the integral functional that arises from the continuity of $\varphi(\cdot,t)$.
\end{proof}

Similar to the continuous case, we consider the power performance under two alternatives such that under these hypotheses, there has no solution for \eqref{eq.convert_Y_dis}. That is, for any $\H_t$, the global alternative $\mH_1^{\mathrm{fix}}$ satisfies the following:
\begin{equation*}
    \mH_1^{\mathrm{fix}}: \mathbb{E}\{\varphi(Y,t)-H(W,t)|X=x\} \neq 0 \text{ for some $t \in \cT$ and some $x\in \mathcal{X}$}. 
\end{equation*}
Besides, we consider a sequence of local alternatives $\mH^\alpha_{1n}$. There exists $$\H^0_t := [H^0(w^{(1)}, t),..., H^0(w^{(|\mathcal{W}|)}, t)]^\top,
$$
such that: 
\begin{equation*}
    \mathbb H^\alpha_{1n}: \mathbb{E}\{\varphi(Y,t)|x\}= \mathbb{E}\{H^0(W,t)|x\}+\frac{r(x,t)}{n^\alpha}, \forall \thinspace t
\end{equation*}
where $0 < \alpha \leq \frac{1}{2}$. To be a valid alternative, $r(X,t)/n^\alpha$ can not be written as $\mE\{H - H^0|X\}$ for any $H$; besides, there exists $t$ and $x$ such that $|r(x,t)| \neq 0$. We define $\r_t := [\r(x^{(1)},t), ..., \r(x^{(|\mathcal{X}|)},t)]^\top$.

\ProjCltAlt*
\begin{proof}
\textbf{(i). The case of $\mH_{1}^{\mathrm{fix}}$.}

Define $\H_t^*=(\Q^\top\Q)^{-1} \Q^\top \q_t$. By theorem \ref{thm:consistency}, we have $\wh{\H}_t\xrightarrow{p}\H_t^*$. Note that
\begin{align*}
    \T_n(t)&=\frac{1}{\sqrt{n}}\sum_{i=1}^n{\widehat{U}(w_i,y_i,t)\mathbf{e}(x_i)}\\
    &=\sqrt{n}\mathbb{P}_n[\{\varphi(Y,t)-\wh{H}(W,t)\}\mathbf{e}(X)]\\
    &=\sqrt{n}\mathbb{P}_n[\{\varphi(Y,t)-H^*(W,t)\}\mathbf{e}(X)]+\sqrt{n}\mathbb{P}_n[\{H^*(W,t)-\wh{H}(W,t)\}\mathbf{e}(X)].
\end{align*}

According to the definition of $\mH_1^{\mathrm{fix}}$, there exists $r(x,t)$ such that $\mathbb{E} \{ \varphi ( Y,t ) -H^*( W,t)|X=x\} = r(x,t)$ for some $t\in\mathcal{T}$ and $x\in\mathcal{X}$, where $r(x,t)$ cannot be written as $\mE\{ H(W,t) -H^*( W,t)|X=x\}$ for any $H$. Without loss of generality, we assume $\mathbb{E} \{ \varphi ( Y,t ) -H^*( W,t)|X=x^{(k)}\} = r(x^{(k)},t)$ for $x^{(k)}\in \mathcal{X}$. Thus, we have
\begin{align*}
    \T^{(k)}_n(t) &= \sqrt{n}\mathbb{P}_n[\{\varphi(Y,t)-H^*(W,t)\}\mathbf{1}(X=x^{(k)})]+\sqrt{n}\mathbb{P}_n[\{H^*(W,t)-\wh{H}(W,t)\}\mathbf{1}(X=x^{(k)})].
\end{align*}

For the first term, since $\mathcal{W}$ and $\mathcal{Y}$ is finite, we can obtain $\mathbf{1}\{X=x^{(k)}\}\{\varphi(Y,t)-H^*(W,t)\}$ is integrable. By WLLN,
\begin{align*}
    \mathbb{P}_n[\{\varphi(Y,t)-\H^*(W,t)\}\mathbf{1}\{X=x^{(k)}\}]&=P(x^{(k)})\cdot \mathbb{E}\{\varphi(Y,t)-H^*(W,t)|X=x^{(k)}\}+o_p(1)\\
    &=P(x^{(k)})\cdot r(x^{(k)},t)+o_p(1).
\end{align*}

For the second term, since $\wh{\H}_t\xrightarrow{p} \H_t^*$, we have
\begin{align*}
    |\mathbb{P}_n[\{H^*(W,t)-\wh{H}_t(W,t)\}\mathbf{1}(X=x^{(k)})]|&\le \max_{w\in \mathcal{W}}|H^*(w,t)-\wh{H}_t(w,t)|\cdot \mathbb{P}_n\{\mathbf{1}(X=x^{(k)})\}\\
    &=o_p(1).
\end{align*}

Combining these results, we have
$\T^{(k)}_n(t)=\sqrt{n}\{ P(x^{(k)}) \cdot r(x^{(k)},t) +o_p(1)\}$, 
which means that $\lim_{n\rightarrow \infty} \max_{t\in \cT} \|\{\T_n(t)\|_{\infty}$ under $\mH_1^{\mathrm{fix}}$.
 
\textbf{(ii). The case of $\mH_{1n}^\alpha$ with 
$0<\alpha<1/2$.} 

Following the first step of theorem \ref{thm:proj-clt_null}, we have
$$
\T_n(t) = \sqrt{n}\wh{\D}(\wh{\q}_t-\wh{\Q}\wh{\H}_t) =\sqrt{n}\wh{\D}(\I-\wh{\P})\wh{\q}_t.
$$
Besides, following the third step of theorem \ref{thm:proj-clt_null}, we have $\sqrt{n}\wh{\D}(\I-\wh{\P})\wh{\q}_t-\sqrt{n}\D(\I-\P)\q_t$ converges weakly to a zero-mean vector-valued Gaussian process, where covariance kernel $\D(\I-\P)\Sigma(\I-\P)^\top \D^\top $. Next, we analyze $\sqrt{n}\D(\I-\P)\q_t$. Since $\q_t=\Q\H_t^0+\r_t/n^\alpha$ and $(\I-\P)\Q=0$, we have $(\I-\P)\q_t=(\I-\P)(\Q\H_t^0-\r_t/n^\alpha)=-(\I-\P)\r_t/n^\alpha$. Combining these results, we have
\begin{align*}
    \T_n(t)&=\sqrt{n}\wh{\D}(\I-\wh{\P})\wh{\q}_t-\sqrt{n}\D(\I-\P)\q_t+\sqrt{n}\D(\I-\P)\q_t.\\
    &= \underset{( \star )}{\underbrace{O_p(1)}} +\lim_{n\rightarrow \infty} \sqrt{n}\left[ \frac{1}{n^{\alpha}}\{\D(\I-\P)\r_t\}\right].
\end{align*}
Since there exists $t$ and $x$, such that $r(x,t) \neq 0$, we have $\lim_{n\to \infty} \max_{t \in \cT}\Vert T_n(t) \Vert_\infty = \infty$ under $\mH_{1n}^{\alpha} (0<\alpha<1/2)$, where $(\star )$ follows from portmanteau theorem and the fact that $\sqrt{n}\wh{\D}(\I-\wh{\P})\wh{\q}_t-\sqrt{n}\D(\I-\P)\q_t$ converges to Gaussian process. 

\textbf{(iii). The case of $\mH_{1n}^\alpha$ with $\alpha=1/2$.}

Following the proof of $\mH_{1n}^\alpha$ with $0<\alpha<1/2$, we have
\begin{align*}
    \T_n(t)&=\sqrt{n}\wh{\D}(\I-\wh{\P})\wh{\q}_t-\sqrt{n}\D(\I-\P)\q_t+\sqrt{n}\D(\I-\P)\q_t.\\
    &= \lim_{n\rightarrow \infty}\sqrt{n}\wh{\D}(\I-\wh{\P})\wh{\q}_t-\sqrt{n}\D(\I-\P)\q_t +\lim_{n\rightarrow \infty} \sqrt{n}\left[ \frac{1}{\sqrt{n}}\{\D(\I-\P)\r_t\}\right]\\
    &\to_d \mG(t)+ \D(\I-\P)\r_t.
\end{align*}
We complete the proof. 
\end{proof}

\section{Proxy Maximum Characteristic Restriction}\label{sec.mmr}

For the sake of completeness, we introduce some preliminary concepts that are necessary to understand the theoretical analysis of our PMCR method. First, in section~\ref{appx.linear_operators}--\ref{appx.rkhs}, we introduce some background knowledge of the linear operators and Reproducing Kernel Hilbert Spaces required in this article. Upon this, we provide details on the derivation of our empirical loss~\eqref{eq:mmr-vstat} in section~\ref{appx.mr}. Section~\ref{appx.operator} rewrites the loss into the Tikhonov regularized form, which serves as the foundation of our theoretical analysis for Theorem~\ref{theorem:null-hypothesis}. 

\subsection{Bounded linear operator}\label{appx.linear_operators}
For two normed linear spaces $\cF$ and $\cG$ over $\mR$, a function $A: \cF\to \cG$ (where $\cF$ and $\cG$ are both normed linear spaces over $\mR$) is called a linear operator if it satisfies the following properties:
\begin{enumerate}
    \item Homogeneity: $A(\alpha f)=\alpha(Af)$, for any $\alpha\in \mR, f\in \cF$;
    \item Additivity: $A(f+g)=Af +Ag$, for any $f,g\in\cF$.
\end{enumerate}
\textbf{Operator Norm and Boundedness.} The operator norm of a linear operator $A: \cF\to \cG$ is defined as 
$$
\|A\|_{\mathrm{op}}=\sup_{f\in\mathcal{F}}\frac{\|Af\|_{\mathcal{G}}}{\|f\|_{\mathcal{F}}}.
$$
A linear operator $A$ is called bounded if there exists a finite constant $C$ such that for all $f\in\cF$, we have
$$
\|Af\|_\mathcal{G}\leq C\|f\|_\mathcal{F}.
$$
In terms of the operator norm, this condition is equivalent to saying that $\|A\|_{\mathrm{op}}<\infty$.

\subsection{Hilbert space}\label{appx.hilbert}
We begin by introducing definitions and basic properties of an inner product space. Based on this, we introduce the Hilbert space.

A function $\langle\cdot,\cdot\rangle_{\mathcal{F}}:\mathcal{F}\times\mathcal{F}\to\mathbb{R}$ is said to be an inner product on $\cF$ if it satisfies the following three properties
\begin{enumerate}
    \item $\langle\alpha_{1}f_{1}+\alpha_{2}f_{2},g\rangle_{\mathcal{F}}=\alpha_{1}\langle f_{1},g\rangle_{\mathcal{F}}+\alpha_{2}\langle f_{2},g\rangle_{\mathcal{F}}$.
    \item $\langle f,g\rangle_{\mathcal{F}}=\langle g,f\rangle_{\mathcal{F}}$.
    \item $\langle f,f\rangle_{\mathcal{F}}\geq0$ and $\langle f,f\rangle_{\mathcal{F}}=0$ if and only if $f=0$.
\end{enumerate}
One can always define a norm induced by the inner product: $\|f\|_{\mathcal{F}}=\langle f,f\rangle_{\mathcal{F}}^{1/2}$. For this norm, the following Cauchy-Schwarz inequality holds, \emph{i.e.},  $|\langle f,g\rangle_{\mathcal{F}}|\leq\|f\|_{\mathcal{F}}\cdot\|g\|_{\mathcal{F}}$. 

A Hilbert space is a complete inner product space. This means, a Hilbert space is an inner product space in which every Cauchy sequence (a sequence where the elements get arbitrarily close to each other) converges to an element within the space. An orthonormal basis of a Hilbert space $\cH$ is a set of vectors $\{e_i\}$, such that $\Vert e_i \Vert_\cH = 1$ for each $i$ and $\langle e_i, e_j \rangle_\cH = 0$ for each $i \neq j$. Besides, each $f\in\cH$ can be expanded in a Fourier series:
$$
\varphi=\sum_{j}\langle f,e_i\rangle_{\mathcal{H}}e_i.
$$
\textbf{Hilbert adjoint operator.} In the context of Hilbert spaces, we can define the adjoint operator. Let $\cH_1$ and $\cH_2$ be Hilbert spaces, and let $A:\cH_1\to \cH_2$ be a linear operator. The adjoint operator $A^*:\cH_2\to \cH_1$ is defined by the property that for all
$$
\langle Af,g\rangle_{\mathcal{H}_2}=\langle f,A^*g\rangle_{\mathcal{H}_1}.
$$
The operator enjoys a number of important properties:
\begin{enumerate}
    \item If $A$ is bounded, so is $A^*$, and $\|A\|_{\mathrm{op}}=\|A^*\|_{\mathrm{op}}$;
    \item $(A^*)^*=A$;
    \item If $A$ is invertible, so is $A^*$, and $(A^{*})^{-1}=(A^{-1})^{*}$.
\end{enumerate}

\subsection{Reproducing Kernel Hilbert Space}\label{appx.rkhs}
For any space $\cW$, let $k_W: \cW \times \cW \to \mathbb{R}$ be a positive semi-definite kernel. A kernel is called \emph{characteristic} if $\mP \mapsto \mE_{W \sim \mP}[k_W(W,\cdot)]$ is injective \citep{fukumizu2004dimensionality}. We denote by $\phi_W$ its associated canonical feature map $\phi_W(w)=k_W(w,\cdot)$ for any $w\in \cW$, and $\cH_W$ its corresponding RKHS of real-valued functions on $W$. The space $\cH_W$ is a Hilbert space with inner product $\langle\cdot\rangle_{\cH_W}$ and norm $\|\cdot\|_{\cH_W}$. It satisfies two important properties:
\begin{enumerate}
    \item $k_W(w,\cdot)\in \cH_W$ for all $w\in \cW$;
    \item reproducing property: for all $f\in \cH_W$ and $w\in \cW$, $f(w)=\langle f,k_W(w,\cdot)\rangle_{\mathcal{H}_{W}}$. 
\end{enumerate}
Since the Reproducing Kernel Hilbert Space (RKHS) is a Hilbert space, it satisfies all properties in section~\ref{appx.hilbert}. Besides, we can define the kernel mean embedding, which helps to take the expectation of a function. Suppose we wish to calculate $\mE\{f(W)\}$ for any $f\in\cH_W$. By the reproducing property and linearity of the inner product, we have
\begin{align*}
    \mathbb{E}\{f(W)\}&=\int f(w)dF(w)=\int\langle f,\phi_W(w)\rangle_{\mathcal{H}_W}dF(w)\\
    &=\left\langle f,\int\phi_W(w)dF(w)\right\rangle_{\mathcal{H}_W}=\langle f,\mu_W\rangle_{\mathcal{H}_W}.
\end{align*}
The object $\mu_W:=\int\phi_W(w)dF(w)$ is called the mean embedding of the distribution $F(w)$. This property of RKHS implies that, to calculate the expectation of a function, it suffices to take the inner product between the function and the mean embedding of the corresponding distribution. Following this property, we can also calculate the expectation $\mE\{f(W)m(X)\}$ for any $f\in\cH_W$
\begin{equation}\label{eq.kernel_mean_embedding}
    \begin{aligned}
        \mathbb{E}\{f(W)m(X)\}&=\int f(w)m(x)dF(w,x)\\
        &=\int\langle f,\phi_W(w)\rangle_{\mathcal{H}_W}m(x)dF(w,x)=\left\langle f,\int m(x)\phi_W(w)dF(w,x)\right\rangle_{\mathcal{H}_W}.
    \end{aligned}
\end{equation}
Finally, we introduce properties for the norm $\|\cdot\|_{\cH_W}$. A function $f\in \cH_W$ if and only if $\|f\|_{\mathcal{H}_W}^2=\langle f,f\rangle_{\mathcal{H}_W}<\infty$. Further, if $k_W(w,\cdot)$ is bounded, we have $\|f\|_{\cL^2\{F(w)\}}\lesssim \|f\|_{\cH_W}$. To see this, note that by Cauchy-Schwarz inequality, for any $f\in\cH_W$, we get:
$$
|f(w)|^2=\langle k_W(w,\cdot),f\rangle_{\cH_W}^2\leq\|k_W(w,\cdot)\|_{\mathcal{H}_W}^2\|f\|_{\mathcal{H}_W}^2.
$$
Therefore, we have
\begin{equation}\label{eq.h_norm}
    \|f\|_{\cL^2\{F(w)\}}\lesssim \|f\|_{\cH_W}.
\end{equation}

\subsection{Validity of optimizing~\eqref{eq:mmr-vstat}}
\label{appx.mr}
Since \eqref{eq.convert_Y} implies $\mathbb{E}[\{\varphi(Y,t)-H(W,t)\}g(X)]=0$ holds for any measurable functions $g:\mathcal{X}\to\mathbb{R}$, we follow \cite{zhang2020maximum,mastouri2021proximal} to take $g$ over a unit-ball of RKHS $\mathcal{H}_X$ with a fixed kernel $k^g$, and minimizes 
\begin{equation}\label{eq:mmr-sup-h}
    R(H)=\underset{g\in \cH_X,\| g\|_{\cH_X} \le 1}{\sup}\left( \mathbb{E} \left[ \{\varphi(Y,t)-H(W,t)\} g(X) \right] \right)^2.
\end{equation}
\cite{mastouri2021proximal} provides an equivalent form of this risk, which is the population version of our empirical loss~\eqref{eq:mmr-vstat}. 
\begin{lemma}[Lemma 2 in \cite{mastouri2021proximal}]\label{lemma: fun-kernel}
    Assume that $\mathbb{E} [ \{\varphi(Y,t)-H(W,t)\}^2k_X(X ,X') ] <\infty$ and denote by $X'$ an independent copy of the random variable $X$. Then $R(H)=\mathbb{E} [\{\varphi(Y,t)-H(W,t)\}\{\varphi(Y',t)-H(W',t)\}k_X(X,X')]$.
\end{lemma}
\cite{zhang2020maximum,mastouri2021proximal} demonstrated that if the kernel function $k_X$ derived from the conditional variable $X$ in the conditional moment equation \eqref{eq.convert_Y} is integrally strictly positive definite (ISPD defined in Asm.~\ref{assum.ISPD kernel}), continuous, and bounded, then the conditional moment equation~\eqref{eq.convert_Y} shares the same solution with $R(H)$. That means, optimizing $R(H)$ ensures us to find the right solution. 

\subsection{Tikhonov regularization}
\label{appx.operator}
In this section, we rewrite our loss~\eqref{eq:mmr-vstat} into the following Tikhonov regularized form, which serves as the foundation to prove Theorem~\ref{theorem:null-hypothesis}. 
\begin{equation}\label{eq:mmr-obj-emp}
    \widehat{R}_{\lambda}(H) = \| \widehat{b}(x,t)-\widehat{A}H(\cdot,t)(x)\|_{\mathcal{H}_X}^{2}+\lambda\| H(w,t)\|_{\mathcal{H}_W}^{2}.
\end{equation}
This can be achieved by reformulating the PMCR into a linear ill-posed inverse problem in the RKHS. Specifically, let $\phi_X(x)(\cdot) := k_X(x,\cdot)$ and $\phi_W(w)(\cdot) := k_W(w,\cdot)$ be the canonical feature maps. For notational simplicity, we omit the brackets in the feature maps. Then, by $\langle \phi_X(x),\phi_X(x')\rangle_{\cH_X}=k_X(x,x')$, $R(H)$ of Lemma~\ref{lemma: fun-kernel} can be rewritten in terms of mean square error:
$$
\begin{aligned}
	R(H) &=\| \mathbb{E} [\{ \varphi (Y,t) -H(W,t)\} \phi_X(X) ] \|_{\mathcal{H}_X}^{2}\\
	&=\| \mathbb{E} \{\varphi (Y,t) \phi_X(X) \}-\mathbb{E} \{H(W,t) \phi_X(X)\} \|_{\mathcal{H}_X}^{2}\\
	&=\| b(X,t)-AH(\cdot,t)(X)\|_{\mathcal{H}_X}^{2},
\end{aligned}
$$
where
\begin{equation}\label{eq.operator_defination}
    b(x',t):=\int{\varphi (y,t) \phi_X(x)(x') p(x,y) dxdy},\quad AH(\cdot,t)(x'):=\int{H(w,t) \phi_X(x)(x') p(x,w) dxdw}.
\end{equation}

Thus, we can treat PMCR as a linear ill-posed inverse problem in the RKHS by the operator $A$. To ensure that $A$ is a bounded linear operator, we require some standard assumptions \citep{zhang2020maximum,mastouri2021proximal}:
\begin{assumption}\label{ass:y_bounded}
There exists a constant $c_Y<\infty$ such that $|\varphi(Y,t)|\le c_Y$ almost surely for all $t$.
\end{assumption} 
\begin{remark}
    If we choose $\varphi(Y,t)=\sin(tY)$ or $\cos(tY)$, then $|\varphi(Y,t)|\le 1$ for all $Y,t$, and condition~\ref{ass:y_bounded} is satisfied without requiring $Y$ itself to be bounded.
\end{remark}
\begin{assumption}\label{ass:kernel_characteristic}
\textbf{(i).} $k_X(x,\cdot)$ and $k_W(w,\cdot)$ are continuous and bounded, \emph{i.e.}, there exists $\kappa>0$ such that:
$$
\sup\limits_{w}\|\phi_W (w) \|_{\mathcal{H}_W}\le \kappa ,\quad\sup\limits_{x}\|\phi_X (x)\|_{\mathcal{H}_{\mathcal{X}}}\le \kappa.
$$
\textbf{(ii).} Feature maps $\phi_W(W)$ and $\phi_X(X)$ are measurable.
\textbf{(iii).} $\phi_W(W)$ and $\phi_X(X)$ are characteristic kernels. 
\end{assumption}
\begin{assumption}\label{assum.ISPD kernel} 
The kernel $k_X(x,x')$ is integrally strictly positive definite (ISPD), \emph{i.e.}, for any function $f$ that satisfies $0<\|f\|_{\cL^2\{F(x)\}}^2<\infty$, we have $\iint f(x)k_X(x,x')f(x')dxdx'>0$.
\end{assumption}

By conditions~\ref{ass:y_bounded} and~\ref{ass:kernel_characteristic}, $b(x,t)\in\mathcal{H}_X$ and $A$ is a bounded linear operator from $\mathcal{H}_{W}$ to $\mathcal{H}_{X}$. Based on the above formulation, we can rewrite $R(H)$ of Lemma~\ref{lemma: fun-kernel} with regularized term as follow:
\begin{equation}\label{eq:mmr-obj}
    R_{\lambda}(H)=\| b(x,t)-AH(\cdot,t)(x)\|_{\mathcal{H}_X}^{2}+\lambda \| H(w,t)\|_{\mathcal{H}_W}^{2}.
\end{equation}
Plugging the estimates of $\wh{b}(x,t)$ and $\wh{A}$ into the loss, we have \eqref{eq:mmr-obj-emp}. Based on the i.i.d. samples $(x_i,w_i,y_i)_{i=1}^n$ and $\phi_X(x_i)$, the estimates $\widehat{b}(x,t)$ and $\widehat{A}$ are given by:
\begin{align}
\label{eq.estimate_f_K}
    \widehat{b}(x,t):=\frac{1}{n}\sum_{i=1}^n{\varphi(y_i,t) k_X(x,x_i)}, \quad
    \widehat{A}H(\cdot,t)(x):=\frac{1}{n}\sum_{i=1}^n{H(w_i,t) k_X(x,x_i)}.
\end{align}

Let $A^*: \mathcal{H}_{X} \to \mathcal{H}_{W}$ be an adjoint operator of $A$ such that $\langle Au,v\rangle_{\mathcal{H}_{X}} = \langle u,A^*v\rangle_{\mathcal{H}_{W}}$ for all $u\in\mathcal{H}_{W}$ and $v\in\mathcal{H}_{X}$. And we denote $\widehat{A}^*$ as an adjoint operator of $\widehat{A}$. By \cite{mastouri2021proximal}, for any $m(w,t)\in\mathcal{H}_{W}$, we have:
\begin{equation}\label{eq.operator_defination_adjoin_copy}
    A^*m(\cdot,t)(w'):=\int{m(x,t) k_W(w,w') p(x,w)dxdw}.
\end{equation}
The estimate $\wh{A}$ is given by its empirical form:
\begin{equation}\label{eq.operator_defination_adjoin}
    \widehat{A}^*m(\cdot,t)(w'):=\frac{1}{n}\sum_{i=1}^n{m( x_i,t) k_W(w_i,w')}.
\end{equation}


\subsection{Ill-posed inverse problem and solutions}
Solving $R(H)$ is generally an ill-posed inverse problem, as it may
not have a unique solution \citep{carrasco2007linear}. We allow the \emph{Conditional} \emph{Characteristic} \emph{Restrictions} \eqref{eq.convert_Y} to be ill-posed and have non-unique solutions. Thus, the set of all solutions is given by 
\begin{equation}\label{eq. all_solution}
    \cH_{W,0}=\{H(\cdot,t)\in\mathcal{H}_W:AH(\cdot,t)(x)=b(x,t)\}=H^0(\cdot,t)+\mathrm{Ker}(A),
\end{equation}
where $\mathrm{Ker}(A)=\{H(\cdot,t):AH(\cdot, t)(x)=0\}$ is the null space of the adjoint operator $A$. A general solution can be represented as the sum of the special solution $H^0(w,t)\in \mathrm{Ran}(A)$, and the element that belongs to the null space.

If the solution exists, we can express the solution in the form of the singular value decomposition of $A$. Let $(s_j,u_j,v_j)_{j}$ be the singular value decomposition of the operator $A$. Then, if we define the orthogonal projection operator $Q: \cH_W\to \mathrm{Ker}(A)$, we have:
$$
H(\cdot, t)=\sum_{j}\langle H(\cdot, t),u_j\rangle_{\cH_W}u_j+QH(\cdot, t)=\sum_{j}\frac{1}{s_j}\langle b(\cdot, t),v_j\rangle_{\cH_X}u_j+QH(\cdot, t).
$$
Thus, we target at the special solution $H^0(W,t)$, which achieves the least norm, \emph{i.e.}, 
\begin{equation}\label{eq._least_norm}
H^0(W,t) =\underset{H(W,t)\in \mathcal{H}_{W,0}}{\mathrm{arg}\min}\|H(W, t)\|_{\cH_W}.
\end{equation}
By solving for $R^{\lambda}(H)$ of Eq.~\eqref{eq:mmr-obj}, we attempt to estimate the minimum norm solution $H^0(W,t)$ in~\eqref{eq._least_norm} via the Tikhonov regularization solutions in respectively the population and in the finite sample regime:
\begin{eqnarray}
    H^{\lambda}(W,t)&:=\underset{H(W,t)\in \mathcal{H}_W}{\mathrm{arg}\min}R_{\lambda}(H) =\{( A^*A+\lambda I)^{-1}A^*b(\cdot,t)\}(W,t), \label{eq:h_pop} \\
    \widehat{H}^{\lambda}(W,t)&:=\underset{H(W,t)\in \mathcal{H}_W}{\mathrm{arg}\min}\widehat{R}_{\lambda}(H) =\{( \widehat{A}^*\widehat{A}+\lambda I)^{-1}\widehat{A}^*\widehat{b}(\cdot,t)\}(W,t). \label{eq:h_emp}
\end{eqnarray}

\section{Proofs of Asymptotic Properties}
In this section, we provide the asymptotic properties of the testing statistics $\Delta_{\varphi,m}$. Since $\Delta_{\varphi,m}$ depends on $T_n(s,t)$ through \eqref{eq.icm-statistics}, we first study the asymptotic properties of $T_n(s,t)$.

\textbf{Notations.} For a generic random vector $W\in\cW$, we use $\cL^2\{F(w)\}$ to denote the space of square integrable functions of $W$ with respect to the cumulative distribution of $W$. For any $f(W),g(W)\in \cL^2\{F(w)\}$, we denote the $\cL^2$-norm by $\|f\|_{\cL^2\{F(w)\}}=\sqrt{\mathbb{E}\{f(W)^2\}}$ and inner product by $\langle f,g\rangle_{\cL^2\{F(w)\}}=\mathbb{E}\{f(W)g(W)\}$. We use $\cH_W$ to denote the reproducing kernel Hilbert spaces of $W$. For any $f(W),g(W)\in \cH_W$, let $\|f\|_{\cH_W}$ denote the $\cH_W$-norm and $\langle f,g\rangle_{\cH_W}$ denote the inner product. Let $\mathbb{P}\{f(W)\}=\int \wh{f}_n(w)dP(w)$ be the expectation with respect to $W$ alone. We differentiate this from $\mathbb{E}\{\wh{f}_n(W)\}$, which we use to denote full expectation with respect to both $W$ and data $w_1,...,w_n$. Thus if $\wh{H}$ depends on the data $w_1,...,w_n$, then $\mathbb{P}\{\wh{f}(W)_n\}$ remains a function of $w_1,...,w_n$ but $\mathbb{E}\{f(W;\wt{H})\}$ is a nonrandom scalar. We use both $\mP_n$ to denote the empirical expectation with respect to $W$ given data $w_1,...,w_n$: $\mP_n\{f(W)\}=\frac{1}{n}\sum_{i=1}^{n}f(W_{i})$. 

For the operator $A$, let $b_t(w):=b(w,t)$ defined in \eqref{eq.operator_defination}, and $A^*$ in~\eqref{eq.operator_defination_adjoin}. The corresponding estimators are given by $\wh{b}_t(w):=\wh{b}(w,t)$ in \eqref{eq.estimate_f_K}, and $\wh{A}^*$ in \eqref{eq.operator_defination_adjoin}. Besides, for the operator $A$, its singular value decomposition is given by $(s_n,u_n,v_n)_{n=1}^{+\infty}$. We denote $H^0_t=H^0(w,t)$ as the least norm solution is defined in~\eqref{eq._least_norm}. The population Tikhonov regularization solution $H^{\lambda}_t(w):=H^{\lambda}(w,t)$ and the empirical Tikhonov regularization solution $\wh{H}^{\lambda}_t(w):=\wh{H}^{\lambda}(w,t)$ are respectively defined in~\eqref{eq:h_pop} and~\eqref{eq:h_emp}. Further, recall that 
\begin{align}
    g_s(\cdot) & := \mathbb{E} \{ m(X,s)\phi_W(W)\} \ (\text{condition~\ref{assum.var}}), \label{eq.gs} \\
    U(W,Y,t) & :=\varphi(Y,t)-H^0(W,t) 
    \ (\text{section~\ref{sec.equivalent}}), \label{eq.U} \\
    \wh{U}(W,Y,t) & := \varphi(Y,t)-\wh{H}^{\lambda}(W,t)  (\text{section~\ref{sec.equivalent}}).
    \label{eq.U-hat}
\end{align}

\subsection{Proof roadmap and key assumptions}
\label{sec.poof_roadmap}

In this section, we present the overview and the required assumptions of our proof. We decompose $T_n(s,t)$ as follows
\begin{equation}\label{eq.Tnst decompose}
    \begin{aligned}
	T_n(s,t)&=\frac{1}{\sqrt{n}}\sum_{i=1}^n{\widehat{U}(w_i,y_i;t)m(x_i,s)}\\ &=\sqrt{n}\mathbb{P}_n\left\{ \widehat{U}(W,Y,t)m(X,s) \right\}\\
	&=\sqrt{n}\mathbb{P}_n\left[ \left\{ \varphi(Y,t) -\widehat{H}^{\lambda}(W,t) \right\} m(X,s) \right]\\
	&=\sqrt{n}\mathbb{P}_n\left[ \left\{ \varphi(Y,t) -H^0(W,t) +H^0(W,t) -\widehat{H}^{\lambda}(W,t)\right\}  m(X,s) \right]\\
	&= \sqrt{n}\mathbb{P}_n\{ U(W,Y,t) m(X,s) \}+\underbrace{\sqrt{n}\mathbb{P} \left[ \left\{ H^0(W,t) -\widehat{H}^{\lambda}(W,t)\right\} m(X,s) \right]}_{\text{Expected risk difference}} \\
	&\qquad+\underset{ \text{Empirical process} }{\underbrace{\sqrt{n}( \mathbb{P}_n-\mathbb{P}) \left[\{ H^0(W,t) -\widehat{H}^{\lambda}(W,t)\} m(X,s) \right] }}.
\end{aligned}
\end{equation}
To derive the asymptotic distribution of $T_n(s,t)$, we first investigate the last two terms in \eqref{eq.Tnst decompose}:
\begin{itemize}
    \item \textbf{Empirical process} (Proposition~\ref{prop.empirical_process}): $(\mathbb{P}_n-\mathbb{P}) \left[ \left\{ H^0(W,t) -\widehat{H}^{\lambda}(W,t)\right\} m(X,s) \right]=o_p(n^{-1/2})$. 
    \item \textbf{Expected risk difference} (Proposition~\ref{prop.expected_risk}): 
    \begin{equation*}
        \sqrt{n}\mathbb{P} \left[ \left\{ H^0(W,t) -\widehat{H}^{\lambda}(W,t)\right\} m(X,s) \right]=-\frac{1}{\sqrt{n}}\sum_{i=1}^n{U(w_i,y_i,t) \{A(A^*A)^{-1}g_s\}(x_i)}+o_{p}(1),
    \end{equation*}
    where $g_s$ is defined in \eqref{eq.gs}. 
\end{itemize}
Lastly, we show that $-n^{-1/2}\sum_{i=1}^nU(w_i,y_i,t) \{A(A^*A)^{-1}g_s\}(x_i)$ plus the remaining term $\sqrt{n}\mathbb{P}_n\{ U(W,Y,t)m(X,s\}$ converges to the zero-man Gaussian process $\mG(s,t)$, \emph{i.e.}, 
\begin{equation*}
    \lim_{n \to \infty} \sqrt{n}\mathbb{P}_n\{ U(W,Y,t) m(X,s) \}-\frac{1}{\sqrt{n}}\sum_{i=1}^n{U(w_i,y_i,t) \{A(A^*A)^{-1}g_s\}(x_i)} \to_d \mathbb{G}(s,t). 
\end{equation*}
Since $\Delta_{\varphi ,m} =\max_{t\in\mathcal{T}}\int_{\mathcal{S}}{| T_n(s,t)| ^2d\mu (s)}$ in \eqref{eq.icm-statistics}, we therefore obtain that $\Delta_{\varphi,m}$ converges to $\max_{t \in T} \int |\mG_{s,t}|^2d\mu(s)$ in Theorem~\ref{theorem:null-hypothesis}. 


Before proving these properties, we first introduce some regularity conditions. Let $\cH_W$ denote the function space such that $H^0(W,t) \in \cH_W$ for each $t$.  

\begin{assumption}\label{assum.empirical_process}
    Let $N_{[\cdot]}\left(\epsilon,\mathcal{H}_W,\|\cdot\|_{\mathcal{H}_W}\right)$ be the bracketing number of size $\epsilon$ of $\mathcal{H}_W$. We assume $\int_{0}^{1}\sqrt{\log N_{[\cdot]}(\epsilon,\mathcal{H}_W,\|\cdot\|_{\cL^2\{F(w)\}}}d\epsilon<\infty $ and $\mathbb{P}(\widehat{H}_t^\lambda\in\mathcal{H}_W)\to1$.
\end{assumption}

\begin{assumption}\label{assum.source condition} 
Let $(s_{j},u_{j},v_{j})_{j}$ be the singular value decomposition of the operator $A$ described in section~\ref{sec.mmr}. Then we assume: 
(a). For some $\eta\geq 2$, $\sum_j{s_{j}^{-2\eta}|\langle g_s,u_{j}\rangle_{\cH_W} |^2}<\infty $; (b) For some $\theta\geq 2$,
$\sum_j{s_{j}^{-2\theta}|\langle H^0_t,u_{j} \rangle_{\cH_W} |^2}<\infty$.
\end{assumption}

Condition~\ref{assum.empirical_process} restricts the complexity of $\cH_W$ and ensures $\cH_W$ is a $P$-Donsker class \citep{vd1998asymptotic}, which was a standard assumption to analyze the empirical process \citep{beyhum2024testing,lapenta2022encompassing}. Our analysis still holds when $N_{[\cdot]}\left(\epsilon,\mathcal{H}_W,\|\cdot\|_{\mathcal{H}_W}\right)$ denotes the entropy in condition~\ref{assum.empirical_process}. According to \cite{hable2012asymptotic}, $\cH_W$ belongs to the $P$-Donsker class if the kernel function is chosen to be the Gaussian kernel. 

Condition~\ref{assum.source condition} is the source condition that is commonly assumed in nonparametric regression \citep{carrasco2007linear,florens2012instrumental}. These have also been employed in \cite{florens2012instrumental, beyhum2024testing} to obtain a faster convergence rate for nonparametric instrumental regression. Here, we require $g_s$ and $H^0_t$ to satisfy the source condition for establishing the asymptotic properties of the statistic in examining the integral equation. Since $g_s:=\mE\{m(X,s)\phi_W(W)\}$, the source condition for $g_s$ puts requirement on the smoothness for the space $\cH_W$ when $m(\cdot,s)$ is chosen properly.

\subsection{Empirical process}

\begin{proposition}\label{prop.empirical_process}
    Under condition~\ref{differentiabilty and integrability}-\ref{assum.bandwidth},~\ref{ass:y_bounded}-\ref{assum.ISPD kernel}, and~\ref{assum.empirical_process}-\ref{assum.source condition}, the empirical process $  \sqrt{n}( \mathbb{P}_n-\mathbb{P} )[ \{ H^0(W,t) -\widehat{H}^{\lambda}(W,t)\} m(X,s)]=o_{p}(1)$.
\end{proposition}
\begin{proof}
    We first proof $\| \{ H^0(W,t) -\widehat{H}^{\lambda}(W,t)\}m(X,s)\|_{\cL^2\{F(x,w)\}}^{2}=o_{p}(1)$. In fact, we have
    $$
\begin{aligned}
	\| \{ H^0(W,t) -\widehat{H}^{\lambda}(W,t)\} m(X,s) \|_{\cL^2\{F(x,w)\}}^{2}&=\int{\{ H^0(W,t) -\widehat{H}^{\lambda}(W,t)\}^2|m(X,s)|^2d\mathbb{P} (W,X)}\\
	&=\int{\{ H^0(W,t) -\widehat{H}^{\lambda}(W,t)\}^2\mathbb{E} \{|m(X,s)|^2|W\} d\mathbb{P} (W)}\\
	&\overset{(1)}{\lesssim} \| H^0(w,t)-\widehat{H}^{\lambda}(w,t) \|_{\cL^2\{F(w)\}}^{2}\\
    &\overset{(2)}{\lesssim } \| H^0(w,t)-\widehat{H}^{\lambda}(w,t)\|_{\mathcal{H}_W}^{2},
\end{aligned}
$$
where (1) follows from condition~\ref{differentiabilty and integrability} and (2) follows from \eqref{eq.h_norm} by condition~\ref{ass:kernel_characteristic}. By Lemma~\ref{Lem: rates for Hhat - H}, we have $\| H^0(w,t)-\widehat{H}^{\lambda}(w,t) \|_{\mathcal{H}_W}^{2} = o_{p}(1)$. Therefore, all conditions in Lemma~\ref{lemma:empirical process} are satisfied, and we obtain
$$
\sqrt{n}( \mathbb{P}_n-\mathbb{P} )[ \{ H^0(W,t) -\widehat{H}^{\lambda}(W,t)\} m(X,s)] = o_{p}(1).
$$
The proof is completed. 
\end{proof}
\subsection{Expected risk difference}
Proposition~\ref{prop.expected_risk} is our main result, and the proof is developed through Lemmas~\ref{lemma:Tn2}-\ref{lemma:Tn1}. 

\begin{proposition}\label{prop.expected_risk}
    Under conditions~\ref{assum.bandwidth},~\ref{ass:y_bounded}--\ref{ass:kernel_characteristic}, and~\ref{assum.source condition}, the expected risk difference term has: 
    \begin{equation*}
        \sqrt{n}\mathbb{P} \left[ \left\{ H^0(W,t) -\widehat{H}^{\lambda}(W,t) \right\} m(X,s) \right] =-\frac{1}{\sqrt{n}}\sum_{i=1}^n{U(w_i,y_i,t) \{A(A^*A)^{-1}g_s\}(x_i)}+o_{p}(1).
    \end{equation*}
\end{proposition}
\begin{proof}
   Based on the interpretation of PMCR as a linear ill-posed problem and the form of Tikhonov regularization solutions in \eqref{eq:h_pop}--\eqref{eq:h_emp}, we have the following decomposition \citep{babii2017completeness,babii2020unobservables}:
\begin{equation}\label{eq: decomposition of phi.hat}
    \widehat{H}^{\lambda}(w,t) - H^0(w,t)= G_1 + G_2 + G_3 + G_4 + G_5,
\end{equation}
where 
\begin{align}
    G_1:=& (\lambda I + A^*A)^{-1}A^* (\widehat{b}_t-\widehat{A}H^0_t); \label{eq.G1}\\
    G_2 := & (\lambda I + A^*A)^{-1}(\widehat{A}^*-A^*)(\widehat{b}_t-\widehat{A}H^0_t); \label{eq.G2} \\
    G_3 := & \left\{ (\lambda I + \widehat{A}^*\widehat{A})^{-1} -(\lambda I + A^*A)^{-1}  \right\}\widehat{A}^*(\widehat{b}_t-\widehat{A}H^0_t); \label{eq.G3}\\
    G_4:= & (\lambda I + \widehat{A}^*\widehat{A})^{-1}\widehat{A}^*\widehat{A}H^0_t-(\lambda I + A^*A)^{-1}A^*b_t; \label{eq.G4} \\
    G_5:= &(\lambda I + A^*A)^{-1}A^*b_t-H^0_t. \label{eq.G5}
\end{align}
Therefore, we have
$$
\sqrt{n}\mathbb{P}[\{\widehat{H}^{\lambda}(W,t)-H^0(W,t)\} m(X,s)] =\sum_{i=1}^5{S_{ni}(s,t)},
$$
where $S_{ni}(s,t)$ is define as $\sqrt{n}\mathbb{P} \{ G_i m(X,s) \}$. By applying Lemmas~\ref{lemma:Tn1},~\ref{lemma:Tn2},~\ref{lemma:Tn3},~\ref{lemma:Tn4} and \ref{lemma:Tn5} to $S_{n1}(s,t)$, $S_{n2}(s,t)$, $S_{n3}(s,t)$, $S_{n4}(s,t)$ and $S_{n5}(s,t)$, respectively, we have:
    $$
    \sqrt{n}\mathbb{P} [ \{ H^0(W,t) -\widehat{H}^{\lambda}(W,t)\} m(X,s) ] =-\frac{1}{\sqrt{n}}\sum_{i=1}^n{U(w_i,y_i,t) \{A(A^*A)^{-1}g_s\}(x_i)}+o_{p}(1).
    $$
    The proof is completed. 
\end{proof}


Next, we provide proofs for Lemmas~\ref{lemma:Tn2}--\ref{lemma:Tn1}.

\begin{lemma}
\label{lemma:Tn2}
Under conditions~\ref{assum.bandwidth}, \ref{ass:y_bounded}, \ref{ass:kernel_characteristic} and~\ref{assum.source condition}, $S_{n2}(s,t)=o_{p}(1)$ as $n\to\infty$. 
\end{lemma}
\begin{proof}
By the reproducing property, $f(w) = \langle f,k_W(w,\cdot)\rangle_{\mathcal{H}_{W}}$ for each $f\in \mathcal{H}_W$. Hence,
$$
(\lambda I+A^*A)^{-1}(\widehat{A}^*-A^*)(\widehat{b}_t-\widehat{A}H^0_t)(w)
= \langle( \lambda I+A^*A )^{-1}(\widehat{A}^*-A^*)(\widehat{b}_t-\widehat{A}H^0_t),k_W(w,\cdot)\rangle_{\mathcal{H}_{W}}.
$$
Therefore, for $S_{n2}(s,t):=\sqrt{n}  \mathbb{P}\{ G_2m(X,s)\}$ we have
\begin{align*}
	| \mathbb{P}\{ G_2m(X,s) \}|
	&= \left| \mathbb{E} \left\{ ( \lambda I+A^*A )^{-1}(\widehat{A}^*-A^*)(\widehat{b}_t-\widehat{A}H^0_t)(W)\cdot m(X,s) \right\} \right| \\
	&= \left| \mathbb{E} \left\{ \langle ( \lambda I+A^*A )^{-1}(\widehat{A}^*-A^*)(\widehat{b}_t-\widehat{A}H^0_t),\phi_W(W)\rangle_{\mathcal{H}_W}\cdot m(X,s) \right\} \right|  \\
	&=\left| \langle ( \lambda I+A^*A )^{-1}(\widehat{A}^*-A^*)(\widehat{b}_t-\widehat{A}H^0_t), \mathbb{E}\{ m(X,s)\phi_W(W)\} \rangle_{\mathcal{H}_W} \right| ,
\end{align*}
where the last equation follows from \eqref{eq.kernel_mean_embedding}.

By $\{( \lambda I+A^*A)^{-1}\}^*=( \lambda I+A^*A )^{-1}$ (see Sec.~\ref{appx.hilbert}) and the Cauchy--Schwarz inequality,
\begin{align*}
	\left| \mathbb{P}\{ G_2m(X,s) \}\right|
	&=\left| \langle (\widehat{A}^*-A^*)(\widehat{b}_t-\widehat{A}H^0_t), ( \lambda I+A^*A )^{-1} \mathbb{E}\{ m(X,s)\phi_W(W)\} \rangle_{\mathcal{H}_W} \right|\\
    &\le \|(\widehat{A}^*-A^*)(\widehat{b}_t-\widehat{A}H^0_t)\|_{\mathcal{H}_W}\cdot \|( \lambda I+A^*A )^{-1}\mathbb{E}\{m(X,s)\phi_W(W)\}\|_{\mathcal{H}_W} \\
	&\le \|\widehat{A}^*-A^*\|_{\mathrm{op}}\cdot \|\widehat{b}_t-\widehat{A}H^0_t\|_{\mathcal{H}_X}\cdot \|( \lambda I+A^*A )^{-1}\mathbb{E}\{m(X,s)\phi_W(W)\}\|_{\mathcal{H}_W} \\
	&= \|\widehat{A}-A\|_{\mathrm{op}}\cdot \|\widehat{b}_t-\widehat{A}H^0_t\|_{\mathcal{H}_X}\cdot \|( \lambda I+A^*A )^{-1}\mathbb{E}\{m(X,s)\phi_W(W)\}\|_{\mathcal{H}_W},
\end{align*}
where the last equality uses $\|A^*\|_{\mathrm{op}}=\|A\|_{\mathrm{op}}$.  

By condition~\ref{assum.source condition} (a) with $g_s:=\mathbb{E}\{m(X,s)\phi_W(W)\}$ and Lemma~\ref{Lemm: Bounds on Compact Operator} (d), we obtain
$$
\|( \lambda I+A^*A )^{-1}\mathbb{E}\{m(X,s)\phi_W(W)\}\|_{\mathcal{H}_W} = O_p\{\lambda^{\frac{\min(\eta,2)}{2}-1}\}=O_p(1).
$$
By Lemmas~\ref{Lem: rates for Khat and fhat}, $\|\widehat{b}_t-b_t\|_{\mathcal{H}_X}=O_p(n^{-1/2})$ and $\|\widehat{A}-A\|_{\mathrm{op}}=O_p(n^{-1/2})$. Combining the above bounds, we get
\begin{equation}\label{eq: G2-front}
    \left|\mathbb{P}\{ G_2m(X,s)\}\right|\le O_p(n^{-1/2})\cdot \|\widehat{b}_t-\widehat{A}H^0_t\|_{\mathcal{H}_X}.
\end{equation}
Thus, by Lemma \ref{eq. f_hat-K_hatH}, we can obtain
\begin{align}\label{eq: G2}
\sqrt{n}\left|\mathbb{P}\{ G_2m(X,s)\}\right|
&\le \sqrt{n}\cdot O_p(n^{-1/2})\cdot O_p(n^{-1/2}) \cdot\notag\\
&= O_p(n^{-1/2})=o_p(1).
\end{align}
We complete the proof.
\end{proof}

\begin{lemma}
\label{lemma:Tn3}
Under conditions~\ref{assum.bandwidth},~\ref{ass:y_bounded}, \ref{ass:kernel_characteristic} and ~\ref{assum.source condition}, $S_{n3}(s,t)=o_{p}(1)$ as $n\to\infty$.
\end{lemma}
\begin{proof}
By Lemma~\ref{operator decomposition}, we have
\begin{align*}
	G_3
	&=\big\{ (\lambda I+\widehat{A}^*\widehat{A})^{-1}-(\lambda I+A^*A)^{-1} \big\} \widehat{A}^*(\widehat{b}_t-\widehat{A}H^0_t) \\
	&=(\lambda I+A^*A)^{-1}(A^*A-\widehat{A}^*\widehat{A})(\lambda I+\widehat{A}^*\widehat{A})^{-1}\widehat{A}^*(\widehat{b}_t-\widehat{A}H^0_t).
\end{align*}
By the reproducing property, $f(w) = \langle f,k_W(w,\cdot)\rangle_{\cH_W}$ for any $f\in \cH_W$. Hence, $(\lambda I+A^*A)^{-1}(A^*A-\widehat{A}^*\widehat{A})(\lambda I+\widehat{A}^*\widehat{A})^{-1}\widehat{A}^*(\widehat{b}_t-\widehat{A}H^0_t)(w)=\langle (\lambda I+A^*A)^{-1}(A^*A-\widehat{A}^*\widehat{A})(\lambda I+\widehat{A}^*\widehat{A})^{-1}\widehat{A}^*(\widehat{b}_t-\widehat{A}H^0_t),k_W(w,\cdot)\rangle_{\mathcal{H}_{W}}$. 
Therefore, for $S_{n3}(s,t):=\sqrt{n}\mP\{G_3m(X,s)\}$,
\begin{align}
	&\left| \mathbb{P} \{ G_3m(X,s) \} \right|\nonumber \\
	&=\left| \mathbb{E} \left[ (\lambda I+A^*A)^{-1}(A^*A-\widehat{A}^*\widehat{A})(\lambda I+\widehat{A}^*\widehat{A})^{-1}\widehat{A}^*(\widehat{b}_t-\widehat{A}H^0_t)(W) m(X,s) \right] \right| \nonumber \\
	&=\left| \mathbb{E} \left\{ \langle (\lambda I+A^*A)^{-1}(A^*A-\widehat{A}^*\widehat{A})(\lambda I+\widehat{A}^*\widehat{A})^{-1}\widehat{A}^*(\widehat{b}_t-\widehat{A}H^0_t),\phi_W(W) \rangle_{\mathcal{H}_W}\cdot m(X,s) \right\} \right| \nonumber \\
	&=\left| \langle (\lambda I+A^*A)^{-1}(A^*A-\widehat{A}^*\widehat{A})(\lambda I+\widehat{A}^*\widehat{A})^{-1}\widehat{A}^*(\widehat{b}_t-\widehat{A}H^0_t),\mathbb{E} \{ m(X,s)\phi_W(W)\} \rangle_{\mathcal{H}_W} \right|, \label{eq.PG3-gs}
\end{align}
where the last equation follows from \eqref{eq.kernel_mean_embedding}.

By the self-adjointness of $(\lambda I+A^*A)^{-1}$ and $(\lambda I+\wh{A}^*\wh{A})^{-1}$ and the Cauchy--Schwarz inequality,
\begin{align*}
	\left| \mathbb{P} \{ G_3m(X,s) \}\right|
    =&\left| \langle \widehat{b}_t-\widehat{A}H^0_t,\widehat{A}(\lambda I+\widehat{A}^*\widehat{A})^{-1}(A^*A-\widehat{A}^*\widehat{A})(\lambda I+A^*A)^{-1}\mathbb{E} \{ m(X,s)\phi_W(W)\} \rangle_{\mathcal{H}_X} \right|\\
    \le&\| \widehat{b}_t-\widehat{A}H^0_t\|_{\mathcal{H}_X}\cdot \| \widehat{A}(\lambda I+\widehat{A}^*\widehat{A})^{-1}(A^*A-\widehat{A}^*\widehat{A})(\lambda I+A^*A)^{-1}g_s \|_{\mathcal{H}_X}\\
	\le&\| \widehat{b}_t-\widehat{A}H^0_t\|_{\mathcal{H}_X}\cdot \| \widehat{A}(\lambda I+\widehat{A}^*\widehat{A})^{-1}\|_{\mathrm{op}}\cdot \| A^*A-\widehat{A}^*\widehat{A}\|_{\mathrm{op}}\cdot \|( \lambda I+A^*A )^{-1}g_s \|_{\mathcal{H}_W}.
\end{align*}

According to the paragraph above equation (97) in \cite{mastouri2021proximal},  $\wh{A}$ is compact. Thus, by Lemma~\ref{Lemm: Bounds on Compact Operator} (c), $\|\wh{A}(\lambda I+\wh{A}^*\wh{A})^{-1}\|_{\mathrm{op}}=O_p(\lambda^{-1/2})$. By conditions~\ref{assum.bandwidth}, \ref{assum.source condition} (a) and Lemma~\ref{Lemm: Bounds on Compact Operator} (d), 
$$
\|(\lambda I+A^*A)^{-1}g_s\|=O_p\{\lambda^{\frac{\min(\eta,2)}{2}-1}\}=O_p(1).
$$
By Lemma~\ref{Lem: rates for Khat g - fhat}, $\| A^*A-\widehat{A}^*\widehat{A}\|_{\mathrm{op}}=O_p(n^{-1/2})$. Combining all bounds, we get
\begin{equation}\label{eq:G3-front}
    \big|\mP\{G_3m(X,s)\}\big|=O_p(n^{-1/2})\cdot O_p(\lambda^{-1/2})\cdot \| \widehat{b}_t-\widehat{A}H^0_t\|_{\mathcal{H}_X}.
\end{equation}
Thus, by Lemma \ref{eq. f_hat-K_hatH} and condition~\ref{assum.bandwidth}, we can obtain
\begin{align}\label{eq: G3}
\sqrt{n}\big|\mP\{G_3m(X,s)\}\big|
&\le O_p(\lambda^{-1/2})\cdot O_p(n^{-1/2})\notag\\
&= O_p(1/\sqrt{n\lambda})=o_p(1).
\end{align}
We complete the proof.
\end{proof}

\begin{lemma}
\label{lemma:Tn4}
Under conditions~\ref{assum.bandwidth},~\ref{ass:y_bounded}, \ref{ass:kernel_characteristic} and \ref{assum.source condition}, $S_{n4}(s,t)=o_{p}(1)$ as $n\to\infty$.
\end{lemma}
\begin{proof}
By the reproducing property, $f(w) = \langle f,k_W(w,\cdot)\rangle_{\cH_W}$ for any $f\in \cH_W$. Hence, $\{ (\lambda I+\widehat{A}^*\widehat{A})^{-1}\widehat{A}^*\widehat{A}H^0_t-(\lambda I+A^*A)^{-1}A^*b_t \} (w)=\langle(\lambda I+\widehat{A}^*\widehat{A})^{-1}\widehat{A}^*\widehat{A}H^0_t-(\lambda I+A^*A)^{-1}A^*b_t,k_W(w,\cdot)\rangle_{\mathcal{H}_{W}}$. Therefore, for $S_{n4}(s,t):=\sqrt{n}\,\mP\{G_4m(X,s)\}$, 
\begin{align*}
	\left| \mathbb{P} \{G_4m(X,s)\} \right| &=\left| \mathbb{E} \left[ \{ (\lambda I+\widehat{A}^*\widehat{A})^{-1}\widehat{A}^*\widehat{A}H^0_t-(\lambda I+A^*A)^{-1}A^*b_t \} (W)\cdot m(X,s) \right] \right|\\
	&=\left| \mathbb{E} \left[ \langle (\lambda I+\widehat{A}^*\widehat{A})^{-1}\widehat{A}^*\widehat{A}H^0_t-(\lambda I+A^*A)^{-1}A^*b_t,\phi_W\left( W \right) \rangle_{\cH_W} \cdot m(X,s) \right] \right|\\
	&\overset{(1)}{=}\left| \langle (\lambda I+\widehat{A}^*\widehat{A})^{-1}\widehat{A}^*\widehat{A}H^0_t-(\lambda I+A^*A)^{-1}A^*b_t,\mathbb{E} \{ m(X,s)\phi_W\left( W \right) \} \rangle_{\cH_W} \right|,
\end{align*}
where (1) follows from \eqref{eq.kernel_mean_embedding}. By boundedness of $m(X,s)$ and the kernel $k_W$, it follows that 
\begin{equation}\label{eq.g_s}
    \begin{aligned}
        \| g_s\|_{\mathcal{H}_W}&=\| \mathbb{E} \{m(X,s)\phi_W(W) \} \|_{\mathcal{H}_W}=\| \mathbb{E} \left[ \mathbb{E} \{ m( X,s) |W \} \phi_W(W) \right] \|_{\mathcal{H}_W}\\
	&\le C\| \mathbb{E} \{\phi_W(W)\} \|_{\mathcal{H}_W}=C\sqrt{\langle \mathbb{E} \{\phi_W(W)\} ,\mathbb{E} \{\phi_W(W)\} \rangle_{\mathcal{H}_W}}\\
	&= C\sqrt{\mathbb{E} \left\{ \langle \phi_W(W) ,\phi_W\left( W' \right) \rangle_{\mathcal{H}_W} \right\}} =C\sqrt{\mathbb{E} \{ k_W(W,W') \}} <\infty.
    \end{aligned}
\end{equation}

\noindent\textbf{Step 1. Spectral representation.}

For the operator $A:\cH_W\to\cH_X$ defined in ~\eqref{eq.operator_defination}, its singular value decomposition given by $(s_n,u_n,v_n)_{n=1}^{+\infty}$. Hence, we have $Au_j=s_jv_j$ and $A^*v_j=s_ju_j$. Define the formal inverse
\begin{equation}
\label{eq.g_s_define}
    \wt g_s:=\sum_j s_j^{-2}\langle g_s,u_j\rangle_{\mathcal{H}_W} u_j.
\end{equation}
Next, we will calculate $\| \wt{g}_s\|_{\mathcal{H}_W}^2$. In fact, we have
\begin{equation}\label{eq.g_s_wt}
    \| \wt{g}_s\|_{\mathcal{H}_W}^2=\left\langle \sum_j{s_j^{-2}}\langle g_s,u_j \rangle_{\mathcal{H}_W} u_j,\sum_j{s_j^{-2}}\langle g_s,u_j \rangle_{\mathcal{H}_W}  u_j \right\rangle_{\mathcal{H}_W} = \sum_j{s_j^{-4}|\langle g_s,u_j \rangle_{\mathcal{H}_W}  |^2}.
\end{equation}
By condition~\ref{assum.source condition} (a), we have that for some $\eta \geq 2$, $\sum_j{s_j^{-2\eta}|\langle g_s,u_j\rangle_{\cH_W} |^2}<\infty$, which means that $\| \wt{g}_s\|_{\mathcal{H}_W}^2<\infty$. 

According to the properties of singular value decomposition, we have
\begin{equation}
\label{eq.g_s_wt_g_s}
A^*A\wt{g}_s=\sum_j s_j^2 s_i^{-2} \langle g_s,u_j \rangle_{\mathcal{H}_W}  u_j =\sum_j{\langle g_s,u_j \rangle_{\mathcal{H}_W}  u_j}=g_s.
\end{equation}


\noindent\textbf{Step 2. Decomposition of the difference.}

Define $P_t:=(\lambda I+\widehat{A}^*\widehat{A})^{-1}\widehat{A}^*\widehat{A}H^0_t-(\lambda I+A^*A)^{-1}A^*b_t$.  By $AH^0_t = b_t$, we can decompose $P_t$ as follows by Lemma~\ref{operator decomposition}
\begin{equation}\label{eq. P_t}
    \begin{aligned}
        P_t = &(\lambda I+\widehat{A}^*\widehat{A})^{-1}\widehat{A}^*\widehat{A}H^0_t-(\lambda I+A^*A)^{-1}A^*AH^0_t\\
	= &\left\{ (\lambda I+\widehat{A}^*\widehat{A})^{-1}(\lambda I+\widehat{A}^*\widehat{A}-\lambda I)-(\lambda I+A^*A)^{-1}(\lambda I+A^*A-\lambda I) \right\} H^0_t\\
	= &\lambda \left\{ (\lambda I+A^*A)^{-1}-(\lambda I+\widehat{A}^*\widehat{A})^{-1} \right\} H^0_t\\
	= &\lambda (\lambda I+\widehat{A}^*\widehat{A})^{-1}\{ \widehat{A}^*\widehat{A}-A^*A \} (\lambda I+A^*A)^{-1}H^0_t.
    \end{aligned}
\end{equation}

\noindent\textbf{Step 3. Bounding $P_t$ and $\widehat{A}P_t$.}

Note that
\begin{align*}\label{eq: bound for Xi4}
	\|P_t\|_{\mathcal{H}_W}&= \|\lambda (\lambda I+\widehat{A}^*\widehat{A})^{-1}(\widehat{A}^*\widehat{A}-A^*A ) (\lambda I+A^*A)^{-1}H^0_t\|_{\mathcal{H}_W}\\
	&\le \|\lambda (\lambda I+\widehat{A}^*\widehat{A})^{-1}\|_{\mathrm{op}}\cdot \|A^*A-\widehat{A}^*\widehat{A}\|_{\mathrm{op}}\cdot \|(\lambda I+A^*A)^{-1}H^0_t\|_{\mathcal{H}_W} 
\end{align*}

Since $\wh{A}$ is a compact operator as stated in the proof of Lemma~\ref{lemma:Tn2}, we have $\|(\lambda (\lambda I + \widehat{A}^*\widehat{A})^{-1}\|_{\mathrm{op}}\leq 
2$ by Lemma \ref{Lemm: Bounds on Compact Operator} (b). By condition~\ref{assum.source condition} (b), we can apply Lemma~\ref{Lemm: Bounds on Compact Operator} (d) to obtain that 
$$\|(\lambda I + A^*A)^{-1}H^0_t\|_{\mathcal{H}_W}=O_p\{\lambda^{\frac{\min(\theta ,2)}{2}-1}\}=O_p(1).
$$
Finally, by Lemma~\ref{Lem: rates for Khat g - fhat}, we have $ \|A^*A-\widehat{A}^*\widehat{A}\|_{\mathrm{op}}=O_p(n^{-1/2})$. Combining all the inequalities, we get
$$
\| P_t\|_{\mathcal{H}_W} =  O_p(n^{-1/2}).
$$
Next we provide the bound for $\| \widehat{A}P_t \|_{\cH_X}$. Note that
\begin{align*}
	\|\widehat{A}P_t\|_{\mathcal{H}_X} &=\|\lambda \widehat{A}(\lambda I+\widehat{A}^*\widehat{A})^{-1}(\widehat{A}^*\widehat{A}-A^*A )(\lambda I+A^*A)^{-1}H^0_t\|_{\mathcal{H}_X}\\
	&\le \lambda \cdot \|\widehat{A}(\lambda I+\widehat{A}^*\widehat{A})^{-1}\|_{\mathrm{op}}\cdot \|\widehat{A}^*\widehat{A}-A^*A\|_{\mathrm{op}}\cdot \|(\lambda I+A^*A)^{-1}H^0_t\|_{\mathcal{H}_W}.
\end{align*}

Since $\widehat{A}$ is a compact operator, by Lemma \ref{Lemm: Bounds on Compact Operator} (c), we have $\|(\lambda I + \widehat{A}^*\widehat{A} )^{-1}\widehat{A}^*\|_{\mathrm{op}}=\|\widehat{A}(\lambda I +\widehat{A}^*\widehat{A} )^{-1}\|_{\mathrm{op}}= O_p(\lambda^{-1/2})$. Under condition~\ref{assum.source condition} (b), Lemma~\ref{Lemm: Bounds on Compact Operator} (d) yields $$\|(\lambda I + A^*A)^{-1}H^0_t\|_{\mathcal{H}_W}=O_p\{ \lambda^{\frac{\min(\theta ,2)}{2}-1}\}=O_p(1).
$$
Finally, by Lemma~\ref{Lem: rates for Khat g - fhat}, we have $ \|A^*A-\widehat{A}^*\widehat{A}\|_{\mathrm{op}}=O_p(n^{-1/2})$. Combining all the inequalities, we get
$$
\| \widehat{A}P_t\| =  O_p(\lambda^{1/2})\cdot O_p(n^{-1/2}).
$$

\noindent\textbf{Step 4. Conclusion.}

By \eqref{eq.g_s_wt_g_s} and the Cauchy-Schwartz inequality, we have:
\begin{align*}
	\left| \mathbb{P} \{G_4m(X,s) \} \right|&=\left| \langle P_t,\mathbb{E} \{ m(X,s)\phi_W(W)\} \rangle_{\mathcal{H}_W}  \right| \\
    & =\left| \langle P_t,A^*A\wt{g}_s \rangle_{\mathcal{H}_W}  \right|\\
	&\le \left| \langle P_t,(A^*-\widehat{A}^*)A\wt{g}_s \rangle_{\mathcal{H}_W}  \right|+\left| \langle P_t,\widehat{A}^*A\wt{g}_s \rangle_{\mathcal{H}_W}  \right|\\
	&=\left| \langle P_t,(A^*-\widehat{A}^*)A\wt{g}_s \rangle_{\mathcal{H}_W}  \right|+\left| \langle \widehat{A}P_t,A\wt{g}_s \rangle_{\mathcal{H}_W}  \right|\\
	&\overset{(1)}{\le}\| P_t\|_{\mathcal{H}_W}  \cdot \| \widehat{A}-A\|_{\mathrm{op}}\cdot \| A\wt{g}_s\|_{\mathcal{H}_X}  +\| \widehat{A}P_t\|_{\mathcal{H}_X}  \cdot \| A\wt{g}_s\|_{\mathcal{H}_X}  ,
\end{align*}
where (1) follows from $\|\widehat{A}^*-A^*\|_{\mathrm{op}}=\|\widehat{A}-A\|_{\mathrm{op}}$. 

By Lemmas~\ref{Lem: rates for Khat and fhat}, $\|\widehat{A}-A\|_{\mathrm{op}}=O_p(n^{-1/2})$. Besides, since $A$ is bounded, we have $\| A\|_{\mathrm{op}}<\infty$. By \eqref{eq.g_s_wt}, we have $\|\wt{g}_s\|_{\mathcal{H}_W}<\infty$. Thus, we get $\| A\wt{g}_s\|_{\mathcal{H}_X}\le\|A \|_{\mathrm{op}}\cdot\|\wt{g}_s\|_{\mathcal{H}_W}<\infty$. Combining these results, we get
\begin{equation}\label{eq: negligibility of inner product for G4}
\sqrt{n}|\mP\{G_4m(X,s)\}|= O_p(n^{-1/2})+O_p(\lambda^{1/2}).
\end{equation}
The last term is $o_p(1)$ under condition~\ref{assum.bandwidth}. 
\end{proof}

\begin{lemma}
\label{lemma:Tn5}
Under conditions~\ref{assum.bandwidth},~\ref{ass:y_bounded}, \ref{ass:kernel_characteristic} and~\ref{assum.source condition}, $S_{n5}(s,t)=o_{p}(1)$ as $n\to\infty$.
\end{lemma}

\begin{proof}
By the reproducing property, we have $\left\{(\lambda I + A^*A)^{-1}A^*b_t-H^0_t \right\}(w)=\langle(\lambda I + A^*A)^{-1}A^*b_t-H^0_t,k_W(w,\cdot)\rangle_{\mathcal{H}_{W}}$. Thus, for $S_{n5}(s,t):=\sqrt{n}\,\mP\{G_5m(X,s)\}$,
\begin{align*}
	\left| \mathbb{P} \{G_5m(X,s)\} \right|&=\left| \mathbb{E} \left[ \{(\lambda I+A^*A)^{-1}A^*b_t-H^0_t\}(W)\cdot m(X,s) \right] \right|\\
	&=\left| \mathbb{E} \left[ \left< (\lambda I+A^*A)^{-1}A^*b_t-H^0_t,\phi_W\left( W \right) \right>_{\mathcal{H}_W}\cdot m(X,s) \right] \right|\\
	&\overset{(1)}{=}\left| \langle (\lambda I+A^*A)^{-1}A^*b_t-H^0_t,\mathbb{E} \{m(X,s)\phi_W(W)\}\rangle_{\mathcal{H}_W} \right|,
\end{align*}
where (1) follows from \eqref{eq.kernel_mean_embedding}.

By condition~\ref{assum.source condition} (b) and $AH^0_t =b_t$, we can apply Lemma~\ref{lem.H_laambda-H_0} to obtain that 
$$\|(\lambda I + A^*A)^{-1}A^*b_t-H^0_t\|_{\mathcal{H}_W}=\|(\lambda I + A^*A)^{-1}A^*AH^0_t-H^0_t\|_{\mathcal{H}_W}=O_p\{\lambda^{\frac{\min(\theta ,2)}{2}}\}.
$$

Combining this rate with the Cauchy-Schwartz inequality and and $\theta \geq 2$ in condition~\ref{assum.source condition} (a), we have 
\begin{equation}\label{eq: inner product of Regularization Bias }
\begin{aligned}
	\sqrt{n}\left| \mathbb{P} \{ G_5m(X,s) \} \right|&\le \sqrt{n}\cdot \| (\lambda I+A^*A)^{-1}A^*b_t-H^0_t\|_{\mathcal{H}_W} \cdot \| g_s\|_{\mathcal{H}_W} \\
    &\overset{(1)}{=}O_{p}(\sqrt{n\lambda^2})\overset{(2)}{=}o_{p}(1),
\end{aligned}
\end{equation}
where $(1)$ follows from $ \| g_s\|_{\mathcal{H}_W}<\infty$ by Eq.~\eqref{eq.g_s} in Lemma~\ref{lemma:Tn4} and $(2)$ follows from condition~\ref{assum.bandwidth}.
\end{proof}

\begin{lemma}
\label{lemma:Tn1}
Under conditions~\ref{assum.bandwidth}, \ref{ass:y_bounded} and \ref{ass:kernel_characteristic}, we have 
$$
S_{n1}(s,t)=\frac{1}{\sqrt{n}}\sum_{i=1}^n{U(w_i,y_i,t) \{A(A^*A)^{-1}g_s\}(x_i)}+o_{p}(1)
$$ 
as $n\to\infty$, where
$g_s(\cdot):=\mathbb{E}\{m(X,s) k_W( W,\cdot )\}$.
\end{lemma}

\begin{proof}
By the reproducing property, we have $(\lambda I+A^*A)^{-1}A^*(\widehat{b}_t-\widehat{A}H^0_t)(W)=\langle (\lambda I+A^*A)^{-1}A^*(\widehat{b}_t-\widehat{A}H^0_t),\phi_W(W)\rangle_{\cH_W}$. Therefore, for $\mathbb{P} \{G_1m(X,s)\}$, we have
\begin{equation}\label{eq.PG1-gs}
    \begin{aligned}
	\mathbb{P} \{G_1m(X,s)\} &=\mathbb{E} \left[ (\lambda I+A^*A)^{-1}A^*(\widehat{b}_t-\widehat{A}H^0_t)(W)m(X,s) \right]  \\
    &=\mathbb{E}\left[\left\langle (\lambda I+A^*A)^{-1}A^*(\widehat{b}_t-\widehat{A}H^0_t), \phi_W(W) \right\rangle_{\mathcal{H}_W}\cdot m(X,s)\right]  \\
	&\overset{(1)}{=} \left\langle (\lambda I+A^*A)^{-1}A^*(\widehat{b}_t-\widehat{A}H^0_t),\mathbb{E} \{ m(X,s)\phi_W(W)\} \right\rangle_{\mathcal{H}_W} \\
	&\overset{(2)}{=}\langle A^*(\widehat{b}_t-\widehat{A}H^0_t),(\lambda I+A^*A)^{-1}g_s \rangle_{\mathcal{H}_W}  \\
	&=\langle A^*(\widehat{b}_t-\widehat{A}H^0_t),\{ (\lambda I+A^*A)^{-1}-(A^*A)^{-1}\} g_s \rangle_{\mathcal{H}_W} \\
    & \qquad \qquad +\langle A^*(\widehat{b}_t-\widehat{A}H^0_t),(A^*A)^{-1}g_s \rangle_{\mathcal{H}_W},
\end{aligned}
\end{equation}
where (1) follows from \eqref{eq.kernel_mean_embedding} and (2) follows from $\{( \lambda I+A^*A )^{-1}\}^*=( \lambda I+A^*A )^{-1}$.

We first analyze the second term in RHS. By \eqref{eq.estimate_f_K}, we obtain,
\begin{align}
	(\widehat{b}_t-\widehat{A}H^0_t)(X)&=\left\{\frac{1}{n}\sum_{i=1}^n{\varphi (y_i,t) \phi_X(x_i)}-\frac{1}{n}\sum_{i=1}^n{H^0(w_i,t) \phi_X(x_i)}\right\}(X) \nonumber \\
	&=\frac{1}{n}\sum_{i=1}^n{U(w_i,y_i,t) k_X( x_i,X)}, \label{eq.bhat_residue}
\end{align}
where $U$ is defined in \eqref{eq.U}. Since $A^*m_t := \int m(X,t)\phi_W(W)dF(X,W)$ in \eqref{eq.operator_defination_adjoin_copy}, we have
\begin{align*}
	 A^*(\widehat{b}_t-\widehat{A}H^0_t)(W)&=\frac{1}{n}\sum_{i=1}^n{U(w_i,y_i,t) \int{k_X( x_i,X) \phi_W(W) dF(X,W)}}\\
	&=\frac{1}{n}\sum_{i=1}^n{U(w_i,y_i,t) A^*\{ k_X( x_i,\cdot)\}(W)}, 
\end{align*}

Therefore, we obtain
\begin{align*}
	&\sqrt{n}\left\langle A^*(\widehat{b}_t-\widehat{A}H^0_t),(A^*A)^{-1}g_s \right\rangle_{\mathcal{H}_W} \\
    = &\sqrt{n} \left\langle (A^*A)^{-1}A^*(\widehat{b}_t-\widehat{A}H^0_t),\mathbb{E} \{ m(X,s)\phi_W(W)\} \right\rangle_{\mathcal{H}_W} \\
	\overset{(1)}{=} &\sqrt{n}\mathbb{E} \left\{(A^*A)^{-1}A^*(\widehat{b}_t-\widehat{A}H^0_t)(W)m(X,s) \right\}\\
	=&\sqrt{n}\mathbb{E} \left[ (A^*A)^{-1}\left\{ \frac{1}{n}\sum_{i=1}^n{U(w_i,y_i,t) A^*\{ k_X( x_i,\cdot ) \}(W)} \right\} m(X,s) \right]\\
	=&\frac{1}{\sqrt{n}}\sum_{i=1}^n{U(w_i,y_i,t) \int{(A^*A)^{-1}A^*\{ k_X( x_i,\cdot )\}(W) m(X,s) dF(X,W)}}\\
	\overset{(2)}{=}&\frac{1}{\sqrt{n}}\sum_{i=1}^n{U(w_i,y_i,t) \left\langle (A^*A)^{-1}A^*\{ k_X( x_i,\cdot )\}(W) ,\mathbb{E} \{ m(X,s)\phi_W(W)\} \right\rangle_{\mathcal{H}_W}}\\
	=&\frac{1}{\sqrt{n}}\sum_{i=1}^n{U(w_i,y_i,t) \left\langle k_X( x_i,\cdot) ,A(A^*A)^{-1}g_s \right\rangle_{\mathcal{H}_X}}\\
	\overset{(3)}{=}&\frac{1}{\sqrt{n}}\sum_{i=1}^n{U(w_i,y_i,t) \{A(A^*A)^{-1}g_s\}(x_i)},
\end{align*}
where $(1),(2),(3)$ follows from reproducing property $f(x)=\langle f, k_X(x,\cdot)\rangle_{\cH_X}$ and $g(w) = \langle g,k_W(w,\cdot)\rangle_{\mathcal{H}_{W}}$ for each $f\in \mathcal{H}_X$ and $g \in \cH_W$. 

Next, we look at the first term in \eqref{eq.PG1-gs}. By the Cauchy-Schwarz inequality, we have
    \begin{align}
	&\left| \left\langle A^*(\widehat{b}_t-\widehat{A}H^0_t),\{ (\lambda I+A^*A)^{-1}-(A^*A)^{-1} \} g_s \right\rangle_{\mathcal{H}_W} \right|\nonumber\\
        & \le \| A^*(\widehat{b}_t-\widehat{A}H^0_t)\|_{\mathcal{H}_W} \cdot\left\| \left\{ (\lambda I+A^*A)^{-1}-(A^*A)^{-1} \right\} g_s \right\|_{\mathcal{H}_W} \nonumber\\
	&\le  \left\| \left\{ (\lambda I+A^*A)^{-1}-(A^*A)^{-1} \right\} g_s \right\|_{\mathcal{H}_W} \cdot  \| A^*(\widehat{b}_t-\widehat{A}H^0_t)\|_{\mathcal{H}_W}. \label{eq: G1-front}
\end{align}
Besides, for $\sqrt{n}\| A^*(\widehat{b}_t-\widehat{A}H^0_t)\|_{\mathcal{H}_W}$, we have:
\begin{align*}
	\| A^*(\widehat{b}_t-\widehat{A}H^0_t)\|_{\mathcal{H}_W} &=\| A^*\widehat{b}_t-A^*b_t+A^*b_t-A^*\widehat{A}H^0_t\|_{\mathcal{H}_W}\\
	&\le \| A^*\widehat{b}_t-A^*b_t\|_{\mathcal{H}_W} +\| A^*AH^0_t-A^*\widehat{A}H^0_t\|_{\mathcal{H}_W}\\
	&\le \| A^*\|_{\mathrm{op}} \cdot \| \widehat{b}_t-b_t\|_{\mathcal{H}_X} +\| A^*\|_{\mathrm{op}} \cdot \| A-\widehat{A}\|_{\mathrm{op}} \cdot \| H^0_t\|_{\mathcal{H}_W}.
\end{align*}

Since $H^0_t \in \cH_W$, we must have $\| H^0_t\|_{\cH_W} < \infty$. Besides, according to Sec.~\ref{appx.hilbert}, we have $\| A^*\|_{\mathrm{op}}=\| A\|_{\mathrm{op}} < \infty$ since $A$ is a bounded linear operator. Therefore, the last term is $O_{p}(n^{-1/2})$ by Lemma~\ref{Lem: rates for Khat and fhat}, which means that $\sqrt{n}\| A^*(\widehat{b}_t-\widehat{A}H^0_t)\|_{\mathcal{H}_W}=O_{p}(1)$.

By $A^*A\wt{g}_s=g_s$ in \eqref{eq.g_s_wt_g_s}, we obtain
\begin{align}
\label{eq.wt_g_s}
    \{(\lambda I + A^* A)^{-1}-(A^*A)^{-1}\}g_s=(\lambda I + A^* A)^{-1}A^*(A\wt{g}_s)-\wt{g}_s.
\end{align}
Note that the operator $(\lambda I + A^*A)^{-1}A^*$ corresponds to the Tikhonov regularization scheme. In fact, Lemma \ref{lem:tikhonov-regularization} confirms that $(\lambda I + A^*A)^{-1}A^*$ qualifies as a regularization scheme. Recall that, by Definition \ref{def:regularization-scheme}, a family of operators $\{R_{\lambda}\}$ is termed a regularization scheme for the operator $A$ if $\lim_{\lambda\to 0} R_{\lambda}Af=f$ holds for suitable $f$. As a direct consequence of this definition and the aforementioned theorem, the right-hand side (RHS) of \eqref{eq.wt_g_s} converges to $0$ as $\lambda \to 0$.
\end{proof}

\subsection{Proofs in section~\ref{sec.asymptotic}}

\nullhypothesis*
\begin{proof}
  By \eqref{eq.Tnst decompose}, we have
$$
T_n(s,t) =\sqrt{n}\mathbb{P}_n\{ U(W,Y,t) m(X,s) \}+\left( \text{Expected risk difference} \right) +\left( \text{Empirical process} \right).
$$

By Propositions~\ref{prop.empirical_process} and \ref{prop.expected_risk}, we have:
$$
T_n(s,t) =\frac{1}{\sqrt{n}}\sum_{i=1}^n{U(w_i,y_i,t)\left[m(x_i,s)- \left\{A(A^*A)^{-1}g_s\right\}(x_i)\right]}+o_{p}(1).
$$
Next, we apply Lemma~\ref{ central limit theorem} to $\left\{ U(w_i,y_i,t)\left[ \left\{m(x_i,s)-A(A^*A)^{-1}g_s\right\}(x_i)\right] \right\}_i$ to show that it converges to Gaussian process.  To this end, we need to verify $U(W,Y,t)[ m(X,s)-\{A(A^*A)^{-1}g_s\}(X)]$ is zero mean and 
\begin{equation}
\label{eq.var-infinity}
    \mathbb{E} \left[ \left\| U(w_i,y_i,t)[m(x_i,s)-\{A(A^*A)^{-1}g_s\}(x_i))] \right\|_{\cL^2\{\cT \times \cT, \mu \times \mu\}} \right] <\infty. 
\end{equation}
Notice that the zero-mean property is ensured by the fact that $\mathbb{E} \{ U(W,Y,t) |X \} =\mathbb{E} \{ \varphi(Y,t) -H^0(W,t) |X \} =0$ under $\mathbb H_0$. Besides, by condition~\ref{assum.var}, we have 
\begin{align*}
    & \operatorname{Var}(U(w_i,y_i,t)[ m(x_i,s)-\{A(A^*A)^{-1}g_s\}(x_i)] ) \\
    & =\mathbb{E}(U(w_i,y_i,t)[ m(x_i,s)-\{A(A^*A)^{-1}g_s\}(x_i)])^2 \\
    & <\mathbb{E}\{U(w_i,y_i,t)^4\}+\mathbb{E}[ m(x_i,s)-\{A(A^*A)^{-1}g_s\}(x_i)]^4<\infty
\end{align*} 
for any $(s,t)$. Therefore, setting $\mu$ to be a probability measure, we get \eqref{eq.var-infinity}. Thus, we have $T_n(s,t)$ converges weakly to $\mathbb{G}(s,t)$ in $\cL^2\{\cT \times \cT, \mu \times \mu\}$, where $\mathbb{G}(s,t)$ is a Gaussian process with zero-mean.

For any fixed $t$ and $T_n(s,t)\in \cL^2\{\cT, \mu \}$, applying the continuous mapping theorem (Theorem 1.3.6 of \cite{wellner2013weak}), we have:
$$
\int{|T_n(s,t)|^2d\mu(s)}\xrightarrow{d}\int{|\mathbb{G} (s,t)|^2d\mu(s)},
$$
by the continuity of the integral functional. Next, we need take the maximum of $\int{|T_n(s,t)|^2d\mu(s)}\in \cL^2\{\cT, \mu\}$ over $t$, and verify the legality of taking the maximum. Note that the part of $\int{|T_n(s,t)|^2d\mu(s)}\in \cL^2\{\cT, \mu\}$ regarding $t$ is determined by $U(w,y,t)=\varphi(y,t)-H^0(w,t)$. To ensure that taking the max operation is meaningful, we need to prove that if $U(w,y,t)\in \cL^2\{F(w,y)\}$ for any $t$, $\underset{t\in T}{\max}|U(w,y,t)|\in\cL^2\{F(w,y)\}$. By \eqref{eq.Ht_in_L2}, we have: 
\begin{align*}
	&\int{|\varphi (y,t)-H(w,t)|^2p(w,y)dwdy}\\
    &\le 2\int{|\varphi (y,t)|^2p(y)dy}+2\int{|H(w,t)|^2p(w)dw}\\
	&\le 2+2\left(\int \|h(w,y\|_{\cL^2\{F(w)\}}dy \right)^2<\infty,
\end{align*}
where the second inequality follows from $(a-b)^2\le 2a^2+2b^2$ and $|e^{ity}|^2=1$. 
Thus, taking max operation on both sides, we have $\underset{t\in T}{\max} \int{| U(w,y,t) |^2p(w,y) dwdy}<\infty$. Next, we prove the continuity of the max functional in metric $d$. Next, we prove the continuity of the max functional. If $d(f_1,f_2)<\delta$ given any $\delta>0$, we have $\underset{t\in T}{\max}|f_1(t)|-\underset{t\in T}{\max}|f_2(t)|\le\underset{t\in T}{\max}|f_1(t)-f_2(t)|=d(f_1,f_2)<\delta$. Applying the continuous mapping theorem to such a continuous metric $\max$, we have:
$$
\underset{t\in T}{\max} \thinspace \Delta \left( t \right) \xrightarrow{d}\underset{t\in T}{\max}\int{|\mathbb{G} (s,t)|^2d\mu(s)}.
$$
The proof is complete. 
\end{proof}

\powerhypothesis*
\begin{proof}
Following the decomposition in \eqref{eq.Tnst decompose}, we can write
\begin{equation}\label{eq.Tnst decompose_H1}
\begin{aligned}
	T_n(s,t)&=\sqrt{n}\mathbb{P}_n\left[ \{\varphi (Y,t)-H^*(W,t)\}m(X,s) \right] -\sqrt{n}\mathbb{P} \left[ \{\widehat{H}^{\lambda}(W,t)-H^*(W,t)\}m(X,s) \right]\\
	&\qquad -\sqrt{n}(\mathbb{P}_n-\mathbb{P} )\left[\{ \widehat{H}^{\lambda}(W,t)-H^*(W,t)\}m(X,s) \right],
\end{aligned}
\end{equation}
where $H^0(W,t)$ in \eqref{eq.Tnst decompose} is replaced by $H^*(W,t)=( A^*A)^{-1}A^*b_t$. We first consider the local alternative and then consider the global alternative.

\noindent\textbf{(1). The case of $\mH_{1n}^\alpha$ with 
$0<\alpha<1/2$.} 

We first decompose the term into $\widehat{H}^{\lambda}(W,t)-H^*(W,t)$ into $\sum_{i=1}^6 G_i$, where $G_1,G_2,G_3,G_4,G_5$ are defined in \eqref{eq.G1}-\eqref{eq.G5} and $G_6:=H^0( w,t ) -H^*( w,t )$. 

Note that under $\mH_{1n}^\alpha$, we have $    \mathbb H^\alpha_{1n}: \mathbb{E}\{\varphi(Y,t)|X\}= \mathbb{E}\{H^0(W,t)|X\}+\frac{r(X,t)}{n^\alpha}$. Thus, applying the operator \ref{eq.operator_defination}, we can obtain $b_t=AH^0_t+\ell(X,t)/n^{\alpha}$, where $\ell(\cdot,t):=\mathbb{E}\{r(X,t)\phi_X(X)\}$.

\textbf{Analysis of $G_2$.} For $\mP\{G_2m(X,s)\}$, applying \eqref{eq: G2-front} in Lemma~\ref{lemma:Tn2},  we obtain
$$
|\mathbb{P}\{ G_2m(X,s)\}| = O_p(n^{-1/2})\cdot \|\widehat{b}_t-\widehat{A}H^0_t\|_{\mathcal{H}_X}.
$$
Note that $\widehat{b}_t-\widehat{A}H^0_t=\widehat{b}_t-b_t+b_t-\widehat{A}H_{t}^{0}=\widehat{b}_t-b_t+AH^0_t+\ell(X,t)/n^{\alpha}-\widehat{A}H_{t}^{0}$. Thus, by Lemmas~\ref{Lem: rates for Khat and fhat}, $\| \ell(X,t)\|_{\mathcal{H}_X}<\infty$ and $\|H_{t}^{0}\|_{\mathcal{H}_W}<\infty$, we can obtain
\begin{equation}
\begin{aligned}\label{eq. f_hat-K_hatH_H1}
    \|\widehat{b}_t-\widehat{A}H^0_t\|_{\mathcal{H}_X}&\le \|\widehat{b}_t-b_t\|_{\mathcal{H}_X}+ \|A-\widehat{A}\|_{\mathrm{op}}\cdot\|H_{t}^{0}\|_{\mathcal{H}_W}+n^{-\alpha}\| \ell(X,t)\|_{\mathcal{H}_X}\\
    & =  O_p(n^{-1/2}+n^{-\alpha}).
\end{aligned}
\end{equation}
Applying such results to the above, we get
$$
\mP\{G_2m(X,s)\}
\le O_p(n^{-1/2})\cdot \|\widehat{b}_t-\widehat{A}H^0_t\|_{\mathcal{H}_X} =  O_p(n^{-\alpha-1/2}).
$$

\textbf{Analysis of $G_3$.} For $\mP\{G_3m(X,s)\}$, applying \eqref{eq:G3-front} in Lemma~\ref{lemma:Tn3}, we obtain
$$
|\mP\{G_3m(X,s)\}|=O_p(1/\sqrt{n\lambda})\cdot \| \widehat{b}_t-\widehat{A}H^0_t\|_{\mathcal{H}_X}.
$$
Similarly, according to \eqref{eq. f_hat-K_hatH_H1}, we obtain
$$
\mP\{G_3m(X,s)\}
= O_p(1/\sqrt{n\lambda})\cdot \| \widehat{b}_t-\widehat{A}H^0_t\|_{\mathcal{H}_X}  \le    n^{-\alpha}\cdot O_p(1/\sqrt{n\lambda}).
$$

\textbf{Analysis of $G_4$ and $G_5$.} Since $b_t=AH^0_t+\ell(X,t)/n^{\alpha}$, we can obtain
\begin{align*}
	G_4+G_5
	&= (\lambda I+\widehat{A}^*\widehat{A})^{-1}\widehat{A}^*\widehat{A}H_{t}^{0}-(\lambda I+A^*A)^{-1}A^*b_t+(\lambda I+A^*A)^{-1}A^*b_t-H_{t}^{0} \\
	&= \underbrace{(\lambda I+\widehat{A}^*\widehat{A})^{-1}\widehat{A}^*\widehat{A}H_{t}^{0}-(\lambda I+A^*A)^{-1}A^*AH_{t}^{0}}_{\overline{G}_4}+ \underbrace{(\lambda I+A^*A)^{-1}A^*AH_{t}^{0}-H_{t}^{0}}_{\overline{G}_5}.
\end{align*}
According to lemma~\ref{lemma:Tn4} and \ref{lemma:Tn5}, we can obtain $\mP\{\overline{G}_4m(X,s)\}=O_{p}(1/n+\lambda^{1/2}/\sqrt{n})$ and $\mP\{\overline{G}_5m(X,s)\}=O_p(\lambda)$, which means that $\mP\{(G_4+G_5)m(X,s)\}=O_{p}(1/n+\lambda^{1/2}/\sqrt{n}+\lambda)$.

\textbf{Analysis of $G_1$.} For $\mP\{G_1m(X,s)\}$, applying \eqref{eq.gs} in Lemma~\ref{lemma:Tn1}, we have
\begin{align*}
	\mP\{G_1m(X,s)\}=& \underbrace{\left\langle A^*(\widehat{b}_t-\widehat{A}H_{t}^{0}),\left\{ (\lambda I+A^*A)^{-1}-(A^*A)^{-1} \right\} g_s \right\rangle_{\cH_W}}_{(\mathrm{I})} \\
    &+ \underbrace{\langle A^*(\widehat{b}_t-\widehat{A}H_{t}^{0}),(A^*A)^{-1}g_s \rangle_{\cH_W}}_{(\mathrm{II})}.
\end{align*}

For the term $\mathrm{(I)}$, applying \eqref{eq: G1-front}, we have
\begin{align*}
    \mathrm{(I)}&\le \left\| \left\{ (\lambda I+A^*A)^{-1}-(A^*A)^{-1} \right\} g_s \right\|_{\mathcal{H}_W} \cdot  \| A^*(\widehat{b}_t-\widehat{A}H^0_t)\|_{\mathcal{H}_W}\\
    & \le \left\| \left\{ (\lambda I+A^*A)^{-1}-(A^*A)^{-1} \right\} g_s \right\|_{\mathcal{H}_W} \cdot  \| A^*\|_{\mathrm{op}}\cdot\|(\widehat{b}_t-\widehat{A}H^0_t)\|_{\mathcal{H}_X}.
\end{align*}
By 
Lemma~\ref{lemma:Tn1}, we have $\| \left\{ (\lambda I+A^*A)^{-1}-(A^*A)^{-1} \right\} g_s \|_{\cH_W}=o_p(1)$. Besides, by \eqref{eq. f_hat-K_hatH_H1} and the fact that $ \| A^*\|_{\mathrm{op}}<\infty$, we can obtain $\mathrm{(I)}\le o_p(n^{-\alpha})$. 

For term $(\mathrm{II})$, following \eqref{eq.bhat_residue} in Lemma~\ref{lemma:Tn1}, we have:
\begin{align*}
	(\mathrm{II}) &= \left\langle A^*\left\{ \frac{1}{n}\sum_{i=1}^n{\varphi (y_i,t)\phi_X(x_i)}-\frac{1}{n}\sum_{i=1}^n{H^0(w_i,t)\phi_X(x_i)} \right\} ,(A^*A)^{-1}g_s \right\rangle_{\cH_W}\\
	&=\left\langle A^*\left\{ \frac{1}{n}\sum_{i=1}^n{\left\{ U'( w_{i},x_i,y_i,t ) +r( x_i,t ) /n^{\alpha} \right\} \phi_X(x_i)} \right\} ,(A^*A)^{-1}g_s \right\rangle_{\cH_W}\\
	&=\left\langle A^*\left\{ \frac{1}{n}\sum_{i=1}^n{\left\{ U'( w_{i},x_i,y_i,t ) \right\} \phi_X(x_i)} \right\} ,(A^*A)^{-1}g_s \right\rangle_{\cH_W}\\
    &\qquad\qquad+\left\langle A^*\left\{ \frac{1}{n}\sum_{i=1}^n{r( x_i,t ) /n^{\alpha}\phi_X(x_i)} \right\} ,(A^*A)^{-1}g_s \right\rangle_{\cH_W}\\
    &=\frac{1}{n}\sum_{i=1}^{n}U'(w_{i},x_i,y_{i},t)\{A(A^{*}A)^{-1}g_{s}\}(x_{i}) \\
    & \qquad \qquad + \underbrace{\left\langle A^*\left\{ \frac{1}{n}\sum_{i=1}^n{r( x_i,t ) /n^{\alpha}\phi_X(x_i)} \right\} ,(A^*A)^{-1}g_s \right\rangle_{\cH_W}}_{(\star)},
\end{align*}
where we define $U'(W,X,Y,t)=\varphi(Y,t)-H^0(W,t)-r(X,t)/n^{\alpha}$. For the term $(\star)$, applying the reproducing property, we have
\begin{align*}
\sqrt{n} (\star)&=\frac{\sqrt{n}}{n^{\alpha}}\left\langle (A^*A)^{-1}A^*\left\{ \frac{1}{n}\sum_{i=1}^n{r( x_i,t )\phi_X(x_i)} \right\} ,g_s \right\rangle_{\cH_W}\\
&=\frac{\sqrt{n}}{n^{\alpha}}\left\langle (A^*A)^{-1}A^*\left\{ \frac{1}{n}\sum_{i=1}^n{r( x_i,t )\phi_X(x_i)} \right\} ,\mathbb{E}\{m(X,s)\phi(W)\} \right\rangle_{\cH_W}\\
&=\frac{\sqrt{n}}{n^{\alpha}}\mathbb{E}\left\langle (A^*A)^{-1}A^*\left\{ \frac{1}{n}\sum_{i=1}^n{r( x_i,t )\phi_X(x_i)} \right\} ,m(X,s)\phi(W) \right\rangle_{\cH_W}\\
    &=\frac{\sqrt{n}}{n^{\alpha}}\mathbb{E} \left[ (A^*A)^{-1}\left\{ \frac{1}{n}\sum_{i=1}^n{r(x_i,t) A^*\{ k_X( x_i,\cdot ) \}(W)} \right\} m(X,s) \right]\\
    &=\frac{\sqrt{n}}{n^{\alpha}}   \frac{1}{n}\sum_{i=1}^n r(x_i,t) \int  (A^*A)^{-1}A^*\{ k_X( x_i,\cdot ) \}(W)   m(X,s)dF(X,W) \\
     &=\frac{\sqrt{n}}{n^{\alpha}}   \frac{1}{n}\sum_{i=1}^n r(x_i,t) \left\langle k_X( x_i,\cdot ) ,   A(A^*A)^{-1}g_s\right\rangle_{\cH_X} \\
    &=\frac{\sqrt{n}}{n^{\alpha+1}}\sum_{i=1}^n{r(x_i,t) \{A(A^*A)^{-1}g_s\}(x_i)}.
\end{align*}

Combining all these results and the fact that $G_6:=H^0(W,t)-H^*(W,t)$, we have:
\begin{align}
    \sqrt{n}\sum_{i=1}^6{\mathbb{P}\{G_im(X,s)\}} & = \frac{1}{\sqrt{n}}\sum_{i=1}^{n}U'(w_{i},x_i,y_{i},t)\{A(A^{*}A)^{-1}g_{s}\}(x_{i})\nonumber\\
    &+\sqrt{n}\mathbb{P}[\{H^0(W,t)-H^*(W,t)\}m(X,s)]\nonumber\\   
&+\frac{\sqrt{n}}{n^{\alpha+1}}\sum_{i=1}^n{r(x_i,t) \{A(A^*A)^{-1}g_s\}(x_i)}+o_p\left(\frac{\sqrt{n}}{n^{\alpha}}\right)+o_p(1) \label{eq.G_1+...+G_6},
\end{align}
where the last inequality follows from condition \ref{assum.bandwidth}.


Besides, for the first term of $T_n(s,t)$ in \eqref{eq.Tnst decompose_H1}, we have
\begin{align*}
	&\mathbb{P}_n[ \{\varphi (Y,t)-H^*(W,t)\}m(X,s) ]\\ &=\mathbb{P}_n[ \{\varphi (Y,t)-H^0(W,t)+H^0(W,t)-H^*(W,t)\}m(X,s) ]\\
	&=\mathbb{P}_n[ \{U'(W,X,Y,t)+r(X,t)/n^{\alpha} +H^0(W,t)-H^*(W,t)\}m(X,s)].
\end{align*}
By \eqref{eq.operator_defination_adjoin_copy} and \eqref{eq.gs}, we have $g_s=A^*m(\cdot,s)$. Combining the above result with \eqref{eq.G_1+...+G_6}, we have
\begin{align*}
	T_n(s,t)&=\sqrt{n}\mathbb{P}_n\left( U'(W,X,Y,t)[ m(X,s)-\{A(A^*A)^{-1}g_s\}(X)] \right)\\
    &+\sqrt{n}\{\mathbb{P}_n-\mP\}[\{H^0(W,t)-H^*(W,t)\}m(X,s)]\\
    &-\sqrt{n}\{\mathbb{P}_n-\mP\}[\{\widehat{H}^\lambda(W,t)-H^*(W,t)\}m(X,s)]\\
    &+\frac{\sqrt{n}}{n^\alpha}\mP_n[\{r(X,t)m(X,s)-r(X,t)\{A(A^*A)^{-1}A^*m(\cdot,s)\}] +o_p\left( \frac{\sqrt{n}}{n^{\alpha}} \right)+o_p(1).
\end{align*}

We apply Lemma~\ref{ central limit theorem} to $\left\{ U'(w_i,x_i,y_i,t)\left[ \left\{m(x_i,s)-A(A^*A)^{-1}g_s\right\}(x_i)\right] \right\}_i$ to obtain that the first term of $T_n(s,t)$ converges weakly to a Gaussian process $\mG(s,t)$, where $\mathbb{G}(s,t)$ is defined in Theorem~\ref{theorem:null-hypothesis}.  To this end, we need to verify $U'(W,X,Y,t)[ \{A(A^*A)^{-1}g_s\}(X)+m(X,s)]$ is zero mean and 
\begin{equation}
\label{eq.var-infinity-H1}
    \mathbb{E} \left[ \left\| U'(w_i,x_i,y_i,t)[m(x_i,s)-\{A(A^*A)^{-1}g_s\}(x_i))] \right\|_{\cL^2\{\cT \times \cT, \mu \times \mu\}} \right] <\infty. 
\end{equation}
Notice that the zero-mean is met by $\mathbb{E} \{\varphi(y_i,t)-H^0(w_i,t)|x_i\}=r(x_i,t)/n^\alpha$ under $\mH_{1n}^\alpha$. Next, we calculate the second moment.
\begin{align*}
	& \mathbb{E} \left(\left\{\varphi(y_i,t)-H^0(w_i,t)-r(x_i,t)/n^\alpha \}\left[m(x_i,s)-\{A(A^*A)^{-1}g_s \right\}(x_i)\right] \right)^2\\
	& = \mathbb{E} (\{\varphi(y_i,t)-H^0(w_i,t)\}[m(x_i,s)-\{A(A^*A)^{-1}g_s\}(x_i)])^2\\
    & +n^{-2\alpha}\mathbb{E} \left( r(x_i,t)^2[m(x_i,s)\{A(A^*A)^{-1}g_s\}(x_i)]^2 \right)\\
	& -2n^{-\alpha}\mathbb{E} \left( \mathbb{E} \{\varphi(y_i,t)-H^0(w_i,t)|x_i\}r(x_i,t)[m(x_i,s)-\{A(A^*A)^{-1}g_s\}(x_i)]^2 \right)\\
	& \overset{(1)}{=} \underset{( \mathrm{I} )}{\underbrace{\mathbb{E} (\{\varphi(y_i,t)-H^0(w_i,t)\}[m(x_i,s)-\{A(A^*A)^{-1}g_s\}(x_i)])^2}}\\
    & -n^{-2\alpha}\underset{( \mathrm{II} )}{\underbrace{\mathbb{E} \left( r(x_i,t)^2[m(x_i,s)-\{A(A^*A)^{-1}g_s\}(x_i)]^2 \right) }},
\end{align*}
where (1) follows from $\mathbb{E} \{\varphi(y_i,t)-H^0(w_i,t)|x_i\}=r(x_i,t)/n^\alpha$ under $\mH_{1n}^\alpha$. Following \eqref{eq.var-infinity} in Theorem~\ref{theorem:null-hypothesis}, we have $(\mathrm{I})<\infty$. Besides, for the second term, we have $( \mathrm{II} ) \leq 2\mathbb{E} \{r(x_i,t)^4\}+2\mathbb{E} \left( [m(x_i,s)-\{A(A^*A)^{-1}g_s\}(x_i)]^4 \right) <\infty$ by inequality $a^2b^2\le(a^4+b^4)/2$ and condition~\ref{assum.var}. As long as the measure $\nu(T)$ is chosen to be finite, \eqref{eq.var-infinity-H1} holds. Besides, as $n \to \infty$, $\mathrm{(II)}$ vanishes. That means, the first term of $T_n(s,t)$ converges weakly to $\mathbb{G}(s,t)$ in $\mathcal{L}^{2}\{\cT \times \cT, \mu \times \mu\}$ by Lemma~\ref{ central limit theorem}. 

For the second term, we have
\begin{align*}
   & \sqrt{n}\{\mathbb{P}_n-\mP\}[\{H^0(W,t)-H^*(W,t)\}m(X,s)]\\
   =&\sqrt{n}\{\mathbb{P}_n-\mP\}[\{H^0(W,t)-(A^*A)^{-1}Ab_t\}m(X,s)]\\
    =&\sqrt{n}\{\mathbb{P}_n-\mP\}([H^0(W,t)-(A^*A)^{-1}A^*\{AH^0(W,t)+\ell(\cdot,t)/n^{\alpha}\}m(X,s)]\\
    =&-\frac{\sqrt{n}}{n^{\alpha}}\{\mathbb{P}_n-\mP\}[\{(A^*A)^{-1}A^*Ar(\cdot,t)\}m(X,s)]=-\frac{\sqrt{n}}{n^{\alpha}}\{\mathbb{P}_n-\mP\}\{r(X,t)m(X,s)\}.
\end{align*}
Since $\|r(X,t)\|_{\cH_X}<
\infty$ and the selected weight function $m$ satisfies $\| m(X,s)\|_{\mathcal{L}^2\{F(x)\}}<\infty$, we have
\begin{align}
\mE\{|r(X,t)m(X,s)|\}&\le \| r(X,t)\|_{\mathcal{L}^2\{F(x)\}}^{1/2}\cdot \| m(X,s)\|_{\mathcal{L}^2\{F(x)\}}^{1/2} \nonumber \\
	&\le \| r(X,t)\|_{\mathcal{H}_X}^{1/2}\cdot \| m(X,s)\|_{\mathcal{L}^2\{F(x)\}}^{1/2}<\infty, \label{eq.rm}
\end{align}
where the last inequality follows from \eqref{eq.h_norm} by condition~\ref{ass:kernel_characteristic}. According to the law of large numbers, we know that $\{\mathbb{P}_n-\mP\}\{r(X,t)m(X,s)\}=o_p(1)$. Thus, we can obtain that the second term is $o_p(\sqrt{n}/n^\alpha)$. For the third term, similar to the proof of Proposition \ref{prop.empirical_process}, we can obtain $\sqrt{n}\{\mathbb{P}_n-\mP\}[\{\widehat{H}^\lambda(W,t)-H^*(W,t)\}m(X,s)]=o_p(1)$. For the last term, we first prove 
$$\left|\mathbb{E}[r(X,t)\{m(X,s)-A(A^*A)^{-1}A^*m(\cdot,s)\}] \right|<\infty.
$$
In fact, it is sufficient to show that $|\mathbb{E} \left\{r(X,t) m(X,s)\right\}|<\infty$ (which holds by \eqref{eq.rm}), and 
\begin{align}
    \left|\mathbb{E} [(r(X,t)\cdot \{A(A^*A)^{-1}A^*m(\cdot,s)\}(X)]\right|<\infty, \label{eq.r_mX}
\end{align}
where \eqref{eq.r_mX} can be ensured by condition~\ref{assum.var}. Thus, by the law of large numbers, we know that $\mP_n[\{r(X,t)m(X,s)-r(X,t)\{A(A^*A)^{-1}A^*m(\cdot,s)\}]$ converges weakly to $\mu(s,t):=\mathbb{E}[r(X,t)\{m(X,s)-A(A^*A)^{-1}A^*m(\cdot,s)\}]$. Besides, similar to the proof of theorem \ref{theorem:null-hypothesis}, for any fixed $t$ and $T_n(s,t)\in \cL^2\{\cT, \mu \}$, we use the continuous mapping theorem (Theorem 1.3.6 of \cite{wellner2013weak}) to obtain that $\max_{t\in \cT} |T_n(s,t)|$ converges weakly to $\max_{t\in \cT} |\mG(s,t)|$. Combining these results, we have
\begin{align*}
     \max_{t\in \cT} |T_n(s,t)|&=\underset{( \star )}{\underbrace{O_p(1)}}+ \frac{\sqrt{n}}{n^{\alpha}}\left\{\max_{t\in \cT} |\mu(s,t)| +o_p(1)\right\} + o_p\left(\frac{\sqrt{n}}{n^\alpha}\right) +o_p(1) \\
    &\to \infty
\end{align*}
for almost all $s$ under $\mH_{1n}^{\alpha} (0<\alpha<1/2)$, where $(\star )$ follows from the Gaussian process.

\noindent\textbf{(2). The case of $\mH_{1n}^\alpha$ with $\alpha=1/2$.}

Following the proof in the case of $\mH_{1n}^\alpha$ with $0<\alpha<1/2$, we have
\begin{align*}	T_n(s,t)&=\sqrt{n}\mathbb{P}_n\left( U'(W,Y,t)\left[ m(X,s)-\{A(A^*A)^{-1}g_s\}(X) \right] \right)\\
    &+\frac{\sqrt{n}}{n^\alpha}\mP_n[r(X,t)\{m(X,s)-A(A^*A)^{-1}A^*m(\cdot,s)\}] +o_p\left( \frac{\sqrt{n}}{n^{\alpha}} \right)+o_p(1).
\end{align*}
Taking $\alpha=1/2$, we obtain
\begin{align*}	T_n(s,t)&=\sqrt{n}\mathbb{P}_n\left( U'(W,Y,t)\left[ m(X,s)-\{A(A^*A)^{-1}g_s\}(x_i) \right] \right)\\
    &+\mP_n[r(X,t)\{m(X,s)-A(A^*A)^{-1}A^*m(\cdot,s)\}]+o_p(1).
\end{align*}
By \eqref{eq.r_mX}, we have $\mP_n[r(X,t)\{m(X,s)-A(A^*A)^{-1}A^*m(\cdot,s)\}] \to \mu(s,t)$. By Slutsky's theorem, we have $T_n(s,t)$ converges weakly to $\mathbb{G}(s,t)+\mu(s,t)$ in $\mathcal{L}^{2}\{\cT \times \cT, \mu \times \mu\}$ under $\mathbb H^\alpha_{1n}$ with $\alpha=1/2$.

\noindent\textbf{(3). The case of $\mH_{1}^{\mathrm{fix}}$.}

We first analyze $\mathbb{P} \left[ \{\widehat{H}^{\lambda}(W,t)-H^*(W,t)\}m(X,s) \right]$. Note that
\begin{align*}
	\widehat{H}^{\lambda}(w,t)-H^*(w,t)&=(\lambda I+\widehat{A}^*\widehat{A})^{-1}\widehat{A}^*\widehat{b}_t-H^*(w,t)\\
	&=(\lambda I+\widehat{A}^*\widehat{A})^{-1}\widehat{A}^*( \widehat{b}_t-b_t) +\{(\lambda I+\widehat{A}^*\widehat{A})^{-1}\widehat{A}^*-(A^*A)^{-1}A^*\}b_{t}.
\end{align*}

For the first term, by the reproducing property and \eqref{eq.kernel_mean_embedding}, we have
\begin{align*}
	\mathbb{P} \left[ \{(\lambda I+\widehat{A}^*\widehat{A})^{-1}\widehat{A}^*( \widehat{b}_t-b_t ) \}m(X,s) \right] 
	&=\langle (\lambda I+\widehat{A}^*\widehat{A})^{-1}\widehat{A}^*( \widehat{b}_t-b_t ) ,\mathbb{E} \{ m(X,s) \phi_W(W) \} \rangle_{\mathcal{H}_W}\\
	&\le \| (\lambda I+\widehat{A}^*\widehat{A})^{-1}\widehat{A}^* \|_{\mathrm{op}}\cdot \| \widehat{b}_t-b_t \|_{\mathcal{H}_X}\cdot \| g_s\|_{\mathcal{H}_W}.
\end{align*}
By Lemma \ref{Lemm: Bounds on Compact Operator} (c), $\|(\lambda I + \widehat{A}^*\widehat{A} )^{-1}\widehat{A}^*\|_{\mathrm{op}}=O_p(\lambda^{-1/2})$. Moreover, Lemma~\ref{Lem: rates for Khat and fhat} gives $\|\widehat{b}_t-b_t\|_{\mathrm{op}}=O_p(n^{-1/2})$. Hence the rate is $O_p(1/\sqrt{n\lambda})$.

Next, consider
$$
\{(\lambda I+\widehat{A}^*\widehat{A})^{-1}\widehat{A}^*-(A^*A)^{-1}A^*\}b_t=S_1+S_2+S_3,
$$
with
\begin{align*}
	S_1:&=\{(\lambda I+\widehat{A}^*\widehat{A})^{-1}\widehat{A}^*-(\lambda I+A^*A)^{-1}\widehat{A}^*\}b_t,\\
	S_2:&=(\lambda I+A^*A)^{-1}(\widehat{A}^*-A^*)b_t,\\
	S_3:&=\big\{ (\lambda I+A^*A)^{-1}-(A^*A)^{-1} \big\} A^*b_t.
\end{align*}

\textbf{Analysis of $S_2$.} For $S_2$, by the reproducing property and \eqref{eq.kernel_mean_embedding}, we have
\begin{align*}
	\mathbb{P} \{S_2m(X,s) \}&=\langle (\lambda I+A^*A)^{-1}(\widehat{A}^*-A^*)b_t,\mathbb{E} \{ m(X,s) \phi_W(W) \} \rangle_{\mathcal{H}_W}\\
	&=\langle (\widehat{A}^*-A^*)b_t,(\lambda I+A^*A)^{-1}g_s \rangle_{\mathcal{H}_W}\\
	&\le \| \widehat{A}^*-A^* \|_{\mathrm{op}}\cdot\| b_t \|_{\mathcal{H}_X}\cdot \| (\lambda I+A^*A)^{-1}g_s\|_{\mathcal{H}_W}.
\end{align*}
By Lemma~\ref{Lemm: Bounds on Compact Operator} (d), $\| (\lambda I+A^*A)^{-1}g_s \|_{\mathcal{H}_W}=O_p \left\{\lambda^{\frac{\min(\eta,2)}{2}-1} \right\}$. Lemma~\ref{Lem: rates for Khat and fhat} implies $\| \widehat{A}^*-A^*\|_{\mathrm{op}}=O_p(n^{-1/2})$. Under condition~\ref{assum.source condition} (a) with $\eta \geq 2$, it follows that $\mathbb{P} \{S_2m(X,s) \}=O_p(n^{-1/2})=o_p(1)$.

\textbf{Analysis of $S_3$.} Since $A^*A\wt{g}_s=g_s$, we can obtain
\begin{align*}
	\mathbb{P} \{S_3m(X,s)\}&=\langle \{ (\lambda I+A^*A)^{-1}-(A^*A)^{-1} \} A^*b_{t},\mathbb{E} \{m(X,s)\phi_W(W)\}\rangle_{\mathcal{H}_W}\\
	&=\langle A^*b_{t},\{ (\lambda I+A^*A)^{-1}-(A^*A)^{-1} \} g_s\rangle_{\mathcal{H}_W}\\
	&=\langle A^*b_{t},(\lambda I+A^*A)^{-1}A^*A\wt{g}_s-\wt{g}_s\rangle_{\mathcal{H}_W}.
\end{align*}
By \eqref{eq.wt_g_s} in Lemma~\ref{lemma:Tn1}, $\|(\lambda I+A^*A)^{-1}A^*A\wt{g}_s-\wt{g}_s\|_{\cH_W}=o_p(1)$. Together with boundedness of $\|A^*\|_{\mathrm{op}}$ and $\|b_t\|_{\cH_X}$, this yields
$$
\mathbb{P} \{S_3m(X,s)\}=o_p(1).
$$

\textbf{Analysis of $S_1$.} Analogous steps yield
\begin{align*}
	\mathbb{P} \{S_1m(X,s) \}&=\langle \{(\lambda I+\widehat{A}^*\widehat{A})^{-1}\widehat{A}^*-(\lambda I+A^*A)^{-1}\widehat{A}^*\}b_t,\mathbb{E} \{ m(X,s) \phi_W(W) \} \rangle_{\mathcal{H}_W}\\
	&=\langle \{(\lambda I+\widehat{A}^*\widehat{A})^{-1}-(\lambda I+A^*A)^{-1}\}\widehat{A}^*b_t,g_s \rangle_{\mathcal{H}_W}\\
	&=\langle (\lambda I+A^*A)^{-1}\{A^*A-\widehat{A}^*\widehat{A}\}(\lambda I+\widehat{A}^*\widehat{A})^{-1}\widehat{A}^*b_t,g_s \rangle_{\mathcal{H}_W}\\
	&=\langle \{A^*A-\widehat{A}^*\widehat{A}\}(\lambda I+\widehat{A}^*\widehat{A})^{-1}\widehat{A}^*b_t,(\lambda I+A^*A)^{-1}g_s \rangle_{\mathcal{H}_W}\\
	&\le \| A^*A-\widehat{A}^*\widehat{A} \|_{\mathrm{op}}\cdot \| (\lambda I+\widehat{A}^*\widehat{A})^{-1}\widehat{A}^* \|_{\mathrm{op}} \cdot \| b_t \|_{\mathcal{H}_X}\cdot \| (\lambda I+A^*A)^{-1}g_s\|_{\mathcal{H}_W}.
\end{align*}
By Lemma~\ref{Lemm: Bounds on Compact Operator} (d), $\| (\lambda I+A^*A)^{-1}g_s \|_{\mathcal{H}_W}=O_p\left\{\lambda^{\frac{\min(\eta,2)}{2}-1}\right\}$. By Lemma \ref{Lemm: Bounds on Compact Operator} (c), we have $\|\widehat{A}(\lambda I +\widehat{A}^*\widehat{A} )^{-1}\|_{\mathrm{op}}=\|(\lambda I + \widehat{A}^*\widehat{A} )^{-1}\widehat{A}^*\|_{\mathrm{op}}=O_p(\lambda^{-1/2})$. By Lemma~\ref{Lem: rates for Khat g - fhat}, we have $ \| A^*A-\widehat{A}^*\widehat{A}\|_{\mathrm{op}}=O_p(n^{-1/2})$. Combining these results and according to conditions~\ref{assum.bandwidth},~\ref{assum.source condition} (a) with $\eta \ge 2$ and, we get $\mathbb{P} \{S_1m(X,s) \}=O_p(1/\sqrt{n\lambda})=o_p(1)$.

Combining these results, we get 
$$
\mathbb{P} \left[ \{\widehat{H}^{\lambda}(W,t)-H^*(W,t)\}m(X,s) \right]=o_p(1).
$$
Similarly, we can also have $\mathbb{P}_n \left|\{\widehat{H}^{\lambda}(W,t)-H^*(W,t)\}m(X,s) \right| = o_p(1)$, and therefore $(\mathbb{P}_n-\mathbb{P} )\left[\{\widehat{H}^{\lambda}(W,t)-H^*(W,t)\}m(X,s) \right] = o_p(1)$.

Besides, for the first term of $T_n(s,t)$, there exists $t$, we have  
\begin{align*}
	\mathbb{P}_n\left[ \{\varphi (Y,t)-H^*(W,t)\}m(X,s) \right] &=\mathbb{P}_n\left[ \{\varphi (Y,t)-H^0(W,t)+H^0(W,t)-H^*(W,t)\}m(X,s) \right].
\end{align*}
According to the definition of $\mH_1^{\mathrm{fix}}$, there exists $r(X,t)$ such that $\mathbb{E} \{ \varphi ( Y,t ) -H^0( W,t)|X\} = r(X,t)$, where $r(X,t)$ cannot be written as $\mE\{ H(W,t) -H^0( W,t)|X\}$.

We need to verify $\mE[|\{r(X,t) +H^0(W,t)-H^*(W,t)\}m(X,s)|]<\infty$. In fact, it is sufficient to show that $|\mathbb{E} \left\{r(X,t) m(X,s)\right\}|<\infty$ (which holds by \eqref{eq.rm}), and
$$
\begin{aligned}
	\mathbb{E} [ |\{H^0(W,t)-H^*(W,t)\}m(X,s)|] &=\mathbb{E} [| \{H^0(W,t)-H^*(W,t)\}\mathbb{E} \{m(X,s)|W\} |]\\
	&\lesssim C\cdot \mathbb{E} \{|H^0(W,t)-H^*(W,t)|\}\\
	&\lesssim  \|H^0(W,t)-H^*(W,t)\|_{H_W}<\infty,
\end{aligned}
$$
where the second inequality follows from condition~\ref{differentiabilty and integrability}, the third inequality follows from \eqref{eq.h_norm} by condition~\ref{ass:kernel_characteristic} and the last inequality follows from $H^0(W,t)-H^*(W,t)\in \cH_W$. Thus, by the law of large numbers, we know that $\mathbb{P}_n\{ \{\varphi (Y,t)-H^*(W,t)\}m(X,s) \}$ converges weakly to $\mE[ \{r(X,t) +H^0(W,t)-H^*(W,t)\}m(X,s)]$.

If $\mH_1^{\mathrm{fix}}$ holds, there exists $t$ such that $\mathbb{E}[ \{r(X,t) +H^0(W,t)-H^*(W,t)\}|X]\neq0$.
Otherwise, $r(X,t)=\mE[ \{H^0(W,t)-H^*(W,t)\}|X]$ for all $t$, which implies $\mE\{ \varphi ( Y,t)|X \}=\mE\{ H^*( W,t )|X \}$, contradicting $\mH_1^{\mathrm{fix}}$. Combining these results, we have:
$$
\lim_{n\rightarrow \infty} \max_{t\in \cT} |T_n(s,t)|=\lim_{n\rightarrow \infty} \sqrt{n}\{ \mathbb{E}\left[ \{r(X,t) +H^0(W,t)-H^*(W,t)\}|X\right] +o_p(1)\} =\infty.
$$ 
for almost all $s$ under $\mH_1^{\mathrm{fix}}$.
\end{proof}

\girdhypothesis*
\begin{proof}
\textbf{(i). $\widehat{\Delta}_{\varphi ,m}$ is weakly convergent to $\max_{t \in \cT}\int{| \mathbb{G}(s,t)|^2d\mu(s)}$.}

    Let $X_n(t):=\int_{\mathcal{T}}{\left\{ T_n(s,t) \right\}^2d\mu (s)}$. By the continuous mapping theorem, $X_n(t)$ weakly converges to $X(t)=\int_{\cT}{\left| \mathbb{G}(s,t) \right|^2d\mu(s)}$. Since
      integral of the Gaussian process $\mathbb{G}(s,t)$ still a Gaussian process with respect to $t$, we can obtain the variance $\int_{\mathcal{T}}{\left| \mathrm{Var}\{\eta (X,W,Y,s,t)\} \right|^2d\mu (s)}$. Besides, for the Gaussian process, $X(t)$ is continuous in probability if and only if its mean and variance are continuous following \cite{seeger2004gaussian}. Since $\varphi(y,t)$ is continuous with respect to $t$, the variance is continuous. Therefore, $X(t)$ is continuous in probability. Assume that we obtains the maximum value at $t_0$, \emph{i.e.} $\max_{t\in T}X(t) =X(t_0) $. Since the process $X(t)$ is continuous in probability, we have that, for any $\varepsilon > 0$, there exists $\delta$ such that as long as $|t - t_0| < \delta$, $P(|X(t) - X(t_0)| > \varepsilon/3) < \varepsilon$. 
      
      Since $\{t_1,...,t_K\}$ are evaluated at a grid of equidistant indices, for any $t_0 \in \cT$, we have $\lim_{K \to \infty} \min_{k} |t_0 - t_k| = 0$. That means, for any $\delta > 0$, there exists $K_0$, such that as long as $K > K_0$, there exists $t_k$ with $1 \leq k \leq K$, $|t_k - t_0| < \delta$. Further, for any finite $t_1,...,t_K$, denote $\cT_K:=\{k: X(t_k) = \max_{j \leq K} X(t_j)\}$ and set $\delta_0:= X(t_{k_0}) - X(t_{k_1})$, where $t_{k_0} \in \cT_K$ and $X(t_{k_1}):=\arg\max_{t_j \not\in \cT_K} X(t_j)$. For such $K$, there exists $N_K$, such that when $n > N_K$, $P\left[|X_n(t_k) - X(t_k)|>\min\{\varepsilon/3, \delta_0/2\}\right] < \frac{\varepsilon}{2K}$ for any $k \leq K$. Therefore, for any $\varepsilon > 0$, there exists $K > K_0$ such that $\min_{k \leq K} |t_k - t_0| < \delta$, and $N_K$ such that for any $n > N_K$, we have:
      \begin{align*}
      \allowdisplaybreaks
          & P(|\max_{k \leq K}X_n(t_k) - X(t_0)| > \varepsilon) \\
          & \leq P(|\max_{k \leq K} X_n(t_k) - X_n(t_{k_0})| > \varepsilon/3) \\
          & + P(|X_n(t_{k_0}) - X(t_{k_0})| > \varepsilon/3) + \mP(|X(t_{k_0}) - X(t_{0})| > \varepsilon/3) \\
          & \leq \varepsilon + P(|\max_{k \leq K} X_n(t_k) - X_n(t_{k_0})| > \varepsilon/3) + P(|X(t_{k_0}) - X(t_{0})| > \varepsilon/3).
      \end{align*}
For $P(|\max_{k \leq K} X_n(t_k) - X_n(t_{k_0})| > \varepsilon/3)$, we have:
\begingroup
\allowdisplaybreaks
\begin{align*}
    & P(|\max_{k \leq K} X_n(t_k) - X_n(t_{k_0})| > \varepsilon/3) \\
    & \leq P(\max_{k \leq K} X_n(t_k) \neq X_n(t_{k_0})) \\
    & \leq P\{\exists t_j \not\in \cT_K, \max_{k \leq K} X_n(t_k) = X_n(t_j)\} \\
    & \leq \sum_{j \leq K} \mP\{\max_{k \leq K} X_n(t_k) = X_n(t_j)\} \\
    & = \sum_{j \leq K} P\{X_n(t_j) - X(t_j) + X(t_j) - X(t_{k_0}) + X(t_{k_0}) - X_n(t_{k_0}) > 0\} \\
    & \leq \sum_{j \leq K} P\{X_n(t_j) - X(t_j) + X(t_{k_0}) - X_n(t_{k_0}) > \delta_0 \} \\
    & \leq \sum_{j \leq K} \left[ P\{|X_n(t_j) - X(t_j)| > \delta_0/2\} +  P\{|X_n(t_{k_0}) - X(t_{k_0})| > \delta_0/2\} \right] \\
    & \leq \sum_{j \leq K} \left( \frac{\varepsilon}{2K} + \frac{\varepsilon}{2K}\right) = \varepsilon. 
\end{align*}
\endgroup
Denote $k':=\arg\min_{k \leq K} |t_k - t_0|$. Then for $P(|X(t_{k_0}) - X(t_{0})| > \varepsilon/3)$, we have:
\begin{align*}
    P(|X(t_{k_0}) - X(t_{0})| > \varepsilon/3) & = P\{X(t_{0}) - X(t_{k_0})  > \varepsilon/3\} \\
    & = P\{X(t_{0}) - X(t_{k'}) + X(t_{k'}) - X(t_{k_0})  > \varepsilon/3\} \\
    & \leq P\{X(t_{0}) - X(t_{k'}) > \varepsilon/3\} \leq \varepsilon. 
\end{align*}
Combining these results together, we have $\lim_{n \to \infty} \lim_{K \to \infty} \max_{k \leq K} X_n(t_k) =_d \max_{t \in \cT} X(t)$. 


\textbf{(ii). Bootstrapped statistics \eqref{eq.bootstrap} is weakly convergent to the $\max_{t \in \cT}\int{| \mathbb{G}(s,t)|^2d\mu(s)}$.}

By Theorem 2.9.2 of \cite{wellner2013weak}, $\wh{T}^b_n(s,t) = \frac{1}{\sqrt{n}} \sum_{i=1}^n \omega^b_i \wh{U}(w_i,y_i,t)m(x_i,s)$ is weakly convergent to $\mG(s,t)$ conditional the original sample. Applying the continuous mapping theorem, $\int{|\wh{T}^b_n(s,t)|^2d\mu(s)}$ is weakly convergent to $\int{|\mathbb{G}(s,t)|^2d\mu(s)}$. Using the proof in \textbf{(i)} again, we can obtain that $\wh{\Delta}^b_{\varphi,m} = \max_{k\in [K]} \int_{\cT} |\wh{T}^b_n(s,t_k)|^2d\mu(s)$ is weakly convergent to $\max_{t \in \cT}\int{| \mathbb{G}(s,t)|^2d\mu(s)}$, conditional the original sample.
\end{proof}

   


\section{Existence of solutions with two proxies}
\label{appx.two}

\subsection{Proof of Theorem~\ref{thm.exist_solution_twoproxies}}
\label{appx.two_exist}

\solutionsexisttwoproxies*
\begin{proof}
Suppose $h(w,y)$ satisfies $p(y|x)=\int{h(w,y)p(w|x)dw}$. Under $\mathbb{H}_0$, we have $X\ind (W,Y)|U$, which leads to:
$$
\begin{aligned}
	\int{p(y|u) p(u|x) du}&=p(y|x) \\
    & =\int{h(w,y) p(w|x) dw}\\
	&=\int{\left\{ \int{h(w,y) p(w|u) dw} \right\} p(u|x) du}.
\end{aligned}
$$
By the completeness in condition~\ref{assum.complete_U_X_Zx} (1), $h(w,y)$ 
solves the following integral equation for all $(u,y)$. 
$$
p(y|u) =\int{h(w,y) p(w|u)dw}.
$$
Since $\mathbb H_0$ holds, we have $Y\ind (Z,X)|U$. Therefore, for any $(x,z)$, taking expectation over $p(u|z,x)$ on both sides, we have:
$$
p(y|z,x) = \int{p(y|u) p(u|z,x) du} =\int{\left\{ \int{h(w,y) p(w|u) dw} \right\} p(u|z,x) du} \overset{(1)}{=} \int h(w,y)p(w|z,x)dw,
$$
where ``(1)" is due to $W\ind (Z,X)|U$. That means, $h(w,y)$ solves the integral equation~\eqref{eq.bridge_Y_new}. 
To prove the contrary, \emph{i.e.}, the solution to \eqref{eq.bridge_Y}, is also the solution to \eqref{eq.bridge_Y_new}, by $W\ind (Z,X)|U$ and $Y\ind Z|(U,X)$, we have
$$
\begin{aligned}
	\int{p(y|u,x) p(u|z,x) du}&=p(y|z,x)\\
	&=\int{h(w,y) p(w|z,x) dw}\\
	&=\int{\left\{ \int{h(w,y) p(w|u) dw} \right\} p(u|z,x) du}.
\end{aligned}
$$
Since the above equation holds for all $x$, it in particular holds for any fixed $x$, by the completeness condition in condition~\ref{assum.complete_U_X_Zx} (2), we obtain
$$
p(y|u,x)=\int{h(w,y) p(w|u)dw}.
$$
Since the right side of the equation is independent of $x$, we get $p(y|u,x)=p(y|u)$, and thus $\mH_0$ holds. 
\end{proof}

\subsection{Discussions of causal inference and causal discovery}
\label{appx.example_intervention}

In this section, we explore the distinction between causal discovery and causal inference, focusing on why the causal relation cannot be identified solely through the causal effect. We begin by presenting a counter-example that demonstrates that even when the intervention distribution for each $x$ is identical, the independence $X \ind Y | U$ may still fail to hold. Following this, we provide an in-depth discussion of the differences between causal inference and causal discovery.

We first introduce the notations. For any discrete variables $X,Y,Z$ with $k$ categories, we denote $P(X):=\left\{P(x_1),...,P(x_k)\right\}^\top$, 
 $P(Y|X) = \{P(y_i|x_j)\}_{i,j}$, and $P(Y=y|X,Z)=\{P(y|x_i,z_j)\}_{i,j}$. 

\begin{example}
   Suppose $U,X,Y$ are binary, and the causal diagram over $(U,X,Y)$ is $U \to X, U \to Y, X \to Y$. The conditional probability matrices $P(U), P(X|U), P(Y|X,U)$ are given by:
$$
P(U)= \begin{pmatrix}
0.4\\0.6
\end{pmatrix}, \ 
P(X|U)= 
\begin{pmatrix}
0.2&0.4\\
0.8&0.6
\end{pmatrix}, \ 
P(Y=0|X,U)=
\begin{pmatrix}
0.5&0.1\\
0.2&0.3
\end{pmatrix}.
$$ 
By the definition, we know $X \not\ind Y|U$. However, the intervention distribution is the same, \emph{i.e.}, $P\{y|do(X=0)\}=P\{y|do(X=1)\}$ for any $y$.
\end{example}

\begin{proof}
   According to the backdoor formula, we have
$$
P\{Y=y|do(X=x)\} =\sum_{u\in\{ 0,1\}}{P(Y=y|U=u,X=x) \mP(U=u)}.
$$
Plugging $P(Y=0|X,U)$ into the formula, we have:
$$
\begin{aligned}
	P\{Y=0|do(X=0)\} &=0.5\times 0.4+0.1\times 0.6=0.26\\
	P\{ Y=0|do(X=1) \} &=0.2\times 0.4+0.3\times 0.6=0.26\\
	P\{ Y=1|do(X=0) \} &=0.5\times 0.4+0.9\times 0.6=0.74\\
    P\{ Y=1|do(X=1) \} &=0.8\times 0.4+0.7\times 0.6=0.74,
\end{aligned}
$$
which implies intervention distributions are equal. However, through data generation, we know $X\not\ind Y|U$. 
\end{proof}

Next, we will verify that in this example, 
\begin{equation*}
    P(Y=y|X=x)\ne \sum_u{P(Y=y|U=u)P(U=u|X=x)},
\end{equation*}
which implies the example contradicts our assumption that there is no solution in  \eqref{eq.bridge_Y} under $\mH_1$. To this end, we need to obtain probability matrix $P(Y|X),P(Y|U)$, and $P(U|X)$. First, by $P(U)$ and $P(X|U)$, we can get the probability matrix $P(X)$ and $P(U|X)$.
$$
P(X) =P(X|U)P(U)=\begin{pmatrix}
0.2&0.4\\
0.8&0.6
\end{pmatrix}\begin{pmatrix}
0.4\\0.6
\end{pmatrix}=\begin{pmatrix}
0.32\\0.68
\end{pmatrix}, \mathbb{P} (U|X) = \begin{pmatrix}
0.25&8/17\\
0.75&9/17
\end{pmatrix}.
$$

Besides, we calculate the probability of $P(y|x)$ for any $y,x$. According to the Bayesian formula, we have
$$
\begin{aligned}
	P(Y=y|X=x) &=\sum_u{P( Y=y|X=x,U=u)  P( U=u|X=x)}\\
	&=\sum_u{P( Y=y|X=x,U=u) \frac{P( X=x|U=u) P ( U=u )}{P( X=x)}}.
\end{aligned}
$$
Therefore, we have
$$
P( Y|X) =\begin{pmatrix}
0.2&43/170\\
0.8&127/170
\end{pmatrix}.
$$
According to the Bayesian formula, we have
$$
P( Y=y|U=u) =\sum_x{P( Y=y|X=x,U=u) P( X=x|U=u)}.
$$
Therefore, we have
$$
P( Y|U)=\begin{pmatrix}
0.26&0.22\\
0.74&0.78
\end{pmatrix}.
$$
Thus, we can verify
\begin{align*}
\allowdisplaybreaks
    P( Y=0|X=0) =0.2 
    & \ne 0.23=0.26\times 0.25+0.22\times 0.75=\sum_u{P(Y=0|U=u)P(U=u|X=0)}\\
	P( Y=0|X=1 ) =\frac{43}{170}&\ne \frac{203}{850}=0.26\times \frac{8}{17}+0.22\times \frac{9}{17}=\sum_u{P(Y=0|U=u)P(U=u|X=1)}\\
	P( Y=1|X=0 ) =0.8&\ne 0.64=0.22\times 0.25+0.78\times 0.75=\sum_u{P(Y=1|U=u)P(U=u|X=0)}\\
	P( Y=1|X=1 ) =\frac{127}{170}&\ne \frac{439}{850}=0.22\times \frac{8}{17}+0.78\times \frac{9}{17}=\sum_u{P(Y=1|U=u)P(U=u|X=1)}.
\end{align*}

\textbf{More discussions about causal discovery and causal inference.} Causal inference and causal discovery address fundamentally different problems \citep{guo2020survey}. Causal inference focuses on quantifying the effects of interventions, often requiring strong assumptions and additional information to ensure accurate estimation. In contrast, causal discovery aims to uncover the underlying causal structure, emphasizing the identification of causal relationships rather than their magnitudes. 

It may not be feasible to infer whether the causal relationship exists from the causal effect. One key reason is that the inference is often complicated by noise in the estimates, making it hard to determine whether a nonzero effect arises from an actual causal relationship or random noise perturbing the estimation. Even if we can estimate a confidence interval for the effect at each treatment value \citep{robins1988confidence,robins2003uniform,calonico2018effect,colangelo2020double}, there are no valid statistics to determine whether the relation exists. Moreover, as shown in the previous example, a causal effect of zero does not necessarily imply the existence of the causal relation. Additionally, estimating causal effects often requires satisfying other conditions. For example, proximal causal inference depends on additional completeness assumptions \citep{miao2018identifying, mastouri2021proximal}. In our scenario, such conditions are assumed on $Z|X,W$ (\emph{i.e.}, for any square-integrable function $g$, $\mE\{g(z)|x,w\} = 0$ almost surely if and only if $g(z) = 0$ almost surely) and $\{X,W\}|\{X,Z\}$ \citep{mastouri2021proximal}.

\subsection{Proof of Proposition~\ref{prop.fail} and Example~\ref{example:linear_gaussian_two_proxy}}
\label{appx.two_fail}

We first prove Proposition~\ref{prop.fail}. 

\begin{proposition}
Suppose that $X, Y, U, W$ satisfy the linear Gaussian model, \emph{i.e.} $U=\varepsilon_U,X =\alpha_U U + \alpha_0+\varepsilon_X,W =\beta_U U +\beta_0 + \varepsilon_W, Y=\gamma_U U +\gamma_X X+\gamma_W W +\gamma_0 +\varepsilon_Y$, where $\varepsilon_X, \varepsilon_W, \varepsilon_Y, \varepsilon_U$ are standard normal. When $\gamma_W = 0$, as long as $|\gamma_X|>\frac{|B|+\sqrt{\Delta}}{2A}$, where $A=1+\frac{1}{\alpha_{U}^{2}}+\frac{2}{\beta_{U}^{2}}+\frac{1}{\alpha_{U}^{2}\beta_{U}^{2}}+\frac{\alpha_{U}^{2}}{\beta_{U}^{2}}$, $B=\frac{2\gamma_U}{\alpha_U}+\frac{2\gamma_U}{\alpha_U\beta_{U}^{2}}+\frac{2\alpha_U\gamma_U}{\beta_{U}^{2}}$ and $\Delta=\frac{4\left( 1+\alpha_{U}^{2}+\beta_{U}^{2} \right) \left( 1+\alpha_{U}^{2}+\gamma_{U}^{2} \right)}{\alpha_{U}^{2}\beta_{U}^{2}}$, the integration equation \eqref{eq.bridge_Y} has no solution. When $\gamma_W \ne 0$, as long as $|\gamma_W|>\frac{|C|+|B||\gamma_X|+A\gamma_X^2}{2|D|}$, where $C=1-\gamma_U^2/\beta_U^2$ and $D=\frac{\gamma_X}{\alpha_U\beta_U}+\frac{\alpha_U}{\beta_U}\gamma_X+\frac{\beta_U}{\alpha_U}\gamma_X+\frac{\gamma_U}{\beta_U}$, \eqref{eq.bridge_Y} has a solution. 
\end{proposition}
\begin{proof}
Based on the data generation structure, we can obtain the joint distribution
$$
\left( \begin{array}{c}
	U\\
	X\\
	W\\
	Y\\
\end{array} \right) \sim \cN\left\{ \left( \begin{array}{c}
	0\\
	\alpha_0\\
	\beta_0\\
	\gamma_0\\
\end{array} \right) ,\left( \begin{matrix}
	1&		\alpha_U&		\beta_{U}&		\mathrm{Cov}(U,Y)\\
	\alpha_U&		1+\alpha_U^{2}&		\alpha_U\beta_{U}&		\mathrm{Cov}(X,Y)\\
	\beta_{U}&		\alpha_U\beta_{U}&		1+\beta_{U}^{2}&		\mathrm{Cov}(W,Y)\\
	\mathrm{Cov}(U,Y)&		\mathrm{Cov}(X,Y)&		\mathrm{Cov}(W,Y)&		\mathrm{Var}(Y)
\end{matrix} \right) \right\},
$$
where covariance $\mathrm{Cov}(U,Y),\mathrm{Cov}(X,Y),\mathrm{Cov}(W,Y)$ and $\mathrm{Var}(Y)$ are respectively
$$
\left\{ \begin{aligned}
	\mathrm{Cov}(U,Y)&=\gamma_{U}+\gamma_X\alpha_U+\gamma_W\beta_{U}\\
	\mathrm{Cov}(X,Y)&=\alpha_U\left( \gamma_{U}+\gamma_W\beta_{U}+\gamma_X\alpha_U \right) +\gamma_X\\
	\mathrm{Cov}(W,Y)&=\beta_{U}\left( \gamma_{U}+\alpha_U\gamma_X+\gamma_W\beta_{U} \right) +\gamma_W\\
	\mathrm{Var}(Y)&=(\gamma_{U}+\gamma_X\alpha_U+\gamma_W\beta_{U})^2+\gamma_X^{2}+\gamma_W^{2}+1.
\end{aligned} \right.
$$
We can therefore derive the explicit form of the conditional distributions $p(w|x)$ and $p(y|x)$:
$$
\begin{aligned}
	W|X=x&\sim \cN\left\{ \mu_W+\frac{\mathrm{Cov}(W,X)}{\mathrm{Var}(X)}(x-\mu_{X}),\mathrm{Var}(W)\left( 1-\frac{\mathrm{Cov}^2(W,X)}{\mathrm{Var}(X)\cdot \mathrm{Var}(W)} \right) \right\}\\
	&\sim \cN\left\{ \mu_{X}^{W|X}x+\mu_{0}^{W|X},\sigma_{W|X}^{2} \right\} 
\end{aligned}
$$
$$
\begin{aligned}
    Y|X=x&\sim \cN\left\{ \mu_Y+\frac{\mathrm{Cov}(Y,X)}{\mathrm{Var}(X)}(x-\mu_{X}),\mathrm{Var}(Y)\left( 1-\frac{\mathrm{Cov}^2(Y,X)}{\mathrm{Var}(X)\cdot \mathrm{Var}(Y)} \right) \right\}\nonumber \\
	&\sim \cN\left\{ \mu_{X}^{Y|X}x+\mu_{0}^{Y|X},\sigma_{Y|X}^{2} \right\},
\end{aligned}
$$
where $\mu_{X}^{W|X},\mu_{0}^{W|X},\sigma_{W|X}^{2}, \mu_{X}^{Y|X}, \mu_{0}^{Y|X}$ and $\sigma_{Y|X}^{2}$ are defined as follows
$$
\begin{cases}
	\mu_{X}^{W|X}=\frac{\alpha_U\beta_{U}}{1+\alpha_U^{2}}\\
	\mu_{0}^{W|X}=\beta_0-\frac{\alpha_0\alpha_U\beta_{U}}{1+\alpha_U^{2}}\\
	\sigma_{W|X}^{2}=1+\beta_{U}^{2}-\frac{(\alpha_U\beta_{U})^2}{1+\alpha_U^{2}}\\
	\mu_{X}^{Y|X}=\frac{\alpha_U\left( \gamma_{U}+\gamma_W\beta_{U}+\gamma_X\alpha_U \right) +\gamma_X}{1+\alpha_U^{2}}\\
	\mu_{0}^{Y|X}=\gamma_0-\frac{\alpha_0\alpha_U\left( \gamma_{U}+\gamma_W\beta_{U}+\gamma_X\alpha_U \right) +\alpha_0\gamma_X}{1+\alpha_U^{2}}\\
	\sigma_{Y|X}^{2}=(\gamma_{U}+\gamma_X\alpha_U+\gamma_W\beta_{U})^2+\gamma_X^{2}+\gamma_W^{2}+1-\frac{(\alpha_U\left( \gamma_{U}+\gamma_W\beta_{U}+\gamma_X\alpha_U \right) +\gamma_X)^2}{1+\alpha_U^{2}}.
\end{cases}
$$
By applying Lemma~\ref{lemma.miao-example}, the solution of \eqref{eq.bridge_Y} is given by:
$$
h(w,y)=\frac{1}{\sqrt{\sigma_{Y|X}^{2}-\left(\mu_{X}^{Y|X}\right)^2\sigma_{W|X}^{2}/\left(\mu_{X}^{W|X}\right)^2}}\phi \left( \frac{y-\left( \mu_{0}^{Y|X}-\mu_{X}^{Y|X}\mu_{0}^{W|X}/\mu_{X}^{W|X} \right) -\mu_{X}^{Y|X}/\mu_{X}^{W|X}w}{\sqrt{\sigma_{Y|X}^{2}-\left(\mu_{X}^{Y|X}\right)^2\sigma_{W|X}^{2}/\left(\mu_{X}^{W|X}\right)^2}} \right),
$$
where $\phi$ is the standard normal distribution's probability density function (pdf).

For $h(w, y)$ to be meaningful, we need $\sigma_{Y|X}^{2}-\left(\mu_{X}^{Y|X}\right)^2\sigma_{W|X}^{2}/\left(\mu_{X}^{W|X}\right)^2>0$, which implies
\begin{equation}\label{eq. solution_h_exist}
\begin{aligned}
	1-\frac{\gamma_{U}^{2}}{\beta_{U}^{2}}-\left( \frac{2\gamma_{U}}{\alpha_U}+\frac{2\gamma_{U}}{\alpha_U\beta_{U}^{2}}+\frac{2\alpha_U\gamma_{U}}{\beta_{U}^{2}} \right) \gamma_X-\left( 1+\frac{1}{\alpha_U^{2}}+\frac{2}{\beta_{U}^{2}}+\frac{1}{\alpha_U^{2}\beta_{U}^{2}}+\frac{\alpha_U^{2}}{\beta_{U}^{2}} \right) \gamma_X^{2}\\
	-2\left( \frac{1}{\alpha_U\beta_{U}}+\frac{\alpha_U}{\beta_{U}}+\frac{\beta_{U}}{\alpha_U} \right) \gamma_X\gamma_W-2\frac{\gamma_{U}}{\beta_{U}}\gamma_W&>0.
\end{aligned}
\end{equation}

We discuss the following two cases: \textbf{(i)} $X \to Y$ ($\gamma_X \neq 0$) and $W \not\to Y$ ($\gamma_W = 0$); \textbf{(ii)} $X \to Y$ ($\gamma_X \neq 0$) and $W \not\to Y$ ($\gamma_W = 0$).

\textbf{(i). $\gamma_X\ne0,\gamma_W=0$.}

We first rewrite \eqref{eq. solution_h_exist} as:
$$
\underset{C}{\underbrace{1-\frac{\gamma_{U}^{2}}{\beta_{U}^{2}}}}\underset{B}{\underbrace{-\left( \frac{2\gamma_U}{\alpha_U}+\frac{2\gamma_U}{\alpha_U\beta_{U}^{2}}+\frac{2\alpha_U\gamma_U}{\beta_{U}^{2}} \right) }}\gamma_X\underset{A}{\underbrace{-\left( 1+\frac{1}{\alpha_{U}^{2}}+\frac{2}{\beta_{U}^{2}}+\frac{1}{\alpha_{U}^{2}\beta_{U}^{2}}+\frac{\alpha_{U}^{2}}{\beta_{U}^{2}} \right) }}\gamma_{X}^{2}>0.
$$

Noting that this is a quadratic function, we can get its discriminant 
$$
\Delta:=B^2-4AC=\frac{4\left( 1+\alpha_{U}^{2}+\beta_{U}^{2} \right) \left( 1+\alpha_{U}^{2}+\gamma_{U}^{2} \right)}{\alpha_{U}^{2}\beta_{U}^{2}}>0.
$$
Besides, we can find $ 1+\frac{1}{\alpha_{U}^{2}}+\frac{2}{\beta_{U}^{2}}+\frac{1}{\alpha_{U}^{2}\beta_{U}^{2}}+\frac{\alpha_{U}^{2}}{\beta_{U}^{2}}>0$. Therefore, this is a quadratic function whose discriminant is always positive and opens downward. When $\gamma_X$ satisfies $\frac{-B+\sqrt{\Delta}}{2A}<\gamma_X<\frac{-B-\sqrt{\Delta}}{2A}$, \eqref{eq.bridge_Y} will have a solution. When $\gamma_X\geq\frac{-B-\sqrt{\Delta}}{2A}$ or $\gamma_X\leq\frac{-B+\sqrt{\Delta}}{2A}$, \eqref{eq.bridge_Y} will have no solution. 

Without loss of generality, we consider the case where $\alpha_U$ and $\gamma_U$ have the same sign. First, we can find $ B=-(\frac{2\gamma_U}{\alpha_U}+\frac{2\gamma_U}{\alpha_U\beta_{U}^{2}}+\frac{2\alpha_U\gamma_U}{\beta_{U}^{2}})<0$ since $\beta_U^2>0$. Thus, we have $|-B-\sqrt{\Delta}|<|-B+\sqrt{\Delta}|$. Thus, when $|\gamma_X|>\frac{-B+\sqrt{\Delta}}{2A}$,  \eqref{eq.bridge_Y} will have no solution. If $\alpha_U$ and $\gamma_U$ have the different sign, we have $ B=-(\frac{2\gamma_U}{\alpha_U}+\frac{2\gamma_U}{\alpha_U\beta_{U}^{2}}+\frac{2\alpha_U\gamma_U}{\beta_{U}^{2}})>0$ since $\beta_U^2>0$. Thus, we have $|-B-\sqrt{\Delta}|>|-B+\sqrt{\Delta}|$. Thus, when $|\gamma_X|>\frac{-B-\sqrt{\Delta}}{2A}$,  \eqref{eq.bridge_Y} will have no solution.  

Combining the two cases, we can obtain that as long as $|\gamma_X|>\frac{|B|+\sqrt{\Delta}}{2A}$, the integration equation \eqref{eq.bridge_Y} has no solution.

\textbf{(ii). $\gamma_X\ne0,\gamma_W\ne0$.} 

We consider the case $|\gamma_X|>\frac{-B+\sqrt{\Delta}}{2A}$ under $\alpha_U\gamma_U>0$, since  \eqref{eq.bridge_Y} have no solution. We can rewrite  \eqref{eq. solution_h_exist} as
\begin{align*}
    & 2\left( \frac{\gamma_X}{\alpha_U\beta_U}+\frac{\alpha_U}{\beta_U}\gamma_X+\frac{\beta_U}{\alpha_U}\gamma_X+\frac{\gamma_U}{\beta_U} \right) \gamma_W \\
    & <1-\frac{\gamma_{U}^{2}}{\beta_{U}^{2}}-\left( \frac{2\gamma_U}{\alpha_U}+\frac{2\gamma_U}{\alpha_U\beta_{U}^{2}}+\frac{2\alpha_U\gamma_U}{\beta_{U}^{2}} \right) \gamma_X-\left( 1+\frac{1}{\alpha_{U}^{2}}+\frac{2}{\beta_{U}^{2}}+\frac{1}{\alpha_{U}^{2}\beta_{U}^{2}}+\frac{\alpha_{U}^{2}}{\beta_{U}^{2}} \right) \gamma_{X}^{2}.
\end{align*}
Thus, if $\frac{\gamma_X}{\alpha_U\beta_U}+\frac{\alpha_U}{\beta_U}\gamma_X+\frac{\beta_U}{\alpha_U}\gamma_X+\frac{\gamma_U}{\beta_U}<0$, we can obtain that \eqref{eq.bridge_Y} may still have a solution, as long as
$$
\gamma_W>\frac{1-\frac{\gamma_{U}^{2}}{\beta_{U}^{2}}-\left( \frac{2\gamma_U}{\alpha_U}+\frac{2\gamma_U}{\alpha_U\beta_{U}^{2}}+\frac{2\alpha_U\gamma_U}{\beta_{U}^{2}} \right) \gamma_X-\left( 1+\frac{1}{\alpha_{U}^{2}}+\frac{2}{\beta_{U}^{2}}+\frac{1}{\alpha_{U}^{2}\beta_{U}^{2}}+\frac{\alpha_{U}^{2}}{\beta_{U}^{2}} \right) \gamma_{X}^{2}}{2\left( \frac{\gamma_X}{\alpha_U\beta_U}+\frac{\alpha_U}{\beta_U}\gamma_X+\frac{\beta_U}{\alpha_U}\gamma_X+\frac{\gamma_U}{\beta_U} \right)}.
$$
We find that if $\frac{\gamma_X}{\alpha_U\beta_U}+\frac{\alpha_U}{\beta_U}\gamma_X+\frac{\beta_U}{\alpha_U}\gamma_X+\frac{\gamma_U}{\beta_U}<0$, the right-hand side of the above inequality is positive. That means, as long as $|\gamma_W|$ is sufficiently large, the solution to  \eqref{eq.bridge_Y} will still exist when $|\gamma_X|>\frac{-B+\sqrt{\Delta}}{2A}$.

If $\frac{\gamma_X}{\alpha_U\beta_U}+\frac{\alpha_U}{\beta_U}\gamma_X+\frac{\beta_U}{\alpha_U}\gamma_X+\frac{\gamma_U}{\beta_U}>0$, we can obtain \eqref{eq.bridge_Y} may still have a solution, as long as
$$
\gamma_W<\frac{1-\frac{\gamma_{U}^{2}}{\beta_{U}^{2}}-\left( \frac{2\gamma_U}{\alpha_U}+\frac{2\gamma_U}{\alpha_U\beta_{U}^{2}}+\frac{2\alpha_U\gamma_U}{\beta_{U}^{2}} \right) \gamma_X-\left( 1+\frac{1}{\alpha_{U}^{2}}+\frac{2}{\beta_{U}^{2}}+\frac{1}{\alpha_{U}^{2}\beta_{U}^{2}}+\frac{\alpha_{U}^{2}}{\beta_{U}^{2}} \right) \gamma_{X}^{2}}{2\left( \frac{\gamma_X}{\alpha_U\beta_U}+\frac{\alpha_U}{\beta_U}\gamma_X+\frac{\beta_U}{\alpha_U}\gamma_X+\frac{\gamma_U}{\beta_U} \right)}.
$$
We find that if $\frac{\gamma_X}{\alpha_U\beta_U}+\frac{\alpha_U}{\beta_U}\gamma_X+\frac{\beta_U}{\alpha_U}\gamma_X+\frac{\gamma_U}{\beta_U}>0$, the right-hand side is negative. That also means,  as long as $|\gamma_W|$ is sufficiently large, the solution to \eqref{eq.bridge_Y} will still exist when $|\gamma_X|>\frac{-B+\sqrt{\Delta}}{2A}$.

If $\alpha_U\gamma_U<0$, the proof is similar. Besides, in the above cases, as long as $|\gamma_W| >\frac{|C|+|B||\gamma_X|+A\gamma_X^2}{2|D|}$ with $D:=\frac{\gamma_X}{\alpha_U\beta_U}+\frac{\alpha_U}{\beta_U}\gamma_X+\frac{\beta_U}{\alpha_U}\gamma_X+\frac{\gamma_U}{\beta_U}$,  \eqref{eq.bridge_Y} has a solution. 
\end{proof}
\begin{remark}
If $\gamma_X=\gamma_W=0$,  \eqref{eq. solution_h_exist} will become $1-\frac{\gamma_{U}^{2}}{\beta_{U}^{2}}>0$. This means that if the strength between $W-U$ is greater than the confounder strength between $W-U$, \eqref{eq.bridge_Y} will have a solution under $\mH_0$. Otherwise, similar to the case when $\gamma_X \neq 0$, if the effect of $W$ on $Y$ is strong enough (\emph{i.e.}, $|\gamma_W|$), the solution exists again. Specifically, if $\gamma_X=0,\gamma_W\ne0$,  \eqref{eq. solution_h_exist} will become $1-\frac{\gamma_{U}^{2}}{\beta_{U}^{2}}-2\frac{\gamma_U}{\beta_U}\gamma_W>0$. If $-2\frac{\gamma_U}{\beta_U}\gamma_W$ is large enough, \eqref{eq.bridge_Y} still have a solution. If we $\gamma_U/\beta_U>0$, we need $\gamma_W$ to be as negative as possible; if$\gamma_U/\beta_U<0$, we need $\gamma_W$ to be as positive as possible.
\end{remark}

\begin{proposition}
Suppose that $X, Y, U,U_1, W$ satisfy the linear Gaussian model, \emph{i.e.} $U=\varepsilon_U,X =\alpha_U U + \alpha_0+\varepsilon_X,	U_1=\mu_0+\varepsilon_{U_1},W =\beta_U U +\beta_{U_1}U_1+\beta_0 + \varepsilon_W, Y=\gamma_U U+\gamma_{U_1}U_1 +\gamma_X X+\gamma_W W +\gamma_0 +\varepsilon_Y$, where $\varepsilon_X, \varepsilon_W, \varepsilon_Y, \varepsilon_U,\varepsilon_{U_1}$ are standard normal. When $\gamma_{U_1} = 0$, as long as $|\gamma_X|>\frac{|B|+\sqrt{\Delta}}{2A}$, where $A= 1+\frac{2}{\beta_{U}^{2}}+\frac{1}{\alpha_{U}^{2}}+\frac{1}{\beta_{U}^{2}\alpha_{U}^{2}}+\frac{\alpha_{U}^{2}}{\beta_{U}^{2}}+\frac{2\beta_{U_1}^{2}}{\beta_{U}^{2}}+\frac{\beta_{U_1}^{2}}{\beta_{U}^{2}\alpha_{U}^{2}}+\frac{\alpha_{U}^{2}\beta_{U_1}^{2}}{\beta_{U}^{2}}$, $B= \frac{2\gamma_U}{\alpha_U}+\frac{2\gamma_U}{\beta_{U}^{2}\alpha_U}+\frac{2\alpha_U\gamma_U}{\beta_{U}^{2}}+\frac{2\beta_{U_1}^{2}\gamma_U}{\beta_{U}^{2}\alpha_U}+\frac{2\alpha_U\beta_{U_1}^{2}\gamma_U}{\beta_{U}^{2}}$ and $\Delta=\frac{4\left( 1+\beta_{U_1}^{2}+\beta_{U}^{2}+\alpha_{U}^{2}\left( 1+\beta_{U_1}^{2} \right) \right) \left( 1+\alpha_{U}^{2}+\gamma_{U}^{2} \right)}{\alpha_{U}^{2}\beta_{U}^{2}}
$, the integration equation \eqref{eq.bridge_Y} has no solution. Further, when $\gamma_{U_1} \ne 0$, if $|\gamma_W| >|C|+|B||\gamma_X|+A\gamma_X^2 $, where $C=1-\gamma_{U}^{2}/\beta_{U}^{2}-\beta_{U_1}^{2}\gamma_{U}^{2}/\beta_{U}^{2}$, \eqref{eq.bridge_Y} has a solution. 
\end{proposition}
\begin{proof}
Based on the data generation structure, we can obtain joint distribution
$$
\left( U,X,U_1,W,Y \right)^\top \sim \cN\left\{ \bm{\mu}, \bm{\Sigma} \right\}, 
$$
where $\bm{\mu}=(0,\alpha_0,\mu_0,\beta_0+\beta_{U_1}\mu_0,\gamma_0+\gamma_X\alpha_0+\gamma_{U_1}\mu_0)^{\top}$ and 
$$
\bm{\Sigma}=\begin{pmatrix}
	1&		\alpha_U&		0&		\beta_U&		\mathrm{Cov}(U,Y)\\
	\alpha_U&		1+\alpha_{U}^{2}&		0&		\alpha_U\beta_U&		\mathrm{Cov}(X,Y)\\
	0&		0&		1&		\beta_{U_1}&		\mathrm{Cov}(U_1,Y)\\
	\beta_U&		\alpha_U\beta_U&		 \beta_{U_1}&		\beta_U^2+\beta_{U_1}^{2}+1&		\mathrm{Cov}(W,Y)\\
	\mathrm{Cov}(U,Y)&		\mathrm{Cov}(X,Y)&		\mathrm{Cov}(U_1,Y)&		\mathrm{Cov}(W,Y)&		\mathrm{Var}(Y)
\end{pmatrix}. 
$$
The covariance $\mathrm{Cov}(U,Y),\mathrm{Cov}(X,Y),\mathrm{Cov}(U_1,Y),\mathrm{Cov}(W,Y)$ and $\mathrm{Var}(Y)$ are respectively
$$
\left\{ \begin{aligned}
	\mathrm{Cov}(U,Y)&=\gamma_U+\gamma_X\alpha_U\\
	\mathrm{Cov}(X,Y)&=\alpha_U\left( \gamma_U+\gamma_X\alpha_U \right) +\gamma_X\\
	\mathrm{Cov}(U_1,Y)&=\gamma_{U_1}\\
	\mathrm{Cov}(W,Y)&=\beta_U \left( \gamma_U+\alpha_U\gamma_X\right) +\gamma_{U_1}\beta_{U_1}\\
	\mathrm{Var}(Y)&=(\gamma_U+\gamma_X\alpha_U)^2+\gamma_{X}^{2}+\gamma_{U_1}^2+1.
\end{aligned} \right. 
$$
We can therefore derive the explicit form of the conditional distributions $p(w|x)$ and $p(y|x)$:
$$
\begin{aligned}
	W|X=x&\sim \cN\left\{ \mu_W+\frac{\mathrm{Cov}(W,X)}{\mathrm{Var}(X)}(x-\mu_{X}),\mathrm{Var}(W)\left( 1-\frac{\mathrm{Cov}^2(W,X)}{\mathrm{Var}(X)\cdot \mathrm{Var}(W)} \right) \right\}\\
	&\sim \cN\left\{ \mu_{X}^{W|X}x+\mu_{0}^{W|X},\sigma_{W|X}^{2} \right\} 
\end{aligned}
$$
$$
\begin{aligned}
    Y|X=x&\sim \cN\left\{ \mu_Y+\frac{\mathrm{Cov}(Y,X)}{\mathrm{Var}(X)}(x-\mu_{X}),\mathrm{Var}(Y)\left( 1-\frac{\mathrm{Cov}^2(Y,X)}{\mathrm{Var}(X)\cdot \mathrm{Var}(Y)} \right) \right\}\nonumber \\
	&\sim \cN\left\{ \mu_{X}^{Y|X}x+\mu_{0}^{Y|X},\sigma_{Y|X}^{2} \right\},
\end{aligned}
$$
where $\mu_{X}^{W|X},\mu_{0}^{W|X},\sigma_{W|X}^{2}, \mu_{X}^{Y|X}, \mu_{0}^{Y|X}$ and $\sigma_{Y|X}^{2}$ are defined as follows
$$
\begin{cases}
	\mu_{X}^{W|X}=\frac{\alpha_U\beta_U}{1+\alpha_{U}^{2}}\\
	\mu_{0}^{W|X}=\beta_0+\beta_{U_1}\mu_0-\frac{\alpha_U\beta_U\alpha_0}{1+\alpha_{U}^{2}}\\
	\sigma_{W|X}^{2}=\beta_U^2+\beta_{U_1}^{2}+1-\frac{\alpha_{U}^{2}\beta_U^2}{1+\alpha_{U}^{2}}\\
	\mu_{X}^{Y|X}=\frac{\alpha_U\left( \gamma_U+\gamma_X\alpha_U \right) +\gamma_X}{1+\alpha_{U}^{2}}\\
	\mu_{0}^{Y|X}=\gamma_0+\gamma_X\alpha_0+\gamma_{U_1}\mu_0-\frac{\alpha_0\{\alpha_U\left( \gamma_U+\gamma_X\alpha_U \right) +\gamma_X\}}{1+\alpha_{U}^{2}}\\
	\sigma_{Y|X}^{2}=(\gamma_U+\gamma_X\alpha_U)^2+\gamma_{X}^{2}+\gamma_{U_1}^2+1-\frac{(\alpha_U\left( \gamma_U+\gamma_X\alpha_U \right) +\gamma_X)^2}{1+\alpha_{U}^{2}}.
\end{cases}
$$
By applying Lemma~\ref{lemma.miao-example}, the solution of \eqref{eq.bridge_Y} is given by:
$$
h(w,y)=\frac{1}{\sqrt{\sigma_{Y|X}^{2}-\left(\mu_{X}^{Y|X}\right)^2\sigma_{W|X}^{2}/\left(\mu_{X}^{W|X}\right)^2}}\phi \left( \frac{y-\left( \mu_{0}^{Y|X}-\mu_{X}^{Y|X}\mu_{0}^{W|X}/\mu_{X}^{W|X} \right) -\mu_{X}^{Y|X}/\mu_{X}^{W|X}w}{\sqrt{\sigma_{Y|X}^{2}-\left(\mu_{X}^{Y|X}\right)^2\sigma_{W|X}^{2}/\left(\mu_{X}^{W|X}\right)^2}} \right),
$$
where $\phi$ is the standard normal distribution's probability density function (pdf).

For $h(w, y)$ to be meaningful, we need $\sigma_{Y|X}^{2}-\left(\mu_{X}^{Y|X}\right)^2\sigma_{W|X}^{2}/\left(\mu_{X}^{W|X}\right)^2>0$. Specifically, this means the following: 
\begin{equation}\label{eq. solution_h_exist_bid}
\begin{aligned}
&1-\frac{\gamma_{U}^{2}}{\beta_U^2}-\frac{\beta_{U_1}^{2}\gamma_{U}^{2}}{\beta_U^2}-\left( \frac{2\gamma_U}{\alpha_U}+\frac{2\gamma_U}{\beta_U^2\alpha_U}+\frac{2\alpha_U\gamma_U}{\beta_U^2}+\frac{2\beta_{U_1}^{2}\gamma_U}{\beta_U^2\alpha_U}+\frac{2\alpha_U\beta_{U_1}^{2}\gamma_U}{\beta_U^2} \right) \gamma_X\\
	&-\left( 1+\frac{2}{\beta_U^2}+\frac{1}{\alpha_{U}^{2}}+\frac{1}{\beta_U^2\alpha_{U}^{2}}+\frac{\alpha_{U}^{2}}{\beta_U^2}+\frac{2\beta_{U_1}^{2}}{\beta_U^2}+\frac{\beta_{U_1}^{2}}{\beta_U^2\alpha_{U}^{2}}+\frac{\alpha_{U}^{2}\beta_{U_1}^{2}}{\beta_U^2} \right) \gamma_{X}^{2}+ \gamma_{U_1}^2>0.
\end{aligned}
\end{equation}

We first show that when $\gamma_{U_1} = 0$, as long as $|\gamma_X| > \frac{|B|+\sqrt{\Delta}}{2A}$, \eqref{eq.bridge_Y} has no solution.

We first rewrite \eqref{eq. solution_h_exist_bid} as:
$$
\begin{aligned}
	&\underset{C}{\underbrace{1-\frac{\gamma_{U}^{2}}{\beta_U^2}-\frac{\beta_{U_1}^{2}\gamma_{U}^{2}}{\beta_U^2}}}-\underset{B}{\underbrace{\left( \frac{2\gamma_U}{\alpha_U}+\frac{2\alpha_U\gamma_U}{\beta_{U}^{2}}+\frac{2\beta_{U_1}^{2}\gamma_U}{\beta_{U}^{2}\alpha_U}+\frac{2\alpha_U\beta_{U_1}^{2}\gamma_U}{\beta_{U}^{2}} \right) }}\gamma_X\\
	&-\underset{A}{\underbrace{\left( 1+\frac{2}{\beta_{U}^{2}}+
    \frac{1}{\alpha_{U}^{2}}+\frac{1}{\beta_{U}^{2}\alpha_{U}^{2}}+\frac{\alpha_{U}^{2}}{\beta_{U}^{2}}+\frac{2\beta_{U_1}^{2}}{\beta_{U}^{2}}+\frac{\beta_{U_1}^{2}}{\beta_{U}^{2}\alpha_{U}^{2}}+\frac{\alpha_{U}^{2}\beta_{U_1}^{2}}{\beta_{U}^{2}} \right) }}\gamma_{X}^{2}>0.
\end{aligned}
$$
Noting that this is a quadratic function, we can get its discriminant 
$$
\Delta :=B^2-4AC=\frac{4\left( 1+\beta_{U_1}^{2}+\beta_{U}^{2}+\alpha_{U}^{2}\left( 1+\beta_{U_1}^{2} \right) \right) \left( 1+\alpha_{U}^{2}+\gamma_{U}^{2} \right)}{\alpha_{U}^{2}\beta_{U}^{2}}>0.
$$
Besides, we can find $ 1+\frac{2}{\beta_{U}^{2}}+\frac{1}{\alpha_{U}^{2}}+\frac{1}{\beta_{U}^{2}\alpha_{U}^{2}}+\frac{\alpha_{U}^{2}}{\beta_{U}^{2}}+\frac{2\beta_{U_1}^{2}}{\beta_{U}^{2}}+\frac{\beta_{U_1}^{2}}{\beta_{U}^{2}\alpha_{U}^{2}}+\frac{\alpha_{U}^{2}\beta_{U_1}^{2}}{\beta_{U}^{2}}>0$. Therefore, this is a quadratic function whose discriminant is always positive and opens downward. When $\gamma_X$ satisfies $\frac{-B+\sqrt{\Delta}}{2A}<\gamma_X<\frac{-B-\sqrt{\Delta}}{2A}$, \eqref{eq.bridge_Y} will have a solution. Otherwise, when $\gamma_X\geq\frac{-B-\sqrt{\Delta}}{2A}$ or $\gamma_X\leq\frac{-B+\sqrt{\Delta}}{2A}$, \eqref{eq.bridge_Y} will have no solution. 
 
That means, as long as $|\gamma_X|>\frac{|B|+\sqrt{\Delta}}{2A}$, the integration equation \eqref{eq.bridge_Y} has no solution.

Next, we show that, if $\gamma_{U_1}\ne 0$, when $|\gamma_X| > \frac{|B|+2\Delta}{2A}$, if $|\gamma_{U_1}| > |C| + |B||\gamma_X| + A\gamma_X^2$, \eqref{eq.bridge_Y} has a solution. 

We only consider the case $|\gamma_X|>\frac{-B+\sqrt{\Delta}}{2A}$ under $\alpha_U\gamma_U>0$, since the proof for the other case (\emph{i.e.}, $\alpha_U \gamma_U<0$) is similar. We can rewrite \eqref{eq. solution_h_exist_bid} as
$$
\begin{aligned}
	\gamma_{U_1}^2&>\left( 1+\frac{2}{\beta_{U}^{2}}+\frac{1}{\alpha_{U}^{2}}+\frac{1}{\beta_{U}^{2}\alpha_{U}^{2}}+\frac{\alpha_{U}^{2}}{\beta_{U}^{2}}+\frac{2\beta_{U_1}^{2}}{\beta_{U}^{2}}+\frac{\beta_{U_1}^{2}}{\beta_{U}^{2}\alpha_{U}^{2}}+\frac{\alpha_{U}^{2}\beta_{U_1}^{2}}{\beta_{U}^{2}} \right) \gamma_{X}^{2}\\
	&+\left( \frac{2\gamma_U}{\alpha_U}+\frac{2\gamma_U}{\beta_{U}^{2}\alpha_U}+\frac{2\alpha_U\gamma_U}{\beta_{U}^{2}}+\frac{2\beta_{U_1}^{2}\gamma_U}{\beta_{U}^{2}\alpha_U}+\frac{2\alpha_U\beta_{U_1}^{2}\gamma_U}{\beta_{U}^{2}} \right) \gamma_X-\left( 1-\frac{\gamma_{U}^{2}}{\beta_{U}^{2}}-\frac{\beta_{U_1}^{2}\gamma_{U}^{2}}{\beta_{U}^{2}} \right).
\end{aligned}
$$
The right-hand side is $\geq 0$ as long as$|\gamma_{U_1}| > |C| + |B||\gamma_X| + A\gamma_X^2$, the solution to \eqref{eq.bridge_Y} will still exist.
\end{proof}
\begin{remark}
If $\gamma_X=\gamma_{U_1}=0$,  \eqref{eq. solution_h_exist_bid} will become $1-\frac{\gamma_{U}^{2}}{\beta_{U}^{2}}-\frac{\beta_{U_1}^{2}\gamma_{U}^{2}}{\beta_{U}^{2}}>0$. If $\beta_{U_1}=0$, the above inequality will become $1-\frac{\gamma_{U}^{2}}{\beta_{U}^{2}}>0$, which is consistent with the result we obtained before. However, if $\beta_{U_1}\ne0$, the above inequality is difficult to satisfy. However, as long as $\beta_{U_1}$ is sufficiently large, the solatability of the integral equation is reduced. Otherwise, similar to the case when $\gamma_X \neq 0$, if the effect of $U_1$ on $Y$ is strong enough (\emph{i.e.}, $|\gamma_{U_1}|$), the solution exists again. Specifically, if $\gamma_X=0,\gamma_W\ne0$,  \eqref{eq. solution_h_exist_bid} will become $1-\frac{\gamma_{U}^{2}}{\beta_{U}^{2}}-\frac{\beta_{U_1}^{2}\gamma_{U}^{2}}{\beta_{U}^{2}}+\gamma_{U_1}>0$. This means that if $\gamma_{U_1}$ is sufficiently large, then \eqref{eq.bridge_Y} still have a solution. 
\end{remark}

Next, we prove the claims in example~\ref{example:linear_gaussian_two_proxy}. We show that as long as the coefficient of $W' \to Y$ is strong enough in example~\ref{example:linear_gaussian_two_proxy}, the solution of the integral equation $p(y|x')=\int h(w',y)p(w'|x')dw'$ exists. As an explanation, we will show that a key condition in Picard's theorem~\ref{lemma.Picard} holds, namely, the series $\sum_{n=1}^\infty \lambda_n^{-2} |\langle p(y|x'),\phi_n \rangle|^2$ converges. 

To compute the series, we need the singular value decomposition of the operator $ T: \mathcal{L}^2\{F(w')\} \to \mathcal{L}^2\{F(x')\} $, where $Th = \mathbb{E}\{ h(W', y) | x'\} = p(y|x') $ for all $(x', y)$. Based on the data-generating process in example~\ref{example:linear_gaussian_two_proxy}, both $ \mathcal{L}^2\{F(w')\}$ and $\mathcal{L}^2\{F(x')\}$ are square-integrable spaces with respect to the standard Gaussian measure. For such spaces, \cite{carrasco2007linear} derived the form of the eigenvectors $\phi_n$, as stated in Lemma~\ref{lem.self-adjoint operator}.
    
Next, we prove the result in Example~\ref{example:linear_gaussian_two_proxy}. 

\examplelineargaussian*
\begin{proof}
    We first show that under $\mH_1$, the integral equation $p(y|x')=\int{h(w',y)p(w'|x')dw'}$ has a solution if and only if the coefficient $\gamma_W$ is large enough. Specifically, since $X'$ and $W'$ are normalized, based on the data generation structure, we have
$$
\left( U,X',W',Y \right)^\top \sim \cN\left\{ \bm{0}_4, \Sigma \right\}, 
$$
where 
\begin{equation*}
    \Sigma := \left( \begin{array}{cccc}
	1&		\frac{2}{\sqrt{5}}&		-\frac{2}{\sqrt{5}}&		-\frac{2}{\sqrt{5}}\gamma_W+1+\frac{2}{\sqrt{5}}\\
	\frac{2}{\sqrt{5}}&		1&		-\frac{4}{5}&		-\frac{4}{5}\gamma_W+1+\frac{2}{\sqrt{5}}\\
	-\frac{2}{\sqrt{5}}&		-\frac{4}{5}&		1&		\gamma_W-\frac{2}{5}( 2+\sqrt{5} ),\\
	-\frac{2}{\sqrt{5}}\gamma_W+1+\frac{2}{\sqrt{5}}&		-\frac{4}{5}\gamma_W+1+\frac{2}{\sqrt{5}}&		\gamma_W-\frac{4}{5}-\frac{2}{\sqrt{5}}&		\gamma^2_W-\frac{4}{5}( 2+\sqrt{5}) \gamma_W+3+\frac{4}{\sqrt{5}}\\
\end{array} \right).
\end{equation*}

We can therefore derive the explicit form of the conditional distributions $p(w'|x')$ and $p(y|x')$: 
\begin{align}
    W'|X'=x' &\sim \cN\left\{ \mu_W+\frac{\mathrm{Cov}(W',X')}{\mathrm{Var}(X')}(x'-\mu_{X'}),\mathrm{Var}(W')\left( 1-\frac{\mathrm{Cov}^2(W',X')}{\mathrm{Var}(X')\cdot \mathrm{Var}(W')} \right) \right\} \nonumber \\
	 &\sim \cN\left( -\frac{4}{5}x',\frac{9}{25} \right); \nonumber \\
     Y|X'=x'&\sim \cN\left\{ \mu_Y+\frac{\mathrm{Cov}(Y,X')}{\mathrm{Var}(X')}(x'-\mu_{X'}),\mathrm{Var}(Y)\left( 1-\frac{\mathrm{Cov}^2(Y,X')}{\mathrm{Var}(X')\cdot \mathrm{Var}(Y)} \right) \right\}, \nonumber \\
	&\sim \cN\left\{ \left( -\frac{4}{5}\gamma_W+1+\frac{2}{\sqrt{5}} \right) x',\frac{9}{25}\gamma_{w}^{2}-\frac{4}{5\sqrt{5}}\gamma_W+\frac{6}{5} \right\}. \label{eq.Y_given_X}
\end{align}

By applying Lemma~\ref{lemma.miao-example}, the solution of \eqref{eq.bridge_Y} is given by:
\begin{equation}\label{eq. true solution_h}
    h(w',y)=\frac{1}{\sqrt{\frac{9+2\sqrt{5}}{10}\gamma_W+\frac{3}{16}-\frac{9\sqrt{5}}{20}}}\phi \left\{ \frac{y-\left( \gamma_W+\frac{2\sqrt{5}-5}{4} \right) w'}{\frac{9+2\sqrt{5}}{10}\gamma_W+\frac{3}{16}-\frac{9\sqrt{5}}{20}} \right\}.
\end{equation}
For $h(w',y)$ to be meaningful, we need $\frac{9+2\sqrt{5}}{10}\gamma_W+\frac{3}{16}-\frac{9\sqrt{5}}{20} > 0$, which implies $\gamma_W>\frac{-15+36\sqrt{5}}{72+16\sqrt{5}}\approx0.61$. 

Next, we need to verify the conditions for the series in Picard's theorem~\ref{lemma.Picard}, which requires proving that $\sum_{n=0}^{+\infty}\lambda_{n}^{-2}|\langle f,\phi_{n}\rangle|^{2}<+\infty $ for the singular system $(\lambda_n,\varphi_n,\phi_{n})_{n=1}^{+\infty}$ associated with the compact operator $Th=f$. In our data generation process, operator $T: \mathcal{L}^2(W',\gamma)\rightarrow  \mathcal{L}^2(X',\gamma)$ satisfies $Th =\mathbb{E}\{ h(W',y)|x'\}=p(y|x')$ for all $(x',y)$ and is characterized by the integral kernel~\eqref{eq.kernel}. Thus, by Lemma~\ref{lem.self-adjoint operator}, we have $T: \mathcal{L}^2(W',\gamma)\rightarrow  \mathcal{L}^2(X',\gamma)$ is a self-adjoint operator and the eigenvalue system of operator $T$ is given by $\varphi_j(w')=\mathrm{he}_{j}(w'),\phi_j(x')=\mathrm{he}_{j}(x'),\lambda_j=\rho_{WX}^{j}$, where $\rho_{WX}$ is the correlation coefficient between $W'$ and $X'$ and $\mathrm{he}_{j}$ \eqref{eq.Hermite_norm} is the normalized Hermite polynomials. Thus, we show that the following series converges, which can explain why the solution may exist if only and if $\gamma_W>\frac{-15+36\sqrt{5}}{72+16\sqrt{5}}$ under $\mH_1$:
$$
\sum_{n=0}^{\infty}{\frac{\left| \langle p( y|x' ) ,\mathrm{he}_n( x' ) \rangle \right|^2}{\rho_{WX}^{2n}}}.
$$
Define the parameters $\mu:= -\frac{4}{5}\gamma_W+1+\frac{2}{\sqrt{5}}$, $\sigma^2:=\frac{9}{25}\gamma_{w}^{2}-\frac{4}{5\sqrt{5}}\gamma_W+\frac{6}{5}$ and the inner product
\begin{align*}
    I_n:&=\langle p(y|x'),\mathrm{he}_n(x')\rangle =\frac{1}{\sqrt{2\pi}}\int{p(y|x')\mathrm{he}_n(x')e^{-(x')^2/2}dx'}\\
    &=\frac{1}{\sqrt{2\pi n!}}\int{p(y|x')\mathrm{He}_n(x')e^{-(x')^2/2}dx'}.
\end{align*}

\noindent\textbf{Step 1. Sufficiency.} 

We first demonstrate that if $\gamma_W > \frac{-15 + 36\sqrt{5}}{72 + 16\sqrt{5}}$, the series converges. We consider two cases: \textbf{(i)} $\gamma_W =\frac{5+2\sqrt{5}}{4} > \frac{-15 + 36\sqrt{5}}{72 + 16\sqrt{5}}$; and \textbf{(ii)} $\gamma_W \neq \frac{5+2\sqrt{5}}{4}$. 

\noindent\textbf{(a). The case of $\gamma_W =\frac{5+2\sqrt{5}}{4}$.}

In this case, the distribution of $p(y|x')$ becomes
$$
Y|X'=x'\sim \cN\left\{ 0,\frac{1}{16}( 29+4\sqrt{5} ) \right\}.
$$
Thus, if we define $\sigma^2_{\mathrm{con}}:=\frac{1}{16}( 29+4\sqrt{5} )$, we have
$$
I_n=\frac{1}{\sqrt{2\pi n!}}\int{\frac{1}{\sqrt{2\pi \sigma_{\mathrm{con}}^{2}}}e^{-\frac{y^2}{2\sigma_{\mathrm{con}}^{2}}}H_n(x')e^{-(x')^2/2}dx'=\frac{e^{-\frac{y^2}{2\sigma_{\mathrm{con}}^{2}}}}{2\pi \sqrt{\sigma^2_{\mathrm{con}}n!}}\int{\mathrm{He}_n(x')e^{-(x')^2/2}dx'}}.
$$
According to Lemma 2.6 in \cite{davis2024general}, the integral of the stretched Hermite polynomial $S_n=\frac{1}{\sqrt{2\pi}}\int{\mathrm{He}_n(\gamma x')e^{-(x')^2/2}dx'}$ is only non-zero for even $n$ and has the value $S_n=(n-1)!!(\gamma^2-1)^{n/2}$. Applying the above results and taking $\gamma=1$, we have $I_n=0$ for all $n\ge 1$. Thus, the series is:
\begin{align*}
    \sum_{n=0}^{\infty}{\left( \frac{I_n}{\rho_{WX}^{n}} \right)^2}=\left( \frac{I_0}{\rho_{WX}^{0}} \right)^2=\left( I_0 \right)^2 & \overset{(1)}{=} \frac{e^{-\frac{y^2}{\sigma_{\mathrm{con}}^{2}}}}{4\pi^2 \sigma_{\mathrm{con}}^{2}}\left\{ \int{e^{-(x')^2/2}dx'} \right\}^2 \\
    & \overset{(2)}{=}\frac{e^{-\frac{y^2}{\sigma_{\mathrm{con}}^{2}}}}{4\pi^2 \sigma_{\mathrm{con}}^{2}} 2\pi =\frac{1}{2\pi \sigma_{\mathrm{con}}^{2}}e^{-\frac{y^2}{\sigma_{\mathrm{con}}^{2}}}<\infty.
\end{align*}
where (1) follows from $\mathrm{He}_0(x)=1$ and (2) follows from $\int{e^{-x^2/2}dx}=\sqrt{2\pi}$. Hence, the series converges.

\noindent\textbf{(b). The case $\gamma_W \neq \frac{5+2\sqrt{5}}{4}$.}

Note that the probabilist's Hermite polynomials $\mathrm{He}_n(x')$ admit the generating function
$$
\sum_{n=0}^{\infty} \frac{\mathrm{He}_n(x')}{n!} t^n = \exp\left(x' t - \tfrac{1}{2} t^2\right).
$$
In particular,
$$
\mathrm{He}_n(x') = n!  [t^n] \exp\left(x' t - \tfrac{1}{2} t^2\right),
$$
where $[t^n] f(t)$ denotes the coefficient of $t^n$ in the power series expansion of $f(t)$. Substituting this expression into the definition of $I_n$, we obtain
\begin{align*}
    I_n &= \frac{1}{\sqrt{2\pi n!}} \int p(y | x') \mathrm{He}_n(x') e^{-(x')^2/2}  dx' \\
    &= \frac{\sqrt{n!}}{\sqrt{2\pi}}  [t^n] \int p(y | x') \exp\left\{-\tfrac{(x')^2}{2} + x' t - \tfrac{1}{2} t^2 \right\} dx'\\
    & \overset{\mathrm{def}}{=} \frac{\sqrt{n!}}{\sqrt{2\pi}}  [t^n] J(y,t).
\end{align*}

Recall that $\mu= -\frac{4}{5}\gamma_W+1+\frac{2}{\sqrt{5}}$ and $\sigma^2=\frac{9}{25}\gamma_{w}^{2}-\frac{4}{5\sqrt{5}}\gamma_W+\frac{6}{5}$. By \eqref{eq.Y_given_X}, we can write
$$
J(y,t) = \frac{1}{\sqrt{2\pi \sigma^2}} \int \exp\left\{-\frac{(y-\mu x')^2}{2\sigma^2} - \frac{(x')^2}{2} + x' t - \frac{t^2}{2} \right\} dx'.
$$
Collecting the quadratic terms in $x'$ yields
$$
-\frac{(y-\mu x')^2}{2\sigma^2} - \frac{(x')^2}{2} + x' t
= -\frac{1}{2}\left(1 + \frac{\mu^2}{\sigma^2}\right) (x')^2 + \left(\frac{\mu y}{\sigma^2} + t\right) x' - \frac{y^2}{2\sigma^2}.
$$
Applying the standard Gaussian integral identity
$$
\int \exp\left\{-\frac{\alpha}{2} (x')^2 + \beta x'\right\} dx'
= \sqrt{\frac{2\pi}{\alpha}} \exp\left(\tfrac{\beta^2}{2\alpha}\right),
\qquad \alpha > 0,
$$
with $\alpha = 1 + \mu^2/\sigma^2$, $\beta = \mu y/\sigma^2 + t$, we obtain
\begin{align*}
    J(y,t) &= \frac{1}{\sqrt{\sigma^2 \alpha}} \exp\left\{ \frac{(\mu y/\sigma^2 + t)^2}{2 \alpha} - \frac{y^2}{2\sigma^2} - \frac{t^2}{2} \right\}\\
    &=C(y)\exp\left\{\frac{\mu y}{\sigma^2 + \mu^2} t - \frac{\mu^2}{2(\sigma^2 + \mu^2)} t^2\right\}:=C(y)\exp\left(b t - \frac{c}{2} t^2\right),
\end{align*}
where $b=\frac{\mu y}{\sigma^2 + \mu^2}$ and $c=\frac{\mu^2}{\sigma^2 + \mu^2}$. Now, by the generating function of Hermite polynomials, we have
$$
\exp\left(b t - \tfrac{c}{2} t^2\right)
= \exp\left\{z \sqrt{c} t - \tfrac{1}{2} (\sqrt{c} t)^2\right\}
= \sum_{n=0}^\infty \frac{\mathrm{He}_n(z)}{n!} (\sqrt{c} t)^n,
$$
where
$
z := \frac{b}{\sqrt{c}}
= \frac{\mu y}{\sigma^2 + \mu^2} \cdot \frac{\sqrt{\sigma^2 + \mu^2}}{|\mu|}
= \operatorname{sign}(\mu)  \frac{y}{\sqrt{\sigma^2 + \mu^2}}.
$
Since $\gamma_W \neq \frac{5+2\sqrt{5}}{4}$, we have $\mu \neq 0$. It then follows that
$
[t^n] J(y,t) = C(y) \frac{\mathrm{He}_n(z)}{n!} (\sqrt{c})^n.
$
Consequently, we have:
$$
I_n = \frac{\sqrt{n!}}{\sqrt{2\pi}} [t^n] J(y,t)
= \frac{\sqrt{n!}}{\sqrt{2\pi}} C(y) \frac{\mathrm{He}_n(z)}{n!} (\sqrt{c})^n.
$$
Hence,
$$
\left( \frac{I_n}{\rho_{WX}^n} \right)^2
= \frac{1}{2\pi} C(y)^2 \frac{\mathrm{He}_n(z)^2}{n!} \left(\frac{c}{\rho_{WX}^2}\right)^n.
$$
To show the convergence, we invoke Mehler’s formula (Lemma \ref{lem:mehler_physicist}). Specifically, it means that the series $\sum_{n=0}^{\infty} \frac{(t/2)^n}{n!} H_n(x) H_n(y)$ is convergent if and only if $|t|< 1$, where $H_n(x'):=(-1)^ne^{(x')^2}\frac{d^n}{dx^n}e^{-(x')^2}$ is physicist's Hermite polynomials. Besides, when $|t|< 1$, we will have:
$$
\sum_{n=0}^{\infty} \frac{(t/2)^n}{n!} H_n(x) H_n(y) 
= \frac{1}{\sqrt{1 - t^2}} 
  \exp\left[\frac{2txy - t^2\{(x')^2 + y^2\}}{1 - t^2} \right],
$$
Since $H_n(x') = 2^{n/2} \mathrm{He}_n(\sqrt{2}x')$, the above equation becomes
$$
\sum_{n=0}^\infty \frac{t^n}{n!}\mathrm{He}_n(x)\mathrm{He}_n(y)
= \frac{1}{\sqrt{1-t^2}}
  \exp\left\{ \frac{txy - \tfrac{t^2}{2}(x^2+y^2)}{1-t^2}\right\}.
\qquad |t|<1.
$$
Thus, we can obtain
\begin{equation}\label{eq:series_convergence}
    \sum_{n=0}^\infty \left( \frac{I_n}{\rho_{WX}^n} \right)^2 = \frac{1}{2\pi} C(y)^2 \sum_{n=0}^{\infty} \frac{\mathrm{He}_n(z)^2}{n!}\left(\frac{c}{\rho_{WX}^2}\right)^n,
\end{equation}
which 
converges if and only if $| c / \rho_{WX}^2 | < 1$. Recall that $c=\frac{\mu^2}{\sigma^2 + \mu^2}$, we have
$$
\frac{\mu^2}{\rho_{WX}^{2}(\sigma^2+\mu^2)}<1, 
$$
which holds if and only if  $\gamma_W>\frac{-15+36\sqrt{5}}{72+16\sqrt{5}}\approx0.61$ by taking $\mu= -\frac{4}{5}\gamma_W+1+\frac{2}{\sqrt{5}}$, $\sigma^2=\frac{9}{25}\gamma_{w}^{2}-\frac{4}{5\sqrt{5}}\gamma_W+\frac{6}{5}$ and  $\rho_{WX} =-\frac{4}{5} $. Thus, if $\gamma_W>\frac{-15+36\sqrt{5}}{72+16\sqrt{5}}$ and $\gamma_W\ne \frac{5+2\sqrt{5}}{4}$, the series converges.

Combine all results, when $\gamma_W > \frac{-15 + 36\sqrt{5}}{72 + 16\sqrt{5}}$, the series converges.

\noindent\textbf{Step 2. Necessity.} 

We now show that if the series converges, then $\gamma_W>\frac{-15+36\sqrt{5}}{72+16\sqrt{5}}$. We will show that $\gamma_W$ either equals to $\frac{5 + 2\sqrt{5}}{4}$, or $>\frac{-15+36\sqrt{5}}{72+16\sqrt{5}}$ but $\neq \frac{5 + 2\sqrt{5}}{4}$. 

\noindent\textbf{(a). The case of $\gamma_W = \frac{5 + 2\sqrt{5}}{4}$.}

When $\gamma_W = \frac{5 + 2\sqrt{5}}{4}$, we have $\mu = 0$. Therefore, $p(y|x') \sim \mathcal{N}\left(0, \sigma_{\text{con}}^2\right)$ with $\sigma_{\text{con}}^2 = \frac{29 + 4\sqrt{5}}{16}$. As shown in Step 1(a), $I_n = 0$ for $n \geq 1$, so
$$
\sum_{n=0}^{\infty}{\left( \frac{I_n}{\rho_{WX}^{n}} \right)^2}
=\frac{1}{2\pi \sigma_{\mathrm{con}}^{2}}e^{-\frac{y^2}{\sigma_{\mathrm{con}}^{2}}}<\infty.
$$
Since $\frac{5 + 2\sqrt{5}}{4} > \frac{-15 + 36\sqrt{5}}{72 + 16\sqrt{5}}$, convergence is consistent with the condition.

\noindent\textbf{(b). The case of $\gamma_W \neq \frac{5 + 2\sqrt{5}}{4}$.}

From \eqref{eq:series_convergence} of step 1 (b), we have
$$
\sum_{n=0}^\infty \left( \frac{I_n}{\rho_{WX}^n} \right)^2 = \frac{1}{2\pi} C(y)^2 \sum_{n=0}^{\infty} \frac{\mathrm{He}_n(z)^2}{n!}\left(\frac{c}{\rho_{WX}^2}\right)^n.
$$
Since $\gamma_W \neq \frac{5 + 2\sqrt{5}}{4}$, we have $\mu \neq 0$. By Mehler's formula, convergence requires $\left| \frac{c}{\rho_{WX}^2} \right| < 1$, as derived in Step 1 (b), which holds if and only if
$$
\gamma_W>\frac{-15+36\sqrt{5}}{72+16\sqrt{5}} \text{ and } \gamma_W \ne  \frac{5 + 2\sqrt{5}}{4}.
$$
We complete the proof. 
\end{proof}

Next, we give the generation details in Fig.~\ref{fig.power_fail}, \emph{i.e.} $U=\varepsilon_U,X =2U + \varepsilon_X,W =-2U +\varepsilon_W$ and $Y= X^2 + U^2+\gamma_W W+\varepsilon_Y,$
where $\varepsilon_U,\varepsilon_Y,\varepsilon_W,\varepsilon_X \sim \cN(0,1)$.

\subsection{Proof of asymptotic properties with two proxies}
\label{appx.two_asymptotic}

\begin{assumption}\label{differentiabilty and integrability_two}
We assume $\mathbb{E}_X\{m(X,Z,s)|W\}$ and $\mathbb{E}_X\{|m(X,Z,s)| ^2|W\}$ are uniformly bounded for all $s$. 
\end{assumption}

\begin{assumption}\label{assum.var_two}
    For any $s,t \in \cT$, $\mathbb{E}\{U(W,Y,t)^4|X\}<\infty$ and $\mathbb{E}(|m(X,Z,s) -\{A(A^*A)^{-1}g_{s}\}(X)|^4)<\infty$, where $g_{s}(\cdot)=\mathbb{E} [ m(X,Z,s) \phi_W(W)](\cdot)$.
\end{assumption}

\begin{theorem}\label{theorem:null-hypothesis_two}
Denote $\overline{\eta}_{s,t}(W,Z,Y,X) :=U(W,Y,t)$ $\left[\{m(Z,X,s) - \{A(A^*A)^{-1}g_{s}\}(X)\right]$, where $g_{s}(\cdot):=$ $\mathbb{E}\{m(Z,X,s)\phi_W(W)\}(\cdot)$. Suppose conditions in Theorem~\ref{theorem:null-hypothesis} hold. If conditions~\ref{assum.complete_U_X_Zx}, \ref{differentiabilty and integrability_two}--\ref{assum.var_two} and \ref{assum.empirical_process}-\ref{assum.source condition} hold, we have that under $\mH_0$, \textbf{(i).} $T^{(Z)}_n(s,t)$ converges weakly to $\mathbb{G}(s,t)$ in $\mathcal{L}^{2}\{\mathcal{T}\times\mathcal{T},\mu\times\mu\}$, where $\mathbb{G}(s,t)$ is a Gaussian process with zero-mean and covariance:
\begin{align*}
    \Sigma\{(s,t),(s',t')\} =\mathbb{E}\{\overline{\eta}_{s,t}(W,Z,Y,X)\overline{\eta}_{s',t'}(W,Z,Y,X)\};
\end{align*}
\textbf{(ii).} $\Delta^{(Z)}_{\varphi,m}$ converges weakly to $\underset{t\in \cT}{\max}\int{| \mathbb{G}(s,t)|^2d\mu(s)}$.
\end{theorem}
\begin{proof}
We need to replace the weight function $m(x,s)$ with $m(z,x,s)$ over $(z,x)$. By  \eqref{eq.Tnst decompose}, we have
$$
T^{(Z)}_n(s,t) =\sqrt{n}\mathbb{P}_n\{ U(W,Y,t) m(Z,X,s) \}+\left( \text{Expected risk difference} \right) +\left( \text{Empirical process} \right).
$$

By Proposition~\ref{prop.empirical_process}, the empirical process term has
$$ \sqrt{n}( \mathbb{P}_n-\mathbb{P} )[ \{ H^0(W,t) -\widehat{H}^{\lambda}(W,t)\} m(Z,X,s)]=o_{p}(1).
$$

By Proposition~\ref{prop.expected_risk}, for fixed $x$, the expected risk difference term has: 
$$
\sqrt{n}\mathbb{P} \left\{ ( H^0(W,t) -\widehat{H}^{\lambda}(W,t)) m(Z,X,s) \right\} =-\frac{1}{\sqrt{n}}\sum_{i=1}^n{U(w_i,y_i,t) \{A(A^*A)^{-1}g_{s}\}(x_i)}+o_{p}(1).
$$
Therefore, combining all the inequalities, we have 
$$
T^{(Z)}_n(s,t) =\frac{1}{\sqrt{n}}\sum_{i=1}^n{U(w_i,y_i,t)\left[m(x_i,z_i,s)- \left\{A(A^*A)^{-1}g_{s}\right\}(x_i)\right]}+o_{p}(1).
$$
Next, we apply Lemma~\ref{ central limit theorem} to $\left\{U(w_i,y_i,t)\left[m(x_i,z_i,s)- \left\{A(A^*A)^{-1}g_{s}\right\}(x_i)\right] \right\}_i$ to obtain $T^{(Z)}_n(s,t)$ converges weakly to $\mathbb{G}(s,t)$ in $\mathcal{L}^{2}\{\mathcal{T}\times\mathcal{T},\mu\times\mu\}$, where $\mathbb{G}(s,t)$ is a Gaussian process with zero-mean and covariance:
\begin{align*}
    \Sigma\{(s,t),(s',t')\} =\mathbb{E}\{\overline{\eta}_{s,t}(W,Z,Y,X)\overline{\eta}_{s,t}(W',Z',Y',X')\}.
\end{align*} 
To show $\mG(s,t)$ is zero-mean, noted that 
$$
\begin{aligned}
	&\mathbb{E} \left\{ U(W,Y,t)[ m(Z,X,s)-\{A(A^*A)^{-1}g_s\}(X)] \right\} \\
    &=\mathbb{E} \{ U(W,Y,t)m(Z,X,s) \} -\mathbb{E} [ U(W,Y,t)\{A(A^*A)^{-1}g_s\}(X) ]\\
	&=\mathbb{E} [ m(Z,X,s)\mathbb{E} \{U(W,Y,t)|Z,X \} ] -\mathbb{E} [\mathbb{E} \{ U(W,Y,t)|X\} \{A(A^*A)^{-1}g_s\}(X) ]\\
	&=0,
\end{aligned}
$$
where the last equation follows from  \eqref{eq.convert_YZ} and \eqref{eq.convert_Y}.

Besides, by condition~\ref{assum.var_two}, we have $\operatorname{Var}(U(w_i,y_i,t)[ m(x_i,z_i,s)-\{A(A^*A)^{-1}g_s\}(x_i)] )=\mathbb{E}(U(w_i,y_i,t)[ m(x_i,z_i,s)-\{A(A^*A)^{-1}g_s\}(x_i)])^2<\infty$ for any $(x,s,t)$. Therefore, by continuous mapping theorem,  we have $\Delta^{(Z)}_{\varphi,m}$ converges weakly to $\underset{t\in \cT}{\max}\int{| \mathbb{G}(s,t)|^2d\mu(s)}$.
\end{proof}

For power analysis, we define the global alternative $\mH^{\mathrm{fix}}_1$
and $\mH^\alpha_{1n}$ ($0 < \alpha \leq 1/2$) of \eqref{eq.convert_YZ}, in terms of $\mathbb{E}\{ \varphi(Y,t) -H(W,t) |Z,X\}$.
\begin{equation*}
    \mH^{\mathrm{fix}}_1: \mathbb{E}\{ \varphi(Y,t) -H(W,t) |Z,X\} \neq 0 \text{ for some $t \in \cT$},
\end{equation*}
for any $H^0(W,t) \in \cH_W$. For the local alternative $\mH^\alpha_{1n}$, there exists $H^0(W,t) \in \cH_W$, such that
\begin{equation*}
    \mH^\alpha_1: \mathbb{E}\{ \varphi(Y,t) |Z,X\} =  \mE\{H^0(W,t)|Z,X\} + \frac{r(Z,X,t)}{n^\alpha}, \ \forall t,
\end{equation*}
where $0 < \alpha \leq 1/2$, and for any $H$, $\frac{r(Z,X,t)}{n^\alpha}$ cannot be written as $\mE\{H(\cdot,t)-H^0(\cdot,t)|Z,X\}$ for some $t$. 

\begin{theorem}
\label{theorem:alternative-hypothesis_two}
Suppose conditions in Theorem~\ref{theorem:null-hypothesis_two} hold. Besides, we assume $\mE\{r(Z,X,t)^4\}<\infty$ for fixed $x$ and any $t$. Then, we have:
\begin{itemize}[noitemsep,topsep=0pt]
    \item[\textbf{\emph{(i)}}] \textbf{Global alternative}.  $\lim_{n \to \infty} \max_{t \in \cT}  |T_n^{(Z)}(s,t)| = \infty$ for almost all $s$ under $\mH_1^{\mathrm{fix}}$. 
    \item[\textbf{\emph{(ii)}}] \textbf{Local alternative} ($\alpha < 1/2$). $\lim_{n \to \infty} \max_{t \in \cT}$ $|T^{(Z)}_n(s,t)| = \infty$ for almost all $s$ under $\mH^\alpha_{1n}$. 
    \item[\textbf{\emph{(iii)}}] \textbf{Local alternative} ($\alpha = 1/2$). $T^{(Z)}_n(s,t)$ converges weakly to $\mathbb{G}(s,t)+\mu(Z,X,s,t)$ in $\mathcal{L}^{2}\{\cT \times \cT, \mu \times \mu \}$ under $\mH^\alpha_{1n}$, where $\mathbb{G}(s,t)$ is defined in Theorem~\ref{theorem:null-hypothesis_two} and $\mu(Z,X,s,t):=\mathbb{E} [r(Z,X,t)m(Z,X,s) -\{A(A^*A)^{-1}A^*m(\cdot,s)\}(X)]$.
\end{itemize}
\end{theorem}
\begin{proof}
    The proof is similar to that of theorem~\ref{theorem:alternative-hypothesis}, with the weight function $m(X,s)$ replaced with $m(Z,X,s)$.
\end{proof}

\section{General Technical Lemmas}
\begin{lemma}[Theorem 15.18 in \cite{kress1989linear}]\label{lemma.Picard}
    Given Hilbert spaces $\cH_1$ and $\cH_2$, a compact operator $T:H_{1}\to H_{2}$ and its adjoint operator $T^{*}:\cH_2\to \cH_1$, there exists a singular system $(\lambda_n,\varphi_n,\phi_{n})_{n=1}^{+\infty}$ of $K$ with nonzero singular values $\{\lambda_{n}\}_{n=1}^{+\infty}$ and orthogonal sequences $\{\varphi_n\in \cH_1\}_{n=1}^{+\infty}, \{\phi_{n}\in \cH_{2}\}_{n=1}^{+\infty}$. Then the equation of the first kind $Th=f$ with $f \in \cH_2$, has a solution if and only if
    \begin{enumerate}
    \item $f\in\mathrm{Ker}(T^{*})^{\perp}$, where $\mathrm{Ker}(T^{*})=\{h:T^{*}h=0\}$ is the null space of the adjoint operator $T^{*}$;
    \item $\sum_{n=1}^{+\infty}\lambda_{n}^{-2}|\langle f,\phi_{n}\rangle|^{2}<+\infty $.
    \end{enumerate} 
\end{lemma}
\begin{lemma}[Theorem 2.32 of \cite{carrasco2007linear}]\label{lem:hs-compact}
    Every Hilbert–Schmidt operator is compact.
\end{lemma}

\begin{lemma}[Theorem 2.34 of \cite{carrasco2007linear}]\label{lem:hs-kernel}
Let $\cL^2(\mR^q,\pi)$ and $\cL^2(\mR^r,\rho)$ denote the Hilbert spaces 
$$
\cL^2(\mR^q,\pi) := \left\{\varphi: \mR^q \to \mR ,\|\varphi\|_{\cL^2(\pi)}^2 := \int |\varphi(s)|^2 \pi(s) ds < \infty \right\},
$$
and similarly for $\cL^2(\mR^r,\rho)$. An operator $K: \cL^2(\mR^q,\pi) \to \cL^2(\mR^r,\rho)$ is Hilbert–Schmidt  if and only if it satisfies the following two conditions:
\begin{enumerate}
    \item It admits a kernel representation as an integral operator $K$ of the form
    $$
    (K\varphi)(\tau) = \int k(\tau, s)\varphi(s)\pi(s) ds.
    $$
    \item Its kernel function $k(\tau, s)$ is square-integrable, satisfying
    $$
    \iint |k(\tau, s)|^2 \pi(s)\rho(\tau)ds d\tau < \infty.
    $$
\end{enumerate}
\end{lemma}

\begin{lemma}
\label{lemma.miao-example}
 If $W|X\sim \cN(\beta_0+\beta_1X,\sigma_{2}^{2})$ and $Y|X\sim \cN(\gamma_0+\gamma_1X,\sigma_{3}^{2})$, then one can verify integral equation $p(y|x)=\int h(w,y)p(w|x)dw$ has a unique solution $h(w,y)$:
\begin{equation}\label{eq.true_solution}
    h(w,y)=\frac{1}{\sigma_{wx}}\phi \left( \frac{y-\gamma_{wx}-\gamma_1/\beta_1 w}{\sigma_{wx}} \right),
\end{equation}
where $\phi$ is the probability density function (pdf) of the standard normal distribution, $\gamma_{wx}=\gamma_{0}-\gamma_{1}\beta_{0}/\beta_{1}$ and $\sigma_{wx}^2=\sigma_3^2-\gamma_1^2\sigma_2^2/\beta_1^2$.
\end{lemma}
\begin{proof}
The proof is similar to that in Example 1 of \cite{miao2018identifying} with $p(w|z, x)$ replaced by $p(w|x)$; and $p(y|z, x)$ replaced by $p(y|x)$.
\end{proof}
\begin{lemma}[Lemma 2.5 of \cite{beyhum2024testing}]\label{Lemm: Bounds on Compact Operator}
Let $(\mathcal{W},\|\cdot\|_{\cW})$ and $(\mathcal{X},\|\cdot\|_{\cX})$ be two Hilbert spaces and $A:\mathcal{W}\to \mathcal{X} $ be a linear compact operator with singular value decomposition given by $(s_n,u_n,v_{n})_{n=1}^{+\infty}$, $\|\cdot\|_{\mathrm{op}}^2$ be operator norm. Let $I: \cW \to \cW$ be the identity operator. For each $\lambda > 0$, we have the following results:
\begin{itemize}
    \item[(a)] $$\|A(\lambda I + A^*A)^{-1}A^*\|_{\mathrm{op}}\leq 1. $$
    \item[(b)] $$\|\lambda(\lambda I + A^*A)^{-1}\|_{\mathrm{op}}\leq 2. $$
    \item[(c)] $$\|(\lambda I + A^*A )^{-1}A^* \|_{\mathrm{op}}=\|A(\lambda I + A^*A )^{-1} \|_{\mathrm{op}}\leq \frac{1}{2\sqrt{\lambda}}.$$
    \item[(d)] For any $\gamma > 0$ and $g \in \cW$ such that $\|g\|_\gamma^2:=\sum_j s_j^{-2\gamma}|\langle g,u_j\rangle|^2<\infty$, there holds: 
    $$
    \|\lambda(\lambda I + A^*A)^{-1}g\|_{\cW}=O\left\{\lambda^{\frac{\min(\gamma,2)}{2}} \right\}.
    $$
\end{itemize} 
\end{lemma}

\begin{lemma}[Lemma 12 of \cite{mastouri2021proximal}]\label{Lem: rates for Khat and fhat}
Suppose conditions~\ref{ass:y_bounded} and \ref{ass:kernel_characteristic} hold for constants $c_Y$ and $\kappa$, respectively. Define $\sigma_f^2$ and $\sigma^2_A$ as follows:
\begin{equation*}
    \sigma^2_f := \mathbb{E}\{\|\varphi(Y,t)\phi_X(X)\|^2\}, \quad 
    \sigma^2_A := \mathbb{E}\{\|\phi_X(X)\|^2\|\phi_W(W)\|^2\}.
\end{equation*}
For $A, f$ defined in Eq.\eqref{eq.operator_defination}, and $A^*$ in \eqref{eq.operator_defination_adjoin}, the estimates $\wh{A}, \wh{f}$ given by \eqref{eq.estimate_f_K} satisfy the following properties with probability at least $1 - \delta$:
\begin{align*}
    \|\widehat{b}_t- b_t\|_{\cH_X}  &\leq \frac{2c_Y\kappa^3\log(2/\delta)}{n} + \sqrt{\frac{2\sigma^2_f\log(2/\delta)}{n}}=O_p\left(\frac{1}{\sqrt{n}}\right) \\
     \|\widehat{A} - A\|_{\mathrm{op}} &\leq \frac{2\kappa^6\log(2/\delta)}{n} + \sqrt{\frac{2\sigma^2_A\log(2/\delta)}{n}}=O_p\left(\frac{1}{\sqrt{n}}\right) \\
     \|\widehat{A}^* - A^*\|_{\mathrm{op}} &\leq \frac{2\kappa^6\log(2/\delta)}{n} + \sqrt{\frac{2\sigma^2_A\log(2/\delta)}{n}}=O_p\left(\frac{1}{\sqrt{n}}\right).
\end{align*}
\end{lemma}

\begin{lemma}\label{eq. f_hat-K_hatH}
Assume the conditions of Lemma~\ref{Lem: rates for Khat and fhat} hold. If $b_t=AH_{t}^0$, we have
$$
\|\widehat{b}_t - \widehat{A}H^0_t\|_{\cH_X} = O_p\left(\frac{1}{\sqrt{n}}\right).
$$
\end{lemma}
\begin{proof}
By Lemma~\ref{Lem: rates for Khat and fhat}, we can obtain $\|\widehat{b}_t-b_t\|_{\cH_X}= O_p(n^{-1/2})$ and $\|\widehat{A}-A\|_{\mathrm{op}}= O_p(n^{-1/2})$. Since $b_t = A H^0_t$, using the triangle inequality and the operator norm bound, we can obtain
\begin{align*}
    \|\widehat{b}_t-\widehat{A}H^0_t\|_{\cH_X}
    &= \|\widehat{b}_t - b_t + (A-\widehat{A})H^0_t\|_{\cH_X} = O_p(n^{-1/2}).
\end{align*}
We complete the proof. 
\end{proof}

\begin{lemma}[Lemma 13 of \cite{mastouri2021proximal}]\label{Lem: rates for Khat g - fhat}
Suppose conditions~\ref{ass:y_bounded} and \ref{ass:kernel_characteristic} hold. For $A, A^*$ defined respectively in \eqref{eq.operator_defination} and \eqref{eq.operator_defination_adjoin}, the estimates $\wh{A}$ given by \eqref{eq.estimate_f_K} satisfies:
$$
\|\widehat{A}^*\widehat{A} - A^*A\|_{\mathrm{op}} =O_p\left(\frac{1}{\sqrt{n}}\right).
$$ 
\end{lemma}

\begin{lemma}[Lemma 2.4 of \cite{beyhum2024testing}]\label{lemma:empirical process}
    For random variables $X, W$, let $m(\cdot)$ be the function such that $\mathbb{E}\{m(X)|W\}$ is bounded. Besides, we denote $\mathcal{F}$ as a class of functions of $W$ such that $\int_{0}^{1}\sqrt{N_{[ ]}(\epsilon,\mathcal{F},\|\cdot\|_{\cL^2\{F(w)\}})} d \epsilon < \infty $, where $N_{[ ]}(\epsilon,\mathcal{F},\|\cdot\|_{\cL^2\{F(w)\}})$ denotes the $\epsilon$-bracketing number under the $\cL^2\{F(w)\}$-norm. If $\|(\widehat{f}-f_{0})m\|_{\cL^2\{F(x,w)\}} = o_{p}(1)$ and $\mathbb{P}(\widehat{f}\in\mathcal{F})\to1$, then
    $$
    \sqrt{n} (\mathbb{P}_{n}-\mathbb{P}) \{(\widehat{f}-f_{0})m\}=o_{p}(1).
    $$
\end{lemma}

\begin{lemma}[Lemma 2.1 of \cite{li2003consistent}]\label{ central limit theorem}
    Let $Z_1(\cdot),\cdots, Z_n(\cdot)$ be independent and identically distributed zero mean random elements on $\mathcal{L}^{2}(\cS,\nu)$ such that $\mathbb \mE\{\|Z_{i}(\cdot)\|_{\mathcal{L}^{2}(\cS,\nu)}^{2}\} := \mE\left\{\int_s Z^2_i(s)  d\nu(s) \right\}<\infty$. Here, $\mathcal{L}^{2}(\cS,\nu)$ is square integrable function space with respect to the measure $\nu$. Then $n^{-1/2}\sum_{i=1}^nZ_i(\cdot)$ converges weakly to a zero mean Gaussian process with the covariance function given by $\Omega(s,s')=\mathbb E\{Z_i(s)Z_i(s')\}$.
\end{lemma}

\begin{lemma}\label{operator decomposition}
    For operators $A$ and $\widehat{A}$ and their adjoint $A^*$ and $\widehat{A}^*$, we have the following transformation:
       $$
    (\lambda I+\widehat{A}^*\widehat{A})^{-1}-(\lambda I+A^*A)^{-1}=(\lambda I+A^*A)^{-1}( A^*A-\widehat{A}^*\widehat{A})(\lambda I+\widehat{A}^*\widehat{A})^{-1}.
    $$
\end{lemma}
\begin{proof}
\begin{align*}
	&(\lambda I+\widehat{A}^*\widehat{A})^{-1}-(\lambda I+A^*A)^{-1}=I\cdot (\lambda I+\widehat{A}^*\widehat{A})^{-1}-(\lambda I+A^*A)^{-1}\cdot I\\
	=&(\lambda I+A^*A)^{-1}(\lambda I+A^*A)(\lambda I+\widehat{A}^*\widehat{A})^{-1}-(\lambda I+A^*A)^{-1}(\lambda I+\widehat{A}^*\widehat{A})(\lambda I+\widehat{A}^*\widehat{A})^{-1}\\
	=&(\lambda I+A^*A)^{-1}\{ (\lambda I+A^*A)-(\lambda I+\widehat{A}^*\widehat{A}) \} (\lambda I+\widehat{A}^*\widehat{A})^{-1}\\
	=&(\lambda I+A^*A)^{-1}(A^*A-\widehat{A}^*\widehat{A} )(\lambda I+\widehat{A}^*\widehat{A})^{-1}.
\end{align*}
\end{proof}

\begin{definition}[Definition 15.5 of \cite{kress1989linear}]\label{def:regularization-scheme}
Let $X$ and $Y$ be normed spaces and let $A:X\to X$ be an injective bounded linear operator. Then a family of bounded linear operators $R_\alpha : Y\to X,\alpha>0$,  with the property of pointwise convergence
$$
\lim_{\alpha\to 0} R_\alpha A\varphi = \varphi, \varphi\in X,
$$
is called a regularization scheme for the operator $A$. The parameter $\alpha$ is called the
regularization parameter.
\end{definition}
\begin{lemma}[Theorem 15.23 of \cite{kress1989linear}]\label{lem:tikhonov-regularization}
Let $A:X\to X$ be a compact linear operator. Then for each $\alpha>0$ the operator $\alpha I+A^*A: X\to X$ has a bounded inverse. Furthermore, if $A$ is injective, then $R_\alpha = (\alpha I+A^*A)^{-1}A^*$ describes a regularization scheme with $\|R_\alpha\|_{\mathrm{op}}\le 1/2\sqrt{\alpha}$.
\end{lemma}

\begin{lemma}\label{Lem: rates for Hhat - H}
Suppose that conditions~\ref{ass:y_bounded},~\ref{ass:kernel_characteristic}, and~\ref{assum.source condition} hold. The PMCR estimator $\widehat{H}^{\lambda}(w,t)$ satisfies
$$
\|\widehat{H}^{\lambda}(w,t)-H^0(w,t)\|_{\mathcal{H}_{W}}=O_p\left\{ \frac{1}{\sqrt{n\lambda}}+ \frac{1}{n\lambda}+\lambda^{\frac{\min(\theta,2)}{2}}\right\}.
$$
In particular, if condition~\ref{assum.bandwidth} holds, we have $\|\widehat{H}^{\lambda}(w,t)-H^0(w,t)\|_{\mathcal{H}_{W}}=o_p(1)$.
\end{lemma}

\begin{proof}
    We decompose the estimation bias into two parts:
    $$
    \|\widehat{H}^{\lambda}(w,t)-H^0(w,t)\|_{\mathcal{H}_{W}}\leq\|\widehat{H}^{\lambda}(w,t)-H^{\lambda}(w,t)\|_{\mathcal{H}_{W}}+\|H^{\lambda}(w,t)-H^0(w,t)\|_{\mathcal{H}_{W}}.
    $$
    We first consider $\|\widehat{H}^{\lambda}(w,t)-H^{\lambda}(w,t)\|_{\mathcal{H}_{W}}$. In fact, following the decomposition \eqref{eq: decomposition of phi.hat}, we have
    $$
    \widehat{H}^{\lambda}(w,t)-H^\lambda(w,t)=G_1+G_2+G_3+G_4,
    $$
    where $G_1,G_2,G_3,G_4$ are defined in \eqref{eq.G1}-\eqref{eq.G4}. For $G_1$, we can apply Lemma~\ref{Lemm: Bounds on Compact Operator} (c) to have $\| (\lambda I+A^*A)^{-1}A^*\|_{\mathrm{op}}=O_p(1/\sqrt{\lambda})$. Besides, according to Lemma \ref{eq. f_hat-K_hatH}, we have $\| \widehat{b}_t-\widehat{A}H^0_t\|_{\mathcal{H}_W}=O_p(1/\sqrt{n})$. Combining these together, we get
    $$
    \|G_1\|_{\mathcal{H}_W}\le \| (\lambda I+A^*A)^{-1}A^*\|_{\mathrm{op}}\cdot \| \widehat{b}_t-\widehat{A}H^0_t\|_{\mathcal{H}_W}= O_p\left( \frac{1}{\sqrt{n\lambda}} \right). 
    $$
    For $G_2$, we apply Lemma~\ref{Lemm: Bounds on Compact Operator} (b) to obtain that $\| (\lambda I+A^*A)^{-1}\|_{\mathrm{op}}=O_p(1/\lambda)$. Besides, according to Lemma \ref{eq. f_hat-K_hatH} and \ref{Lem: rates for Khat and fhat}, we have $\| \widehat{b}_t-\widehat{A}H^0_t\|_{\mathcal{H}_W}=O_p(1/\sqrt{n})$ and $\| \widehat{A}^*-A^*\|_{\mathrm{op}}=O_p(1/\sqrt{n})$. Combining these inequalities together, we have:
    $$
\| G_2\|_{\mathcal{H}_W}\le \| (\lambda I+A^*A)^{-1}\|_{\mathrm{op}}\cdot \| \widehat{A}^*-A^*\|_{\mathrm{op}}\cdot \| \widehat{b}_t-\widehat{A}H^0_t\|_{\mathcal{H}_W}= O_p\left( \frac{1}{n\lambda} \right).
$$
For $G_3$, we have:
\begin{align*}
	\|G_3\|_{\mathcal{H}_W}&\le \| \{(\lambda I+\widehat{A}^*\widehat{A})^{-1}-(\lambda I+A^*A)^{-1}\}\widehat{A}^*\|_{\mathrm{op}}\cdot \| \widehat{b}_t-\widehat{A}H^0_t\|_{\mathcal{H}_W}\\
	&= \| (\lambda I+\widehat{A}^*\widehat{A})^{-1}\widehat{A}^*-(\lambda I+A^*A)^{-1}A^*-(\lambda I+A^*A)^{-1}(\widehat{A}^*-A^*)\|_{\mathrm{op}}\cdot \| \widehat{b}_t-\widehat{A}H^0_t\|_{\mathcal{H}_W}\\
	&\le \| (\lambda I+\widehat{A}^*\widehat{A})^{-1}\widehat{A}^*-(\lambda I+A^*A)^{-1}A^*\|_{\mathrm{op}}\cdot \| \widehat{b}_t-\widehat{A}H^0_t\|_{\mathcal{H}_W}\\
	& \qquad \qquad +\| (\lambda I+A^*A)^{-1}\|_{\mathrm{op}}\cdot \| \widehat{A}^*-A^*\|_{\mathrm{op}}\cdot \| \widehat{b}_t-\widehat{A}H^0_t\|_{\mathcal{H}_W}.
\end{align*}
Since $\widehat{A}$ and $A$ are compact operators, we can apply Lemma~\ref{Lemm: Bounds on Compact Operator} (b), (c) to obtain that $\| (\lambda I+\widehat{A}^*\widehat{A})^{-1}\widehat{A}^*-(\lambda I+A^*A)^{-1}A^*\|_{\mathrm{op}}=O_p(1/\lambda)$ and $\| (\lambda I+A^*A)^{-1}\|_{\mathrm{op}}=O_p(1/\lambda)$. Besides, according to Lemma \ref{Lem: rates for Khat and fhat} and \ref{eq. f_hat-K_hatH}, we have $\| \widehat{b}_t-\widehat{A}H^0_t\|_{\mathcal{H}_W}=O_p(1/\sqrt{n})$ and $\| \widehat{A}^*-A^*\|_{\mathrm{op}}=O_p(1/\sqrt{n})$. Combining all the inequalities, we get
$$
\|G_3\|_{\mathcal{H}_W}= O_p\left( \frac{1}{\sqrt{n\lambda}} \right)+O_p\left( \frac{1}{n\lambda} \right).
$$
For $G_4$, we have:
\begin{align*}
	\| G_4\|_{\mathcal{H}_W}&=\| (\lambda I+\widehat{A}^*\widehat{A})^{-1}\widehat{A}^*\widehat{A}H^0_t-(\lambda I+A^*A)^{-1}A^*AH^0_t\|_{\mathcal{H}_W}\\
	&\overset{(1)}{=} \| \lambda (\lambda I+\widehat{A}^*\widehat{A})^{-1}\{\widehat{A}^*\widehat{A}-A^*A\}(\lambda I+A^*A)^{-1}H^0_t\|_{\mathcal{H}_W}\\
	&=\| \lambda (\lambda I+\widehat{A}^*\widehat{A})^{-1}\{\widehat{A}^*(\widehat{A}-A)+(\widehat{A}^*-A^*)A\}(\lambda I+A^*A)^{-1}H^0_t\|_{\mathcal{H}_W}\\
	&\le \| \lambda (\lambda I+\widehat{A}^*\widehat{A})^{-1}\widehat{A}^*(\widehat{A}-A)(\lambda I+A^*A)^{-1}H^0_t\|_{\mathcal{H}_W}\\
	& \qquad \qquad +\| \lambda (\lambda I+\widehat{A}^*\widehat{A})^{-1}(\widehat{A}^*-A^*)A(\lambda I+A^*A)^{-1}H^0_t\|_{\mathcal{H}_W}\\
	&\le \| (\lambda I+\widehat{A}^*\widehat{A})^{-1}\widehat{A}^*\|_{\mathrm{op}}\cdot\| \widehat{A}-A\|_{\mathrm{op}}\cdot\| \lambda (\lambda I+A^*A)^{-1}\|_{\mathrm{op}}\cdot \| H^0_t\|_{\mathcal{H}_W}\\
	& \qquad \qquad +\| \lambda (\lambda I+\widehat{A}^*\widehat{A})^{-1}\|_{\mathrm{op}}\cdot\| \widehat{A}^*-A^*\|_{\mathrm{op}}\cdot\| A(\lambda I+A^*A)^{-1}\|_{\mathrm{op}}\cdot \| H^0_t\|_{\mathcal{H}_W},
\end{align*}
where (1) follows from \eqref{eq. P_t}. Since $\widehat{A}$ and $A$ are compact operators, we can apply Lemma~\ref{Lemm: Bounds on Compact Operator} (b), (c) to obtain that $\| (\lambda I+\widehat{A}^*\widehat{A})^{-1}\widehat{A}^*\|_{\mathrm{op}}= O_p(1/\sqrt{\lambda})$, $\| (\lambda I+A^*A)^{-1}A^*\|_{\mathrm{op}}=O_p(1/\sqrt{\lambda})$, $\| \lambda (\lambda I+A^*A)^{-1}\|_{\mathrm{op}} \le2$, $\| \lambda (\lambda I+\widehat{A}^*\widehat{A})^{-1}\|_{\mathrm{op}}\le2$. Besides, according to Lemma~\ref{Lem: rates for Khat and fhat}, we have $\| \widehat{A}^*-A^*\|_{\mathrm{op}}=\| \widehat{A}-A\|_{\mathrm{op}}=O_p(1/\sqrt{n})$. Combining all the inequalities, we get:
$$
\| G_4\|_{\mathcal{H}_W}= O_p\left( \frac{1}{\sqrt{n\lambda}}\right).
$$
Combining these results for $G_1$ to $G_4$, we have
$$
\|\widehat{H}^{\lambda}(w,t)-H^{\lambda}(w,t)\|_{\mathcal{H}_{W}}=O_p\left( \frac{1}{\sqrt{n\lambda}} + \frac{1}{n\lambda} \right).
$$
Next, we consider $\|H^{\lambda}(w,t)-H^0(w,t)\|_{\mathcal{H}_{W}}$. By condition~\ref{assum.source condition} (b), we can employ Lemma~\ref{lem.H_laambda-H_0} to obtain that:
$$
\|H^{\lambda}(w,t)-H^0(w,t)\|_{\mathcal{H}_{W}}= O_p\left(\lambda^{\frac{\min(\theta,2)}{2}} \right).
$$
Thus, we have
$$
\|\widehat{H}^{\lambda}(w,t)-H^0(w,t)\|_{\mathcal{H}_{W}}=O_p\left\{ \frac{1}{\sqrt{n\lambda}}+ \frac{1}{n\lambda}+\lambda^{\frac{\min(\theta,2)}{2}}\right\}.
$$
By condition~\ref{assum.bandwidth}, we have $n\lambda \to \infty$ and $\lambda \to 0$, which gives $\|\widehat{H}^{\lambda}(w,t)-H^0(w,t)\|_{\mathcal{H}_{W}}=o_p(1)$. 
\end{proof}

\begin{lemma}\label{lem.H_laambda-H_0}
If $H^0(w,t)$ is the least norm solution to the linear inverse problem and satisfies condition~\ref{assum.source condition} (b), then the solution to the Tikhonov regularization $H^\lambda(w,t)$ satisfies that:
$$
\|H^\lambda(w,t)-H^0(w,t)\|^2_{\mathcal{H}_W}\leq O_p\{\lambda^{\min(\theta,2)}\}.
$$
\end{lemma}
\begin{proof}
For the operator $A:\cH_W\to\cH_X$ defined in ~\eqref{eq.operator_defination}, its singular value decomposition given by $(s_n,u_n,v_n)_{n=1}^{+\infty}$. Thus, we have $H^0_t=\sum_{j}\langle H^0_t,u_j\rangle_{\cH_W}u_j$. Besides, according to $Au_n=s_nv_n$ and $A^*v_n=s_nu_n$, we have $H^{\lambda}_t=( A^*A+\lambda I)^{-1}A^*b_t=\sum_{j}\frac{s^2_j}{s^2_j+\lambda}\langle H^0_t,u_j\rangle_{\cH_W}u_j$. Thus, we have
\begin{align*}
	\| H^{\lambda}(w,t)-H^0(w,t)\|_{\mathcal{H}_W}^{2}&=\left\| \sum_j{\left( \frac{s_j^2}{s_j^2+\lambda}-1 \right) \langle H_{t}^{0},u_j\rangle_{\mathcal{H}_W}u_j} \right\|_{\mathcal{H}_W}^{2}\\
	&=\sum_j{\left\{ \left( \frac{s_j^2}{s_j^2+\lambda}-1 \right) \langle H_{t}^{0},u_j\rangle_{\mathcal{H}_W} \right\}^2}\\
	&=\sum_j{\frac{\lambda^2s_{j}^{2\theta}}{( s_j^2+\lambda)^2}\frac{| \langle H_{t}^{0},u_j\rangle_{\mathcal{H}_W}|^2}{s_{j}^{2\theta}}}\\
	&\le \underset{j}{\sup}\left( \frac{\lambda s_{j}^{\theta}}{s_j^2+\lambda} \right)^2\sum_j{\frac{| \langle H_{t}^{0},u_j\rangle_{\mathcal{H}_W}|^2}{s_{j}^{2\theta}}}.
\end{align*}
Applying condition~\ref{assum.source condition} (b) for $\theta \ge 2$,  and the maximum singular value of the operator equals $\|A\|_{\mathrm{op}}<\infty$, we have 
$$
\underset{j}{\mathrm{sup}}\left( \frac{\lambda s_{j}^{\theta}}{s_j^2+\lambda} \right)^2=\lambda^2\underset{j}{\mathrm{sup}}\left( \frac{s_{j}^{\theta}}{s_j^2+\lambda} \right)^2\le \lambda^2\underset{j}{\mathrm{sup}}s_{j}^{2\theta-4} =  O(\lambda^2).
$$
For $0<\theta<2$, we define $x=\lambda_j^2$ and $f(x) =\frac{\lambda^2x^{\theta}}{( x+\lambda)^2}$. Noted that $f(x)$ is maximized (by using the first order condition) at $x=\lambda\theta(2-\theta)^{-1}$. Thus, the maximum value of $f(x)$ is
$$
\frac{x^{\theta}\lambda^2}{(x+\lambda )^2}\le \lambda^{\theta}\frac{\theta^{\theta}(2-\theta )^{2-\theta}}{4}\le O( \lambda^{\theta}). 
$$
The proof is complete.
\end{proof}

\textbf{Hermite polynomial.} We introduce the concept of Hermite polynomial, which is defined in the square-integrable function space with respect to the standard Gaussian measure. Specifically, we say that a function $f: \mathbb{R}\to \mathbb{R}$ is square integrable w.r.t. the standard Gaussian measure $\gamma=e^{-x^2/2}/\sqrt{2\pi}$ if $\mathbb{E}_{x\sim \cN(0,1)}\{f^2(x)\}<\infty$. We denote by $\mathcal{L}^2\{\Phi(X)\}$ the space of all such functions, whose basis functions are characterized by probabilist’s Hermite polynomials
\begin{equation}
\label{eq.Hermite}
    \mathrm{He}_{n}(x):=(-1)^ke^{x^2/2}\frac{d^k}{dx^k}e^{-x^2/2}.
\end{equation}
The first three Hermite polynomials are $\mathrm{He}_{0}(x)=1,\mathrm{He}_{1}(x)=x,\mathrm{He}_{2}(x)=x^2-1$. Let 
\begin{equation}
\label{eq.Hermite_norm}
    \mathrm{he}_{k}(x):=\frac{\mathrm{He}_{k}(x)}{\sqrt{k!}}
\end{equation}
denote the normalized Hermite polynomials, which form a complete orthonormal basis in $\mathcal{L}^2\{\Phi(X)\}$. Thus, the Hermite expansion of a function $f\in \mathcal{L}^2\{\Phi(X)\}$ is given by
$$
f(x)=\sum_{k=1}^{\infty}\mu_{k-1}(f)\mathrm{he}_{k-1}(x), \ \mu_{k-1}(f)=\mathbb{E}_{X\sim\mathcal{N}(0,1)}\{f(X)\mathrm{he}_{k-1}(X)\}.
$$
Besides, Hermite polynomials can be equivalently defined by identifying 
\begin{equation}\label{eq.Hermite_generate}
    e^{xt-t^2/2}=\sum_{k=0}^\infty\frac{\mathrm{He}_{n}(x)}{k!}t^k.
\end{equation}
We are now ready to introduce the eigenvalue system of the operator $T: \mathcal{L}^2\{\Phi(W)\} \rightarrow  \mathcal{L}^2\{\Phi(X)\}$ derived by \cite{carrasco2007linear}. 
\begin{lemma}[\cite{carrasco2007linear}]\label{lem.self-adjoint operator}
    Let $T: \mathcal{L}^2\{\Phi(W)\}\rightarrow  \mathcal{L}^2\{\Phi(X)\}, Tf =\mathbb{E}\{ f(W)|X=\cdot\}$, where $\mathcal{L}^2(\cdot)$ is square integrable space with respect to the standard Gaussian measure, \emph{i.e.}, $(W, X)$ is jointly Gaussian with zero mean, unit variance, and correlation $\rho_{WX}$. We have $T$ is a self-adjoint operator, and the eigenvalue system for $T$ is given by $\varphi_j(w)=\mathrm{he}_{j}(w),\phi_j(x)=\mathrm{he}_{j}(x),\lambda_j=\rho_{WX}^{j}$, where $\rho_{WX}$ is the correlation coefficient between $W$ and $X$ and $\mathrm{he}_{j}$ is the normalized Hermite polynomials.    
\end{lemma}
\begin{lemma}[\cite{nevai2006orthogonal}]\label{lem:mehler_physicist}
Let $ H_n(x) $ denote the physicist's Hermite polynomials, and let $ x, y \in \mathbb{R} $. For $ w \in \mathbb{R}$, consider the series
$$
\sum_{n=0}^\infty \frac{H_n(x) H_n(y)}{n!} \left( \frac{w}{2} \right)^n.
$$
The following hold:
\begin{enumerate}
    \item \textbf{Convergence.} The series converges absolutely if and only if $ |w| < 1 $.
    \item \textbf{Closed Form.} For $ |w| < 1 $, the series has the closed-form expression
    $$
    \sum_{n=0}^\infty \frac{H_n(x) H_n(y)}{n!} \left( \frac{w}{2} \right)^n = \frac{1}{\sqrt{1 - w^2}} \exp \left\{ \frac{2xy w - (x^2 + y^2) w^2}{1 - w^2} \right\}.
    $$
\end{enumerate}
\end{lemma}
\section{Additional experiments}
\label{appx.add_experiment}
In this section, we evaluate the effectiveness of our procedures in other settings. In section~\ref{appx.discrete}, we consider randomized setting in the discrete case, where probability distribution varies for each time. Next, we evaluate our method in the presence of observed covariates in section~\ref {appx.observed}. 
Finally, in section~\ref{appx.two-proxy}, we examine the benefits of leveraging additional NCE in a nonlinear setting. 


\subsection{Discrete setting}
\label{appx.discrete}
We first evaluate our method in the setting where all variables are discrete. 

\textbf{Data generation.} Suppose $X,U,W,Y$ are discrete variables with $|\cW|=5,|\cU|=5,|\cX|=7,|\cY|=4$, and their generations follow from $U \to X$, $U \to W$, $U \to Y$, and additionally $X \to Y$ if $\mH_1$ holds. We then generate samples from specified $P(U), P(W|U), P(X|U)$, and $P(Y|U)$ (\emph{resp}. $P(Y|U,X)$) under $\mH_0$ (\emph{resp}. $\mH_1$). To mitigate the effect of randomness, we repeat the process $20$ times, where each time has a different probability specification. At each time, we generate 100 replications under each $\mathbb{H}_0$ and $\mathbb{H}_1$, and record the average type-I error rate and power rate.

\begin{figure}[htbp]
\centering
\includegraphics[width=0.95\linewidth]{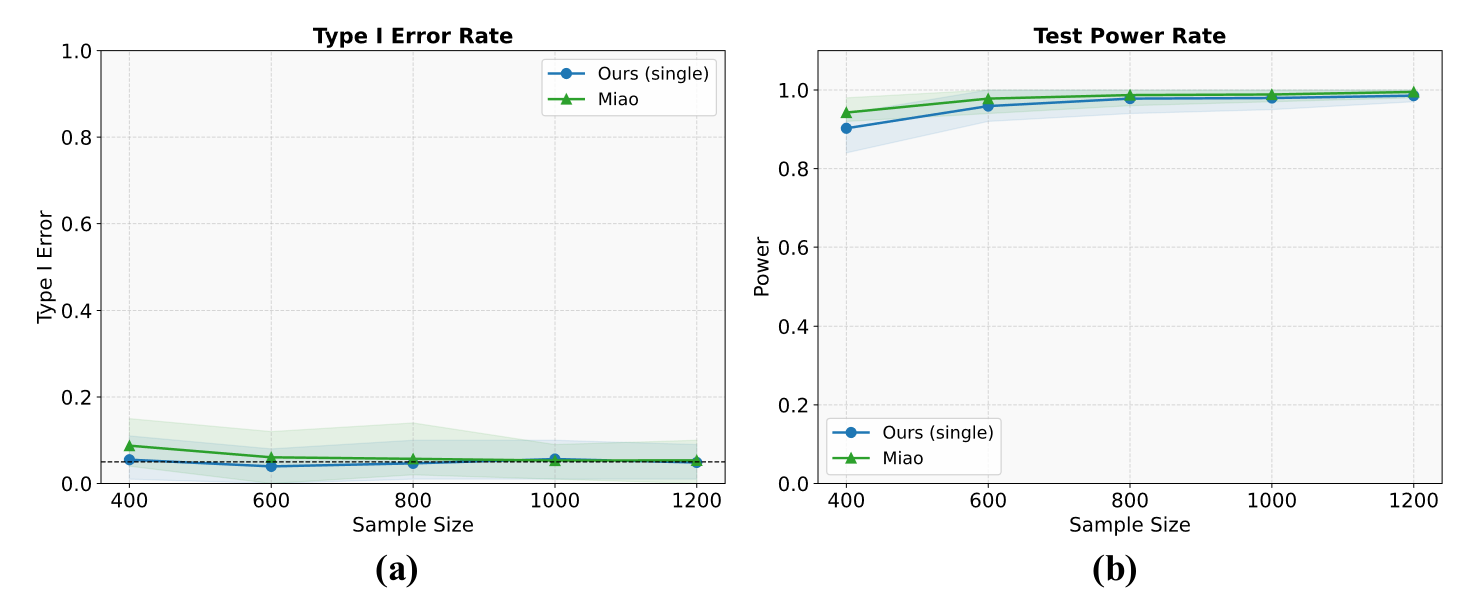}
\caption{Type-I error rate (left) and power rate (right) of our procedure and the Miao's method in discrete random setting.}
\label{fig.random}
\end{figure}

\textbf{Type-I error and power.} In Figure~\ref{fig.random}, we present the average type-I error rate and power rate for our testing procedure and others. As shown, our power approximates one as $n$ increases. Besides, the type-I error closely approximates the significance level (\emph{i.e.}, $0.05$) as $n$ increases.

\begin{figure}[htbp]
\centering
\includegraphics[width=0.6\linewidth]{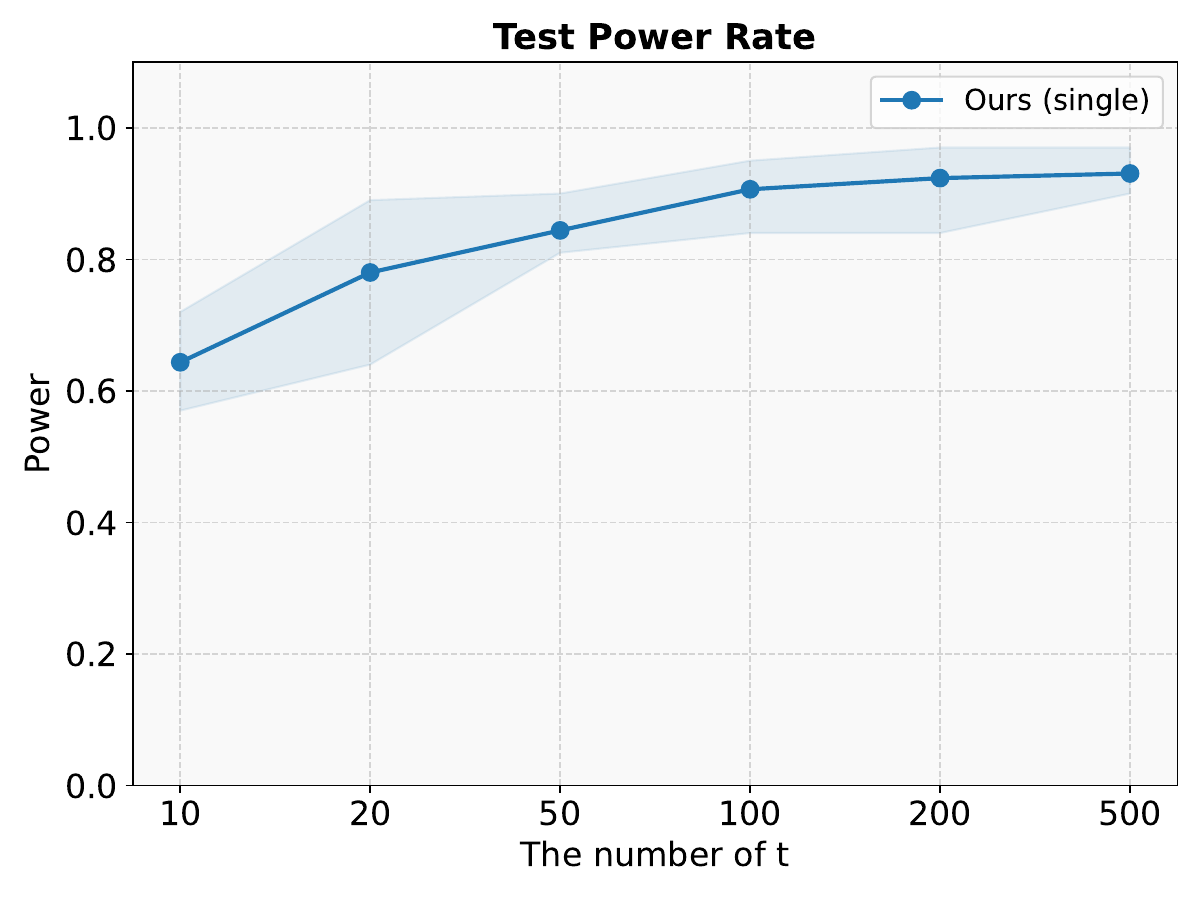}
\caption{Power of $\Delta_\varphi$ \eqref{eq.icm-statistics_dis} with respect to the number of evaluation points $t$ under the discrete setting.}
\label{fig.power_t}
\end{figure}

\textbf{Effect of the number of $t$ in computing $\Delta_\varphi$ \eqref{eq.icm-statistics_dis}.} To further examine how the number of evaluation points $t$ in $\varphi(y,t)$ affects the test statistic, we assess the empirical power under $\mH_1$ for $t \in \{10, 20, 50, 100, 200, 500\}$, while keeping the sample size fixed at $400$. As illustrated in Figure~\ref{fig.power_t}, the test power increases as $t$ grows.

\subsection{Observed covariates setting}\label{appx.observed}
Next, we further evaluate the performance of different methods in the presence of observed covariates. Data are generated from the following model, where we need to adjust for a covariate $V$ for testing the null hypothesis. Since the implementation of \textbf{Liu} does not support covariate adjustment, we omit it from the comparison.

\textbf{Data generation.} Following \cite{ying2025generalized}, we generate dateset by $X=0.5+U+0.3U^2+0.5V+\varepsilon_1$, $Y=-1+U+0.4U^2+V+\delta X+\varepsilon_2$, and $W=1+U+0.5V+\varepsilon_3$, where $(U,V,\varepsilon_1,\varepsilon_2,\varepsilon_3)\sim N(0,I_5)$. Under $\mH_0$, $\delta=0$; otherwise, $\delta=1$. We repeat the process $20$ times, where we generate 100 replications at each time under each $\mathbb{H}_0$ and $\mathbb{H}_1$. 

\begin{figure}[htbp]
\centering
\includegraphics[width=0.95\linewidth]{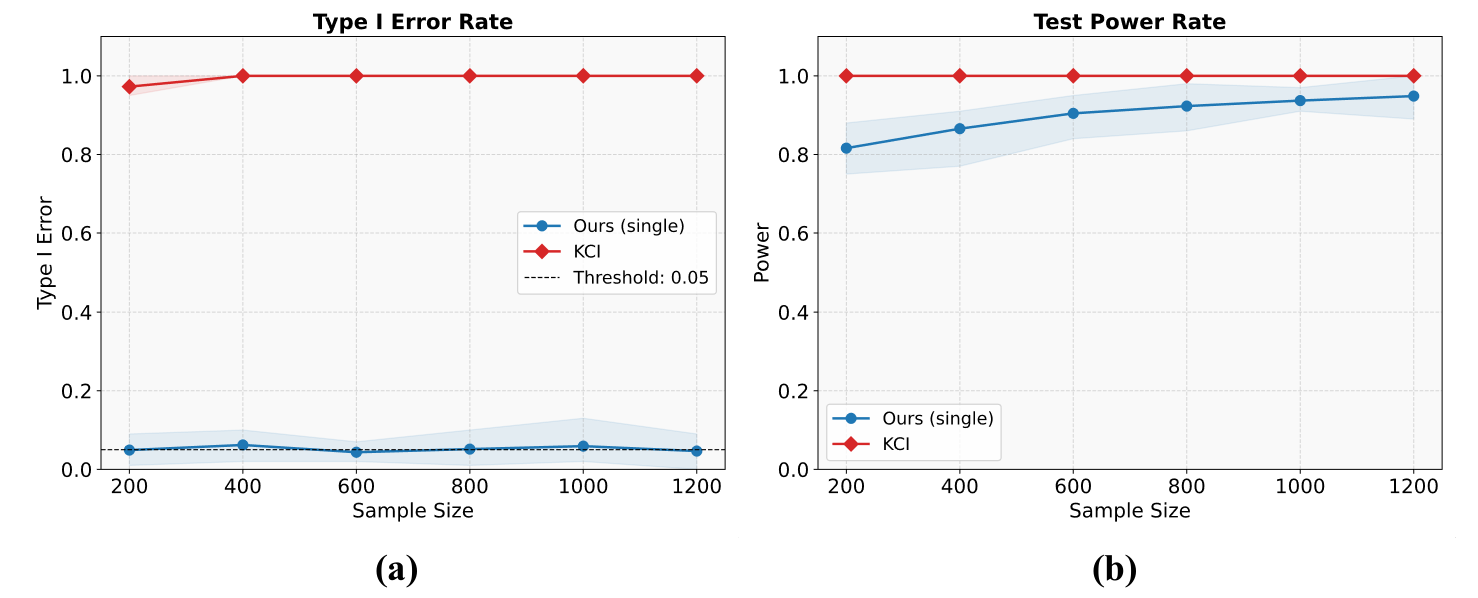}
\caption{Type-I error rate (left) and power rate (right) of our procedure and the KCI's method in observed setting.}
\label{fig.observed}
\end{figure}

\textbf{Type-I error and power.} The results are presented in Figure~\ref{fig.observed}, which are similar to that in \ref{sec.exp-single} without observed covariates. Our testing statistics can approximately control the type I error, and have power approaching to one as the sample size increases. 

\subsection{Two-Proxy procedure in the nonlinear setting}
\label{appx.two-proxy}

Finally, we evaluate our two-proxy procedure to a nonlinear setting, where $W\to Y$ and both $W$ and $Z$ are available. 


\textbf{Data generation.} We generate $U$ via $U\sim \cN(0,1)$. For negative controls, we generate data from $W=-2\sin (U) +\varepsilon_W$ and $Z=2\sin (U) +\varepsilon_Z$. The treatment assignment mechanism follows the generation process: $X=2\sin (U) +\varepsilon_X$. Under $\mH_1: X\not\ind Y|U$, the outcome is generated from $Y=X+\sin (U) +2W^2+\varepsilon_Y$; while under $\mH_0: X\ind Y|U$, the outcome is generated from $Y=\sin (U) +2W^2+\varepsilon_Y$.  In both hypotheses, the noise terms $\varepsilon_X,\varepsilon_Z,\varepsilon_W,\varepsilon_Y$ are independently drawn from a standard normal distribution. We repeat the process 20 times, where at each time we generate 100 replications under $\mH_0$ and $\mH_1$.

\begin{figure}[htbp]
\centering
\includegraphics[width=0.95\linewidth]{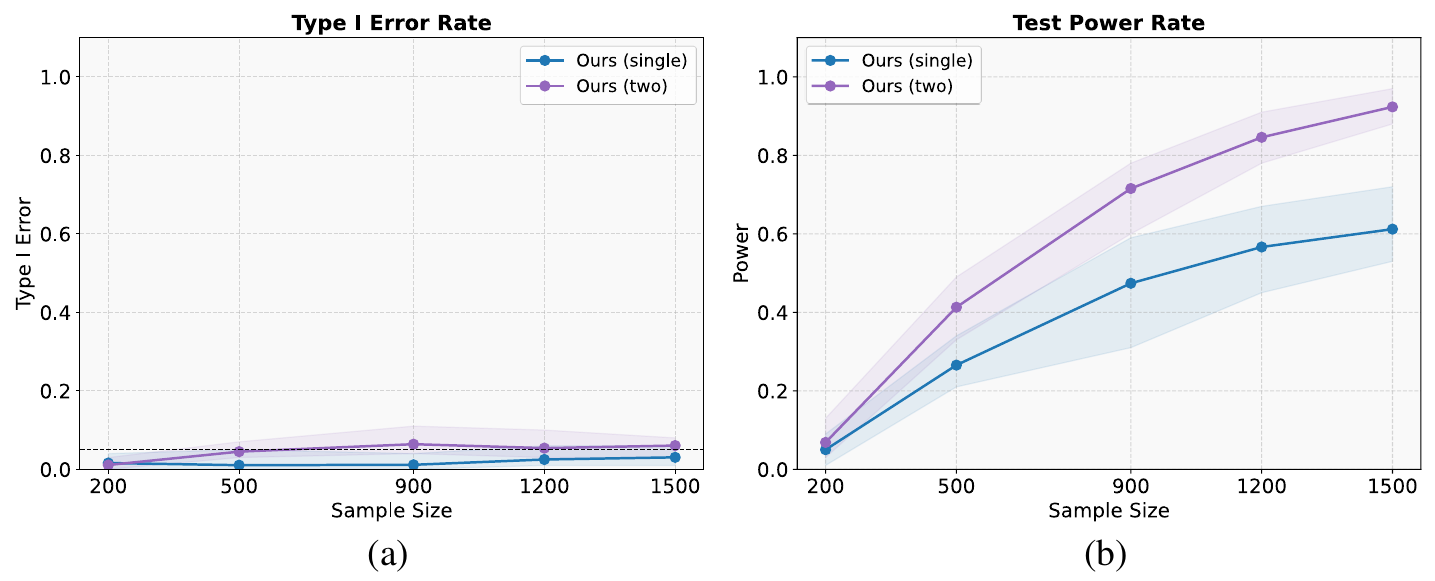}
\caption{Type-I error rate (left) and power rate (right) of our procedure and baselines in the nonlinear setting with two proxies. }
\label{fig.cmp_baselines_two_nonlinear}
\end{figure}

\textbf{Type-I error and power.} The average results are presented in Figure~\ref{fig.cmp_baselines_two_nonlinear}. As observed, while our single-proxy procedure effectively controls the type-I error, it exhibits low power in identifying causal relationships. By incorporating additional restriction from the NCE, the power improves significantly.

\end{document}